\documentclass[12pt]{article}%
\pdfoutput=1

\usepackage{rotating}
\usepackage{cite}
\usepackage[all]{xy}
\usepackage{amsmath,amssymb,amsthm,amsfonts,mathrsfs}
\theoremstyle{definition}

\usepackage{mathtools}
\usepackage{amsfonts}
\usepackage{amssymb}
\usepackage{authblk}
\usepackage{bbm}
\usepackage{bm}
\usepackage{caption}
\usepackage{epsfig}
\usepackage{epstopdf}
\usepackage{heck}
\usepackage[margin=1in]{geometry}
\usepackage{tabularx}
\usepackage[usenames,dvipsnames]{xcolor}
\usepackage{graphicx}
\numberwithin{equation}{section}
\usepackage{tikz}
\usetikzlibrary{decorations.pathreplacing}
\usetikzlibrary{calc}

\usepackage{hyperref}
\usepackage[nameinlink,capitalize]{cleveref}

\makeatletter
\newcommand*{\diff}{\@ifnextchar^{\DIFF}{\DIFF^{}}}
\def\DIFF^#1{\mathop{\mathrm{\mathstrut d}}\nolimits^{#1}\gobblespace}
\newcommand*{\bigdiff}{\@ifnextchar^{\BIGDIFF}{\BIGDIFF^{}}}
\def\BIGDIFF^#1{\mathop{\mathrm{\mathstrut \mathcal{D}}}\nolimits^{#1}\gobblespace}
\def\gobblespace{\futurelet\diffarg\opspace}
\def\opspace{%
	\let\DiffSpace\!%
	\ifx\diffarg(%
		\let\DiffSpace\relax
	\else
		\ifx\diffarg[%
			\let\DiffSpace\relax
		\else
			\ifx\diffarg\{%
				\let\DiffSpace\relax
			\fi
		\fi
	\fi
	\DiffSpace
}
\makeatother

\newcommand{\orb}{{\mathbf X}_\Gamma}
\newcommand{\TM}{\mathcal T_{\orb}}
\newcommand{\Hom}{\mathrm{Hom}}
\newcommand{\bbC}{\mathbb{C}}
\newcommand{\bbD}{\mathbb{D}}
\newcommand{\bbZ}{\mathbb{Z}}

\begin{document}

\date{January 2022}

\title{Higher Symmetries of 5d Orbifold SCFTs}

\institution{UPPSALA}{\centerline{$^{1}$Mathematics Institute, Uppsala University, Box 480, SE-75106 Uppsala, Sweden}}
\institution{UPPSALAphys}{\centerline{$^{2}$Department of Physics and Astronomy, Uppsala University, Box 516, SE-75120 Uppsala, Sweden}}
\institution{PENN}{\centerline{$^{3}$Department of Physics and Astronomy, University of Pennsylvania, Philadelphia, PA 19104, USA}}
\institution{PENNmath}{\centerline{$^{4}$Department of Mathematics, University of Pennsylvania, Philadelphia, PA 19104, USA}}
\institution{SISSA}{\centerline{$^{5}$Theoretical Particle Physics group, SISSA, Via Bonomea 1, 34100 Trieste, Italy}}

\authors{
Michele Del Zotto\worksat{\UPPSALA,\UPPSALAphys}\footnote{e-mail: \texttt{michele.delzotto@math.uu.se}},
Jonathan J. Heckman\worksat{\PENN,\PENNmath}\footnote{e-mail: \texttt{jheckman@sas.upenn.edu}},\\[4mm]
Shani Nadir Meynet\worksat{\UPPSALA,\SISSA}\footnote{e-mail: \texttt{smeynet@sissa.it}},
Robert Moscrop\worksat{\UPPSALA}\footnote{e-mail: \texttt{robert.moscrop@math.uu.se}}, and
Hao Y. Zhang\worksat{\PENN}\footnote{e-mail: \texttt{zhangphy@sas.upenn.edu}}
}

\abstract{We determine the higher symmetries of 5d SCFTs engineered from M-theory on a $\mathbb{C}^3 / \Gamma$
background for $\Gamma$ a finite subgroup of $SU(3)$. This resolves a longstanding question as to how to extract this data when the resulting singularity is non-toric (when $\Gamma$ is non-abelian) and/or not isolated (when the action of $\Gamma$ has fixed loci).
The BPS states of the theory are encoded in a 1d quiver quantum mechanics gauge theory which determines the possible 1-form and 2-form symmetries. We also show that this same data can also be extracted by a direct computation of the corresponding defect group associated with the orbifold singularity. Both methods agree, and these computations do not rely on the existence of a resolution of the singularity. We also observe that when the geometry faithfully captures the global 0-form symmetry, the abelianization of $\Gamma$ detects a 2-group structure (when present). As such, this establishes that all of this data is indeed intrinsic to the superconformal fixed point rather than being an emergent property of an IR gauge theory phase.}

\maketitle

\setcounter{tocdepth}{2}

\tableofcontents

\newpage

\section{Introduction}

Higher-form symmetries \cite{Gaiotto:2014kfa} provide a powerful way to constrain the non-perturbative data of a quantum field theory \cite{Kapustin:2013uxa,Sharpe:2015mja,Gaiotto:2017yup,Gaiotto:2017tne,Cordova:2018cvg,Cordova:2020tij,Brennan:2020ehu,Kaidi:2021gbs,Lee:2021crt,Choi:2021kmx,Kaidi:2021xfk}.
This is especially valuable in the case of $d > 4$ superconformal field theories since all known examples are intrinsically strongly
coupled. Indeed, the main method to construct such examples proceeds by taking a singular limit of a string / M-theory / F-theory compactification. With this in mind, it is important to extract the corresponding data of higher-form symmetries for
such systems directly from the singular geometry of a string compactification \cite{DelZotto:2015isa,Heckman:2017uxe,Eckhard:2019jgg,GarciaEtxebarria:2019caf,Albertini:2020mdx,Morrison:2020ool,Dierigl:2020myk,Closset:2020scj,DelZotto:2020esg,Apruzzi:2020zot,Bhardwaj:2020phs,Closset:2020afy,Heidenreich:2020pkc,DelZotto:2020sop,Gukov:2020btk,Bah:2020uev,Bhardwaj:2021pfz,Apruzzi:2021mlh,Apruzzi:2021phx,Hosseini:2021ged,Apruzzi:2021vcu,Bhardwaj:2021wif,Bhardwaj:2021zrt,Closset:2021lwy,Heidenreich:2021xpr,Buican:2021xhs,Cvetic:2021maf, Debray:2021vob,Apruzzi:2021nmk,Braun:2021sex,Bah:2021brs,Bhardwaj:2021mzl,Cvetic:2020kuw,Cvetic:2021sxm}.

In this paper we determine the higher-form symmetries for 5d superconformal field theories (SCFTs) which originate from an orbifold singularity $\orb = \mathbb{C}^3 / \Gamma$ for $\Gamma$ a finite subgroup of $SU(3)$. We denote the resulting 5d SCFTs as $\TM$.
There is a full classification of finite subgroups of $\Gamma$ (including their group actions) which result in
Gorenstein Calabi-Yau threefold singularities \cite{yau1993gorenstein} (see also \cite{watanabe1982invariant}).
It also gives rise to a large class of well-known 5d SCFTs. For example, the trinion theory $T_N$ with flavor symmetry algebra
$\mathfrak{su}(N)^3$ arises from the singularity $\mathbb{C}^{3} / \mathbb{Z}_N \times \mathbb{Z}_N$ (see \cite{Benini:2009gi}).
Recently the physics and geometry of many such singularities were studied in references \cite{Tian:2021cif, Acharya:2021jsp}. For
further discussion of higher-form symmetries in 5d SCFTs, see in particular \cite{Albertini:2020mdx,Morrison:2020ool,BenettiGenolini:2020doj,Apruzzi:2021vcu,Genolini:2022mpi}. For additional
background on geometric engineering and 5d SCFTs, see \cite{Seiberg:1996bd, Katz:1996xe, Witten:1996qb, Morrison:1996xf, Douglas:1996xp, Katz:1996fh, Intriligator:1997pq, Aharony:1997ju, Aharony:1997bh, Diaconescu:1998cn, Bergman:2012kr, Bergman:2013koa} as well as \cite{DelZotto:2017pti, Jefferson:2018irk,Closset:2018bjz, Bhardwaj:2018yhy, Bhardwaj:2018vuu, Bhardwaj:2019fzv, Apruzzi:2019vpe, Apruzzi:2019opn, Apruzzi:2019enx}.

The higher symmetries of 5d SCFTs that have gauge theory phases can be determined directly from the corresponding Lagrangian description, exploiting standard techniques \cite{Gaiotto:2014kfa} --- with the subtlety that it can happen that 5d instantons are charged with respect to the center symmetry in the presence of a non-zero CS level. There are, however, many 5d SCFTs which do not have a gauge theory phase, and instead are defined purely by singular geometry. A pivotal example of this type is the famous $E_0$ theory \cite{Seiberg:1996bd}, which is realized as the
singular limit of the local Calabi-Yau threefold $O(-3) \rightarrow \mathbb{P}^2$, namely the orbifold $\mathbb C^3 / \mathbb Z_3$ \cite{Morrison:1996xf}. For these theories an alternative route to compute the corresponding higher form symmetries is given by exploiting the defect group of M-theory on the corresponding singularity \cite{DelZotto:2015isa,Albertini:2020mdx,Morrison:2020ool}. For instance proceeding in this way one can show that for the $E_0$ theory, the defect group is:
\begin{equation}
    \mathbb D(E_0) \supset (\mathbb Z_3)^{(1)}_e \oplus (\mathbb Z_3)^{(2)}_m\,,
\end{equation}
where the subscripts and
the superscripts refer to the fact that we have an electric 1-form symmetry and a magnetic 2-form symmetry. The one-form electric symmetry arises from M2-branes wrapped on two-cycles, and the two-form magnetic symmetry similarly arises from wrapped M5-branes on four-cycles. The two are related to different choices of global structures for the $E_0$ theory \cite{Albertini:2020mdx}.

So long as the singularity is isolated, it is straightforward to read off the corresponding electric one-form symmetry via the abelianization $\mathrm{Ab}[\pi_{1}(\partial \mathbb{C}^3 / \Gamma)]$, much as was done in the case of the 6d defect group in \cite{DelZotto:2015isa}. If, however, the group action $\Gamma$ results in a non-isolated singularity, then the boundary $\partial \mathbb{C}^{3} / \Gamma$ will also have singularities. For toric singularities, this problem was resolved in \cite{Albertini:2020mdx}.
For more general orbifold singularities, however, it is still an open question as to how to read off the resulting higher-form symmetries
directly from the singularity.

Our aim in this paper will be to present two complementary solutions to the computation of higher-form symmetries for
such 5d orbifold SCFTs. First of all, there is a well-defined notion of the fundamental group $\pi_{1}(\partial \mathbb{C}^3 / \Gamma) = \pi_{1}(S^5 / \Gamma)$ even when the group action by $\Gamma$ has fixed points. We use this to directly extract the electric one-form symmetry of such theories.

Second of all, we can directly exploit the fact that the higher-form symmetries are closely related to extended defects of the 5d SCFT and that the 1-form and 2-form symmetries above, upon circle reduction, give both rise to 1-form symmetries for the corresponding 4d KK theory. The defect group of the 4d KK theory is then captured by the screening of the latter by BPS particles, which is in turn  specified by the BPS quiver of the 5d SCFT \cite{Closset:2019juk}, the supersymmetric quantum mechanics (SQM) which encodes the dynamics on the worldline of the BPS particles of the 4d KK theory. Indeed, since compactification on a further circle takes us to type IIA on the same singularity, the resulting quiver is just the one obtained from a D0-brane probing $\mathbb{C}^3 / \Gamma$. From the 5d BPS quiver analysis, we expect that the one-form symmetry part of the defect group of the 4d KK theory $D_{S^1}\TM$ has the form
\begin{equation}\label{eq:4defecto}
   \mathbb D(D_{S^1} \TM)^{(1)} = \mathbb G^{(1)} \oplus \mathbb G^{(1)}
\end{equation}
where
\begin{equation}
    \mathbb G \simeq \bigoplus_{\ell = 1}^r \mathbb Z_{n_\ell},.
\end{equation}
In equation \eqref{eq:4defecto} there are two identical factors of $\mathbb G$ that denote respectively the possible electric and magnetic 1-form symmetries that are controlled by a choice of global structure for the 4d KK theory. The positive integers $n_\ell$ can be completely determined by via a standard 't Hooft screening argument \cite{tHooft:1977nqb}  --- see e.g. \cite{Caorsi:2017bnp}. Moreover, the quiver also captures the Weyl pairing \cite{Caorsi:2017bnp} (or linking pairing \cite{GarciaEtxebarria:2019caf}) from which the resulting Heisenberg algebra of non-commuting fluxes \cite{ Freed:2006ya,Freed:2006yc} that governs the global structure of the theories \cite{Aharony:1998qu,Witten:2009at} can be reconstructed \cite{INAKI}. Knowing the 1-form defect group of the 4d KK theory, it is easy to recover the corresponding factors of the defect group of the associated 5d SCFT:
\begin{equation}
    \mathbb D (\TM) \supseteq \mathbb G^{(1)}_e \oplus \mathbb G^{(2)}_m\,.
\end{equation}

Whenever the 5d SCFT has a global structure which allows for a 1-form symmetry as well as a 0-form symmetry, the two can mix, and this can result in a non-trivial global 2-group symmetry -- see e.g. \cite{Kapustin:2013uxa,Cordova:2018cvg,Cordova:2020tij}.\footnote{See also  \cite{Sati:2008eg,Baez:2005sn,Fiorenza:2012tb,Fiorenza:2010mh,Sati:2009ic} for foundational work on higher group gauge symmetries.} As a further result in this short note we begin exploring the 2-group symmetries of some orbifold 5d SCFTs with a Lagangian description \cite{Apruzzi:2021vcu}, reproducing the known features of such systems in terms of the abelianization of the orbifolding group $\Gamma$. Our result indicates that the 2-group structure is indeed a feature of the 5d SCFT rather than an emergent IR artifact.

The result of this paper is organized as follows. In section \ref{sec:PRESCRIPTION}, after a brief review of the defect group
and its use in determining the higher-form symmetries of a 5d SCFT, we give a general prescription
for computing the higher-form symmetries of the 5d SCFT, both via a direct analysis of $\pi_1(S^5 / \Gamma)$,
and via the corresponding 5d BPS quiver. In section \ref{sec:EXAMP} we turn to a collection of examples, illustrating how our method works in
practice. In section \ref{sec:2group} we turn to a preliminary analysis of 2-group structures in such theories, and in particular its (conjectural) relation to the abelianization of $\Gamma$. We present our conclusions and potential future directions in section \ref{sec:CONC}. The appendices contain some additional technical details as well as instructions for reproducing the relevant quiver and group theory computations.

\textbf{Note added:} \textit{While this paper was in preparation and after our results were announced during various online seminars in November and December 2021, we learned that v3 of \cite{Tian:2021cif} (posted Dec. 27, 2021) contains some new material which has some overlap with the present note. To a large extent, the results are consistent.}

\section{Defect Groups and Higher Symmetries in 5d} \label{sec:PRESCRIPTION}

In this section we discuss the interplay between the defect group and higher-form symmetries, with a particular emphasis on 5d theories.
Recall that the defect group is a general way to capture the spectrum of defects with charges which cannot be screened by dynamical states of the theory. This notion was first introduced in reference \cite{DelZotto:2015isa} in the context of 6d SCFTs, but it has far wider applicability, especially when combined with flux non-commutativity \cite{Freed:2006ya,Freed:2006yc}, as exploited, for example in references \cite{GarciaEtxebarria:2019caf,Albertini:2020mdx, Morrison:2020ool}. It is especially helpful in the context of higher-dimensional quantum field theories specified by a compactification of string theory, and we will mainly focus on this case in what follows.

In the context of string compactification, we obtain supersymmetric defects by wrapping branes on non-compact cycles of a local geometry. Branes of the same codimension which are wrapped on compact cycles amount to dynamical degrees of freedom which can screen the charges associated with these defects. Indeed, in many quantum field theories, the corresponding collection of defects needs to be supplemented by a choice of global structure which restricts the spectrum of extended objects \cite{Aharony:2013hda, Kapustin:2014gua, Gaiotto:2014kfa, DelZotto:2015isa}. This can happen whenever the corresponding torsional fluxes do not commute \cite{Aharony:1998qu, Freed:2006ya,Freed:2006yc, Tachikawa:2013hya, Monnier:2014txa, DelZotto:2015isa, Heckman:2017uxe, Monnier:2017klz}. Our conventions and treatment will follow that presented in \cite{Albertini:2020mdx}, to which we refer the interested reader for further details.

In any geometric engineering setup, the BPS spectrum of the resulting quantum field theory is captured by branes of various dimensions that are wrapping on shrinking cycles of a non-compact geometry $\mathbf{X}$, which in our case is a Calabi-Yau threefold singularity CY$_3$. When a $p$-brane wraps a compact $k$-cycle, it describes a $p-k+1$ dimensional BPS excitation. While, $p$-branes wrapped on non-compact $k$-cycles describe $p-k+1$ dimensional defects operators. Branes are charged with respect to flux operators, which can be used to construct the  corresponding quasi-topological symmetry defects that describe the charges of the extended objects. Of course, we can have a generalized 't Hooft screening, due to the possibility of defects to end on dynamical BPS objects, which breaks the associated higher-form symmetry. The remaining symmetry is captured by the defect group:
\begin{equation}
    \bbD := \bigoplus_n \bbD^{(n)} \quad \text{where}\ \  \bbD^{(n)} =  \bigoplus_{p\text{-branes}}\left(\bigoplus_{k \text{ s.t. } \newline p-k+1=n} \left(\frac{H_k(\mathbf{X}, \partial \mathbf{X})}{H_{k}(\mathbf{X})}\right)\right)
\end{equation}
In other words, the defect group $\mathbb{D}$ is the group of charges of higher symmetries acting on defects modulo screening. Moreover, together with the corresponding Heisenberg algebra of non-commuting fluxes, it encodes the quantum data of the Hilbert space at the boundary of the non-compact geometry. This construction captures all possible global structures realized by the geometry of the string compactification. We note that in principle, there could be additional emergent higher-form symmetries in the deep infrared of such a system, which would in turn signal the existence of additional defects. Our operating assumption---which is well-supported in practice--- is that such subtleties will not arise in the analysis to follow.

In this paper our focus is on the 1-form and the 2-form symmetry parts of the defect group in the context of a geometric engineering of M-theory on a Calabi-Yau singularity $\orb$
\begin{equation}
\mathbb D(M/\orb) \supset \mathbb D^{(1)}_{M2} \oplus \bbD^{(2)}_{M5}
\end{equation}
where the subscripts denote the associated branes, and the superscript indicates that the M2-branes are associated with a one-form and the M5-branes with a two-form generalized symmetry.

Since M2s and M5s are mutually non-local and in general the singularity $\orb$ might have some torsional flux, one naturally expects to find examples with non-trivial higher symmetries and a non-trivial global structure. To see this, consider a geometry $\mathcal{M}_{11}=\mathcal{M}_5 \times \orb$ where for ease of exposition we take $\mathcal M_5$ compact and torsion free. The geometry $\mathcal{M}_{11}$ has a boundary at infinity given by $\partial \mathcal{M}_{11} = \mathcal M_5 \times \partial \orb$, to which we associate an Hilbert space $\mathcal{H}(\partial \mathcal{M}_{11})$. The resulting Hilbert space has selection sectors that can be thought of as states in a quantum mechanics where the role of operators is played by torsional fluxes, organized by a generalized cohomology group, $\mathbb{E}(\mathcal{M}_{11})$. The presence of a non-trivial torsion for the generalized cohomology,
\begin{gather}
    \mathrm{Tor}\,\mathbb{E}(\mathcal{M}_{11})=\bigoplus_i H^{i+1}(\mathcal{M}_5) \otimes \mathrm{Tor }\, \mathbb{D}^i,
\end{gather}
might cause the flux operators to form a non-commutative algebra \cite{Freed:2006ya,Freed:2006yc}. Indeed, fluxes corresponding to the M2 and M5-branes that contribute to the 1-form symmetry and the 2-form symmetry part of the defect groups satisfy the following relation
\begin{gather}
    \Psi_2 \Phi_5=\text{exp}\left( 2 \pi i  L(l_1,l_2) \int_{\mathcal{M}_5} \omega_1 \wedge \omega_2 \right)\Phi_5 \Psi_2,
\end{gather}
where the term in the exponential is a pairing of cocycles in $\rm{Tor}\,\mathbb{E}(\mathcal{M}_{11})$, $\omega_{1,2}$ represents the forms dual to the cycle where the extended objects have supports, $l_{1,2}$ are elements of $\mathbb{D}^i$ and $L(\cdot,\cdot)$ is the linking form on $\partial \orb$. To fully specify the quantum system, we need to select a maximal set of mutually commuting fluxes as a base for our Hilbert space.

This construction can be made more rigorous and general \cite{Albertini:2020mdx}. In particular, it is known how to compute the defect group from exact sequences in homology \cite{DelZotto:2015isa,Albertini:2020mdx,Morrison:2020ool}. In the next section we will review this result and use it to compute the defect group of orbifold singularities. Moreover, we will confirm the same result exploiting the corresponding 5d BPS quivers, building on \cite{Hosseini:2021ged,INAKI}.

The rest of this section is organized as follows. Again specializing to the case of 5d SCFTs obtained from M-theory on an orbifold singularity $\orb = \mathbb{C}^3 / \Gamma$, we show how to extract the defect group directly from the fundamental group of $S^5 / \Gamma$, which we refer to as the ``algebraic topology approach''. After this, we turn to a physical realization of the same data in terms of the 5d BPS quiver defined by the 5d SCFT. The physical interpretation of the quiver in terms of the Dirac pairing for BPS particles of the 4d KK theory provides a complementary method for extracting the same data on higher-form symmetries. We turn to examples later in section \ref{sec:EXAMP}.

\subsection{Algebraic Topology Approach}

Let us now turn to a computation of the defect group directly via the corresponding singular geometry specified by the orbifold group $\mathbb{C}^{3} / \Gamma$. We start by considering M-theory on $\mathcal M_5 \times \orb$. In order to capture the defect group one has to consider the long exact sequence of relative homology of $(\orb, S^5/\Gamma)$:
\begin{equation}
    \dots \rightarrow H_2(S^5/\Gamma) \overset{\imath_2}{\rightarrow} H_2(\orb) \overset{\jmath_2}{\rightarrow} H_2(\orb, S^5/\Gamma) \overset{\partial_2}{\rightarrow} H_1(S^5/\Gamma) \overset{\imath_1}{\rightarrow} \underbrace{H_1(\orb)}_{ = 0} \rightarrow \dots.
\end{equation}
Strictly speaking, some of the quantities in the above exact sequence may not involve smooth spaces, for example if $\Gamma$ has fixed points.
In the present context, we can always assume the existence of a crepant resolution, and work in terms of the resolved geometry. Since, however, our answer will be independent of a given choice of a resolution, there is a precise sense in which these objects ought to make sense even without an explicit blowup, consistently with the remark of \cite{Morrison:2020ool} that the higher form symmetries are independent from flop transitions in the resolved geometry. Indeed, we will shortly give a precise definition of $H_{1}(S^5/\Gamma)$ as the abelianization of $\pi_{1}(S^5 / \Gamma)$,
even when $\Gamma$ has a fixed point locus on the $S^5$.

Now, we expect $H_1(\orb)$ to vanish due to the fact that $\orb$ is a Calabi-Yau space, and, moreover, by definition \cite{Albertini:2020mdx,Morrison:2020ool}:
\begin{equation}
\begin{aligned}
    \mathbb D^{(1)} &\equiv \frac{H_2(\orb, S^5/\Gamma)}{\jmath_2(H_{2}(\orb))} \\
    &\simeq H_1(S^5/\Gamma)\\
    &\simeq \text{Ab}[\pi_1(S^5/\Gamma)]
\end{aligned}
\end{equation}
Using both Poincaré duality and the Universal Coefficient Theorem it can be shown that $\mathbb D^{(1)}_{M2} \simeq \mathbb D^{(2)}_{M5}$ whenever $H_3(\orb)$ vanishes, which is indeed the case for the orbifold singularities we are considering.

The defect group is then fully captured by $\pi_1(S^5/\Gamma)$, since the abelianization of this group is just $H_1(S^5 / \Gamma)$. An important subtlety here is that in general, $\Gamma$ may have a fixed point locus which complicates the analysis. In the special case where there are no fixed points, we have $\pi_{1}(S^5 / \Gamma_{\mathrm{no-fixed}}) = \Gamma_{\mathrm{no-fixed}}$. To extend this to the more general case which can include fixed points, we use a result proved by Armstrong in 1967 \cite{armstrong1968fundamental}:

\textit{Let $\Gamma$ be a discontinuous group of homeomorphisms of a path connected, simply connected, locally compact metric space $X$, and let $H$ be the normal subgroup of $\Gamma$ generated by those elements which have fixed points. Then the fundamental group of the orbit space $X/\Gamma$ is isomorphic to the factor group $\Gamma/H$.}

In other words, to compute $\pi_{1}(S^5 / \Gamma)$, we just need to enumerate the generators of $\Gamma$ which might have a fixed point locus on $S^5$. Specifying the particular group action induced via $f_\Gamma: \Gamma \rightarrow \text{Homeo}(\bbC^3)$, we denote by $H_{\Gamma, f_{\Gamma}} \trianglelefteq \Gamma$ the resulting normal subgroup of $\Gamma$. The end result is that $\pi_1(S^5/\Gamma) = \Gamma/H_{\Gamma, f_\Gamma}$, so the one-form symmetry part of the defect group is just:
\begin{equation}
\mathbb{D}^{(1)} = \mathrm{Ab}[\Gamma / H_{\Gamma, f_{\Gamma}}].
\end{equation}
This also shows that the higher-form symmetry is independent of a choice of resolution, and moreover,
provides a systematic way to determine this data without specifying a blowup.
We give explicit examples of this procedure in section \ref{sec:EXAMP}.

\subsection{Quiver Approach}

A complementary way to extract the same information on the higher-form symmetry is to determine the corresponding quiver associated with a given orbifold singularity. In physical terms, this arises from the 5d BPS quiver of the theory \cite{Closset:2019juk}. 5d BPS quivers are the quiver of supersymmetric quantum mechanics that capture the BPS spectrum of particles of the 4d KK theory $D_{S^1}\TM$. Exploiting the Kaluza-Klein (KK) circle as an M-theory circle, it is clear that the 4d KK theory associated to $\orb$ is obtained from type IIA on the same Calabi-Yau threefold \cite{Lawrence:1997jr}. The 5d BPS quivers are therefore captured from the BPS quivers of IIA on $\orb$ \cite{Closset:2019juk} (see also \cite{Duan:2020qjy}). For the case at hand, the structure of the quivers can be reproduced from the D0-brane probe of this singularity. The resulting supersymmetric quiver quantum mechanics follows from the general prescription of Douglas and Moore \cite{Douglas:1996sw}. The resulting 3d McKay quivers were obtained in \cite{Hanany:1998sd,Lawrence:1998ja}. For additional details on how to implement this procedure, see Appendix \ref{app:3dmckay}. In many of our quiver figures, we present an explicit indexing of the nodes, and in particular this does not refer to the rank of each gauge group.

The crucial part needed for our analysis is the physical interpretation of the BPS quivers (see e.g. \cite{Fiol:2000wx,Denef:2002ru,Cecotti:2010fi,Cecotti:2011rv,Cecotti:2011gu,Alim:2011ae,Alim:2011kw}). The nodes of the BPS quivers are in one-to-one correspondence with a basis of generators of the charge lattice of the theory. Let us denote the corresponding charges $\gamma_1,...,\gamma_N$. The states corresponding to the charges of the generators are viewed as a collection of elementary constituents, out of the bound states of which the whole spectrum of the theory can be reconstructed. Since all the states in the spectrum are formed by these bound states, their charges are integer multiples of the charges of the elementary constituents, and we can completely determine the 't Hooft screening in terms of the latter \cite{Caorsi:2017bnp}. The charges of the line defects are in turn valued in a dual lattice of charges, where the duality is determined by the Dirac pairing \cite{Gaiotto:2010be,Aharony:2013hda}. For Lagrangian theories, the relevant Dirac pairing is determined by a simple computation in the Coulomb phase of the theory, for non-Lagrangian theories, however, geometry is needed. Here another aspect of the structure of the BPS quiver quantum mechanics is crucial, namely that the adjacency matrix which determines the structure of the BPS quiver quantum mechanics, is indeed captured by the Dirac pairing among the charges of the elementary constituents
\begin{equation}
\langle \gamma_i,\gamma_j\rangle_D = B_{ij}\,.
\end{equation}
For this reason the relevant quotient, which captures the defect group from the IR \cite{INAKI}, is also reproduced by the cokernel of the Dirac pairing. The 4d KK theory is obtained from type IIA on the same orbifold singularity $\orb$ and the torsional generators of the defect group are:
\begin{equation}\label{eq:5dBPSoneform}
\text{Tor }\mathbb D^{(1)}(\mathrm{IIA}/\orb) = \text{Tor}( \text{coker} (B)) = \mathbb G^{(1)}_e \oplus \mathbb G^{(1)}_m
\end{equation}
where
\begin{equation}
\mathbb G^{(1)}_e \simeq  \mathbb G^{(1)}_m \simeq \bigoplus_{\ell=1}^r \mathbb Z_{n_{\ell}} \,.
\end{equation}
By construction the integers $n_\ell$ can be recovered by the Smith normal form of the matrix $B$  \cite{Caorsi:2017bnp}. The fact that $B$ is antisymmetric entails that one gets the two identical electric and magnetic factors in equation \eqref{eq:5dBPSoneform}.

Whenever we choose a global form for the 5d theory $\TM$ with a magnetic 2-form symmetry, the latter gives rise to a magnetic 1-form symmetry for the 4d KK theory, by wrapping the corresponding surface defects on the KK circle. For this reason we identify
\begin{equation}
\mathbb D^{(2)}(M/\orb) \simeq \mathbb G^{(1)}_m
\end{equation}
above. This strategy gives rise to interesting consistency checks for the entire construction.

\section{Examples} \label{sec:EXAMP}

In section \ref{sec:PRESCRIPTION} we presented a general prescription for how to extract the higher-form symmetries from 5d SCFTs defined by M-theory on the background $\mathbb{C}^3 / \Gamma$. Our plan in this section will be to show how this works in practice, illustrating with a number of examples that both methods produce the same result, and agree with previously established results available in the literature, including v1 and v2 of \cite{Tian:2021cif}.

To frame the discussion to follow, we first divide the subgroups of $SU(3)$ into three main families:
\begin{itemize}
\item Family 1: The abelian subgroups;
\item Family 2: The subgroups of $SU(3)$ induced from finite non-abelian subgroups of $U(2)$;
\item Family 3: The complement of families 1 and 2.
\end{itemize}
We proceed by way of example, illustrating how our method works in each of these cases. In the case of finite abelian subgroups of $SU(3)$, we find agreement with the results of \cite{Albertini:2020mdx, Morrison:2020ool} as well as those of \cite{Tian:2021cif} which  involved a direct analysis of the resolved geometry. The case of family 2 does not appear to have been treated in the existing literature, but again we find examples which contain non-trivial higher-form symmetries. In the case of family 3, all examples we considered have too
many elements in $\Gamma$ whose action on $\mathbb{C}^3$ contains a fixed point locus.
The resulting normal subgroup generated by such elements is so large that $\mathrm{Ab}[\pi_{1}(S^5 / \Gamma)]$ is trivial, and as such they all produce a trivial higher-form symmetry. This is also in agreement with the results of v1 and v2 of \cite{Tian:2021cif}.

\subsection{Abelian Subgroups of \texorpdfstring{$SU(3)$}{SU(3)}}
We now turn to the case of $\Gamma$ a finite abelian subgroup of $SU(3)$. In this case, we have a
diagonal group action on the holomorphic coordinates $(x,y,z)$ of $\mathbb{C}^3$.
Since the maximal torus of $SU(3)$ is just $U(1)^2$, the most general orbifold action is given by
\begin{equation}
(x,y,z) \mapsto (\omega^{a_1} \xi^{b_1} x, \omega^{a_2} \xi^{b_2}y , \omega^{a_3} \xi^{b_3} z),
\end{equation}
with $\omega$ and $\xi$ primitive $m^{\mathrm{th}}$ and $n^{\mathrm{th}}$ roots of unity, and integers $a_i$ and $b_i$ with
$\sum_i a_i=0 \, \mathrm{mod} \, m$ and $\sum_i b_i=0 \, \mathrm{mod} \, n$.

We now consider two specific set of actions that highlight the main properties of these orbifolds. More general actions can be considered, but a complete analysis is left for future work.

\begin{figure}
    \centering
    \begin{tabular}{cc}
    \includegraphics[scale=0.2]{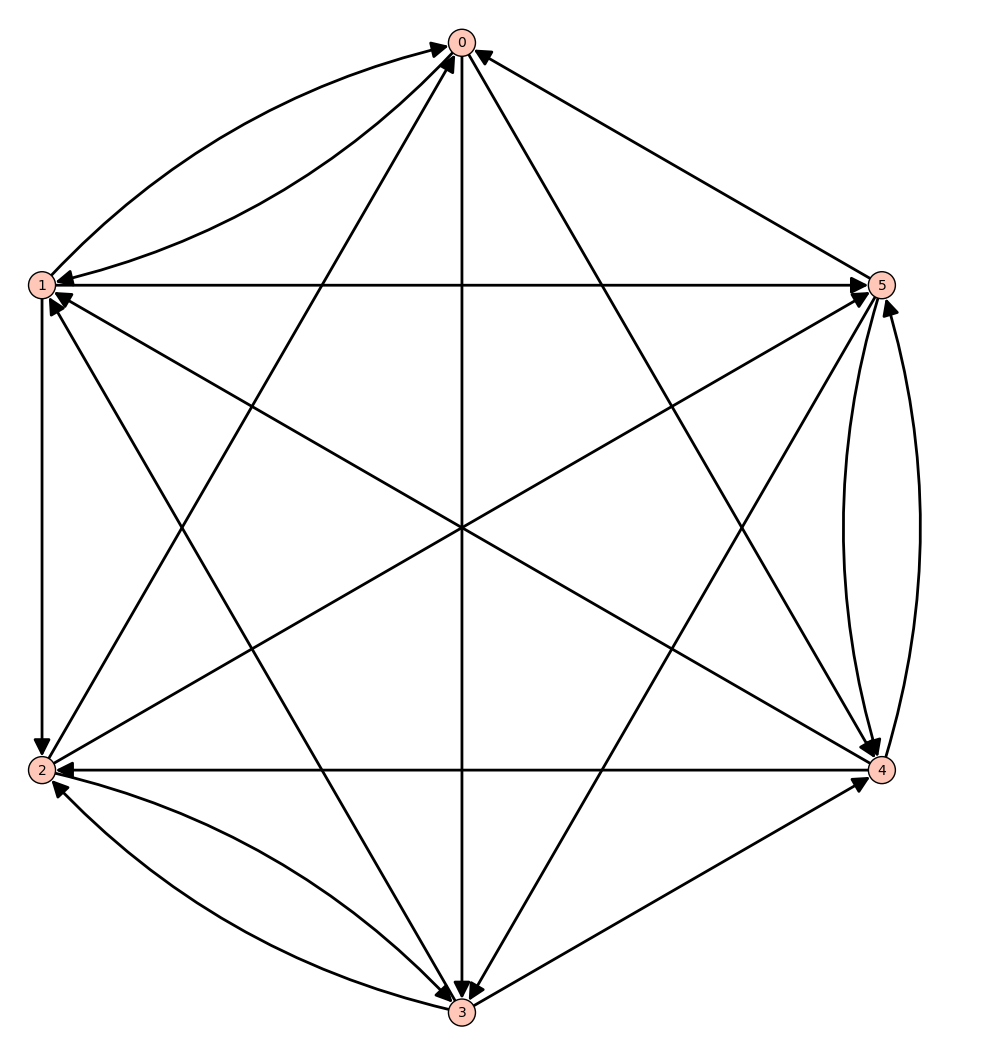}&\includegraphics[scale=0.23]{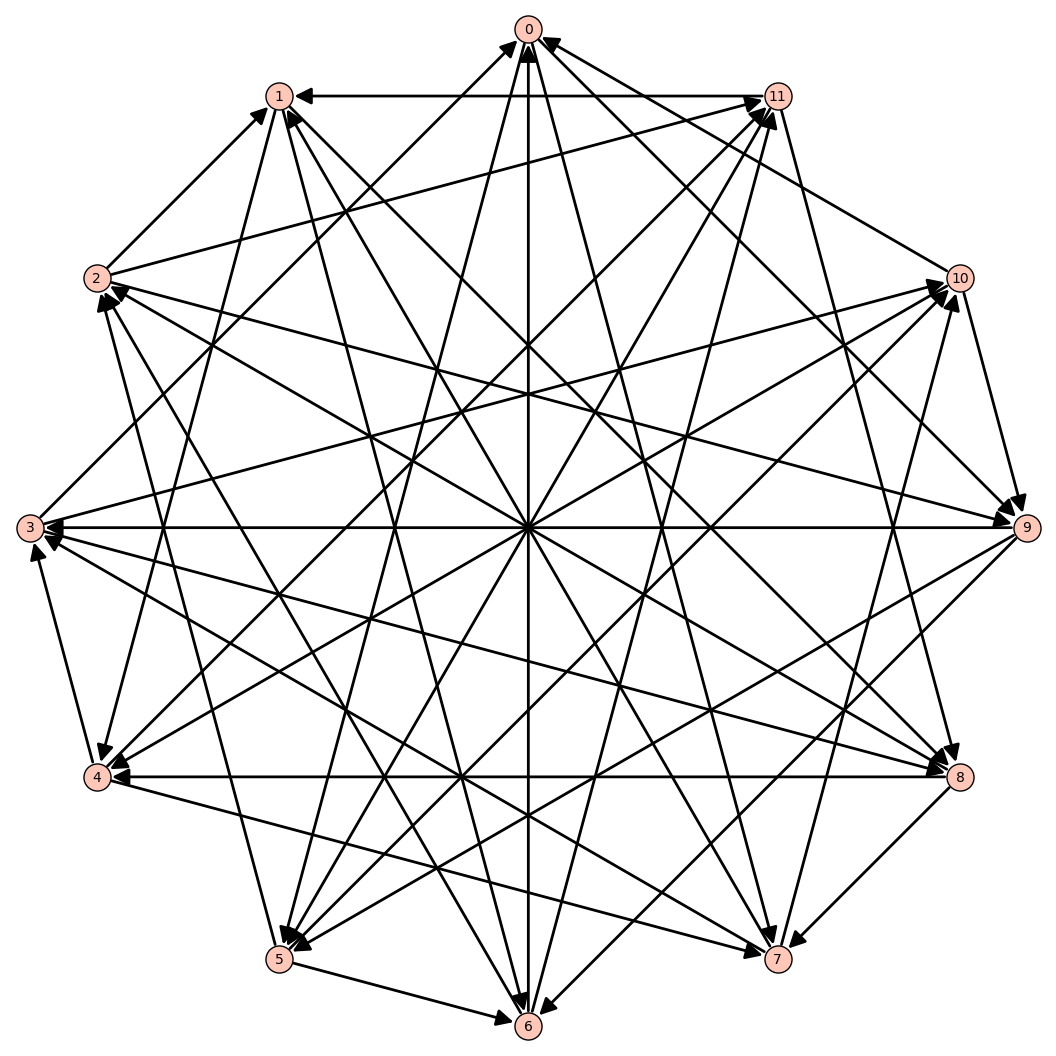}\\
    \end{tabular}
    \caption{\textsc{Left:} the orbifold quiver for the $\bbC^3/\mathbb{Z}_6$ theory generated by $g = \frac{1}{6}(1,2,3)$. \textsc{Right:} The orbifold quiver for the $\bbC^3/\mathbb{Z}_{12}$ theory generated by $g = \frac{1}{12}(1,2,9)$.}
    \label{fig:z6_123}
\end{figure}

The first case we consider is when $n=1$, thus $\Gamma = \mathbb{Z}_m$. Here, the generators are of the form $\frac{1}{m}(1,a_2,a_3)$, with $1+a_2+a_3=0 \mod m$.
In this case the choice of group action dictates the fixed point locus. For these examples  we can state a general rule:
\begin{align}
    H_{\Gamma, f_\Gamma} = \bbZ_{gcd(m,a_2)} \times \bbZ_{gcd(m,a_3)}
\end{align}
Let us check this formula on few examples:
\begin{itemize}
\item One generator for all fixed points.

The subgroup of fixed points is generated by some unique power of the generator of the full group, i.e. $g^k$ has fixed points, with $g$ the generator of $\Gamma = \bbZ_m$. In these cases, $H_{\Gamma, f_\Gamma} = \bbZ_{m/k}$, and thus
\begin{equation}
\mathbb D^{(1)}_{M2} = \text{Ab}[\pi_1(S^5/\Gamma)] = \bbZ_{m}/\bbZ_{m/k} \cong \bbZ_k
\end{equation}

\textbf{Example 1}: For $g = \frac{1}{10}(1, 1, 8)$, only $k = 5$ gives $g^k = \frac{1}{2}(1, 1, 0)$ with fixed loci $|z| = 1$. So we have $\mathbb D^{(1)}_{M2}  = \bbZ_5$.\footnote{In particular, $g = \frac{1}{10}(1, 2, 7)$ is also such that $\mathbb D^{(1)}_{M2} = \bbZ_5$. We have confirmed by quiver computation that the statement of ``$\Lambda_{\text{el.}} = \bbZ_2$" in v1,v2 and v3 of \cite{Tian:2021cif} is a typo.}

\item More generators for all fixed points.

If there are more powers of the generator of the orbifold that lead to a fixed point, the normal subgroup is just the direct product of those factors.

\medskip

\textbf{Example 2}: For $\Gamma = \mathbb{Z}_6$ generated by $g = \tfrac{1}{6}(1, 2, 3)$, both $k = 2$ and $k=3$ have fixed points. $k=2$ leads to a $\bbZ_3$ subgroup, while $k=3$ to a $\bbZ_2$. So $H_{\Gamma, f_\Gamma} = \bbZ_3 \times \bbZ_2 = \Gamma$, and thus $\mathbb D^{(1)}_{M2} = 0$. We have also directly verified this by constructing the corresponding 5d BPS quiver --- see Figure \ref{fig:z6_123} (left).

\medskip

\textbf{Example 3}: For $\Gamma = \mathbb{Z}_{12}$ generated by $g=\tfrac{1}{12}(1, 2, 9)$, $g^4$ generates a $\bbZ_3$ subgroup and $g^6$ a $\bbZ_2$ one. So $H_{\Gamma, f_\Gamma} = \bbZ_2 \times \bbZ_3 \subset \bbZ_{12} = \Gamma$, and thus $\mathbb D^{(1)}_{M2}  = \bbZ_2$. The corresponding 5d BPS quiver can be found in Figure \ref{fig:z6_123} (right).
\end{itemize}

The second case we consider is $\Gamma = \mathbb{Z}_m \times \mathbb{Z}_n$, where we mainly consider the family of generators $\tfrac{1}{m}(1, m-1, 0)$ for $\bbZ_m$ and $\tfrac{1}{n}(0, 1, n-1)$ for $\bbZ_n$ that has been treated in \cite{Tian:2021cif}. In this case, we can rule out any non-trivial higher symmetry using the algebraic approach. Each of these manifestly have a fixed circle ($|z| = 1$ and resp., $|x| = 1$). So they both have $H_{\Gamma, f_\Gamma} = \Gamma$ and so Armstrong's theorem tells us that $\pi_1(S^5/\Gamma) = 0$. This result can also be established using BPS quivers. For example, for $n=m$, the 5d BPS quiver corresponding to models in this class all have a box-product form 
 \cite{Alim:2011kw}\footnote{\, Given two acyclic quivers $Q_1$ and $Q_2$ with adjacency matrices $B_i = S_i^t - S_i$ where $S_i$ are upper triangular 2d/4d $S$-matrices and $i=1,2$, the quiver $Q_1 \boxtimes Q_2$ is a quiver with adjacency matrix $B_\boxtimes = (S_1 \otimes S_2)^t - (S_1 \otimes S_2)$. If the quivers are not acyclic, which is the case in equation \eqref{eqn:AnBoxAn}, one can still define a $\boxtimes$ operation at the level of the corresponding path algebras: the path algebra of the quiver $Q_1\boxtimes Q_2$ is the tensor product of $\mathbb C Q_1$ and $\mathbb C Q_2$ with extra lagrange multipliers which implements the commutativity relation on the corresponding squares --- the quiver $Q_1\boxtimes Q_2$ also comes with a superpotential, which to a first approximation is a superposition of the superpotentials of $Q_1$ and $Q_2$ together with extra Lagrange multipliers which implement the commutativity of the tensor product operation on the path algebra --- see e.g. \cite{Cecotti:2010fi} for a review of the $\boxtimes$ operation for $Q_1$ and $Q_2$ acyclic.}
\begin{equation}\label{eqn:AnBoxAn}
\widehat A(n,0) \boxtimes \widehat A(n,0)
\end{equation}
where $\widehat A(n,0)$ is the quiver corresponding to a loop with $n$ arrows oriented clockwise and superpotential given by $\text{tr}(\prod_i \psi_i)$. A direct case by case analysis for $0\leq n \leq 30$ reveals that the cokernel is trivial in all these cases. Similar considerations hold for $n\neq m$, though in this case we do not have a single concise expression as in line (\ref{eqn:AnBoxAn}), and the quivers have to be extracted from the general procedure summarized in Appendix \ref{app:3dmckay}.

In addition, we want to point out that there are more possible actions of $\bbZ_m \times \bbZ_n$ than those considered above. Here we give an example of such action with a non-trivial 1-form symmetry.

\textbf{Example 4}: For $\omega = \frac{1}{9}(1, 1, 7)$ and $\xi = \frac{1}{3}(1, 2, 0)$.  The fixed loci are generated by $3\omega$ and $3\omega + \xi$, spanning a group of $H_{\Gamma, f_\Gamma} = \bbZ_3 \times \bbZ_3$. So we have $D_{M2}^{(1)} = \bbZ_3$.

\subsection{Examples Induced from Subgroups of \texorpdfstring{$U(2)$}{U(2)}}

Let us now turn to subgroups of $SU(3)$ induced from finite non-abelian subgroups of $U(2)$. These groups are obtained by taking some finite $\hat{\Gamma} \subset U(2)$ and mapping the elements $\hat{g}\in\hat{\Gamma}$ to
\begin{equation}
    g = \begin{pmatrix}
        \hat{g} & 0 \\
        0 & (\det \hat{g})^{-1}
    \end{pmatrix}.
\end{equation}
The finite subgroups of $U(2)$ are well known (see \cite{cohen_U2}, for example) and given by certain cyclic extensions of finite $SU(2)$ subgroups. In principle, this gives us an exhaustive method for generating such finite $SU(3)$ subgroups.

The first set of $U(2)$ derived groups we consider are those found in \cite{Tian:2021cif, watanabe1982invariant}. These groups form a special subclass of $SU(3)$ groups in the sense that their invariant subrings are complete intersection rings. Using the generators provided in \cite{Tian:2021cif}, we easily see that many elements have fixed points regardless of the group. In fact, in all of these cases we get that $H \cong \Gamma$ and hence
\begin{equation}
    \pi_1 (S^5/\Gamma) = 0.
\end{equation}
We list our findings in \cref{tbl:u2}.

\begin{table}[]
\centering
$\begin{array}{|c|c|c|c|}
     \hline\hline
    \Gamma & |\Gamma| & \mathbb D^{(1)}_{M2} & \text{Ab}[\Gamma] \\ \hline
    G_m &  8m      &  0 &  \bbZ_2 \times \bbZ_2 \times \bbZ_2 (2|m);\ \ \bbZ_2 \times \bbZ_2 (2 \!\!\not | m)  \\ \hline
    G_{p, q} & 8p q^2     & 0 & \bbZ_2^2 \times \bbZ_{2q} (2|p);\ \ \bbZ_2 \times \bbZ_{2q} (2\!\!\not | p) \\ \hline
    G'_{m}  & 8m      &  0  &  \bbZ_2 \times \bbZ_2 (2|m);\ \ \bbZ_4 \times \bbZ_2 (2 \!\!\not| m)\\ \hline\hline
    E^{(1)} & 72 & 0 & \bbZ_3 \times \bbZ_3\\ \hline
    E^{(2)} & 24 & 0 & \bbZ_3 \\ \hline
    E^{(3)} & 96 & 0 & \bbZ_2\times\bbZ_2 \\ \hline
    E^{(4)} & 48 & 0 & \bbZ_2 \\ \hline
    E^{(5)} & 96 & 0 & \bbZ_4 \\ \hline
    E^{(6)} & 48 & 0 & \bbZ_2 \times \bbZ_3 \\ \hline
    E^{(7)} & 144 & 0 & \bbZ_2\times\bbZ_3 \\ \hline
    E^{(8)} & 192 & 0 & \bbZ_2\times\bbZ_4 \\ \hline
    E^{(9)} & 240 & 0 & \bbZ_2 \\ \hline
    E^{(10)} & 360 & 0 & \bbZ_3 \\ \hline
    E^{(11)} & 600 & 0 & \bbZ_5 \\ \hline\hline
\end{array}$\caption{Data for orbifold theories derived from subgroups of $U(2)$ which have complete intersection invariant subrings \cite{watanabe1982invariant}.}\label{tbl:u2}
\end{table}

However, our approach means we can consider more general subgroups of $SU(3)$ derived from $U(2)$. An important class which we can consider are those derived from {\it small}\,\footnote{A group $G \subset GL(n,\mathbb{C})$ is small if there are no elements $g\in G$ with exactly $n-1$ many eigenvalues equal to $1$. In other words, $G$ contains no reflections.} subgroups of $U(2)$ \cite{yau1993gorenstein}. The ``small-ness'' condition restricts the number of elements with fixed points in $\Gamma$, and as such can potentially have a larger defect group when compared with a ``larger'' subgroup of $SU(3)$.

\begin{table}[t!]
\centering
$\begin{array}{|c|c|c|c|} \hline\hline
    \Gamma \phantom{\Big|}& |\Gamma| & \mathbb D^{(1)}_{M2} & \text{Ab}[\Gamma] \\ \hline
    T_m  \phantom{\Big|}& 24m &  \bbZ_{3m}\ (3|m)\ \ \bbZ_{m}\ (3\!\!\not| m) &  \bbZ_{3m} \\ \hline
    O_m  \phantom{\Big|}& 48m & \bbZ_m & \bbZ_2 \times \bbZ_m \\ \hline
    I_m  \phantom{\Big|}& 120m & \bbZ_m & \bbZ_m  \\ \hline
    D_{n, q} \phantom{\Big|}&   4qm\ (m = n-q)  & \bbZ_{2m}\ (2|m), \ \ \bbZ_{m}\ (2\!\!\not| m) &  \bbZ_{4m} (2 \!\!\not| q);\ \ \bbZ_{2m} \times \bbZ_2 (2 | q)  \\ \hline\hline
\end{array}$
\caption{Data for orbifold theories derived from small subgroups of $U(2)$. Note that the entries in this table depend are quite sensitive to the divisibility properties of $m,n$ and $q$, as discussed in \cref{app:conv}. To get a sense of the size of the normal subgroup with a fixed point locus, we have also listed the abelianization of $\Gamma$.}\label{tbl:smallu2}
\end{table}

Our findings in Table \ref{tbl:smallu2} can be confirmed again exploiting 5d BPS quivers. A more systematic study of the properties of the BPS categories of 5d orbifold SCFTs will appear elsewhere. For all these cases, we have a non-trivial defect group owing to the fact that they are built from small $U(2)$ subgroups. The analysis of these cases is uniform, and we carried out many consistency checks in this family. For the sake of brevity we report here only few salient examples. We refer to appendix \ref{app:conv} for our conventions about the representations we exploit in the analysis.

\medskip

\begin{figure}
    \centering
    \includegraphics[scale=0.6]{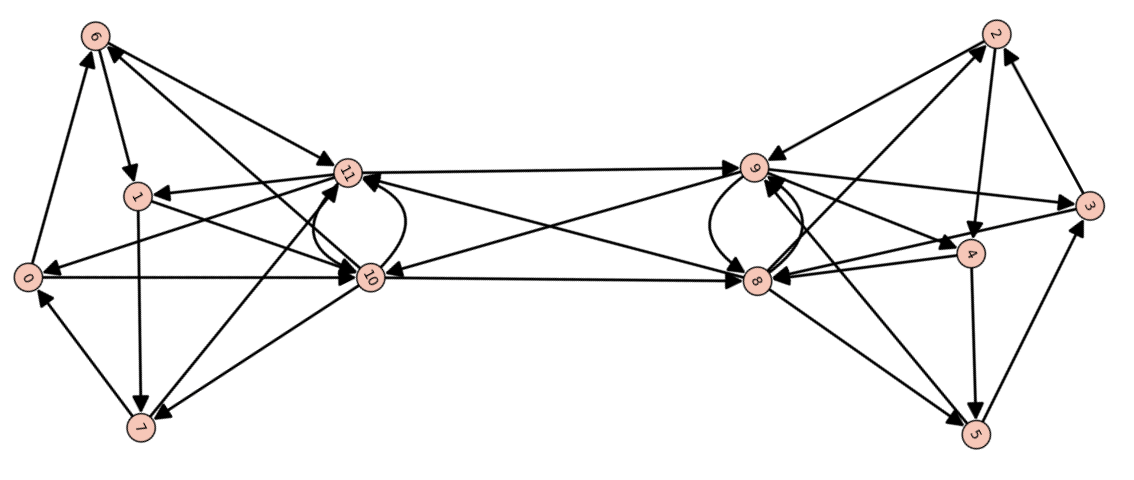}
    \caption{The orbifold quiver for the $\bbC^3/D_{5,3}$ theory.}
    \label{fig:D53}
\end{figure}

\noindent\textbf{The $D_{5,3}$ orbifold SCFT}. Let us consider the lowest rank theory we find with non-trivial defect group -- this happens to be the $D_{5,3}$ orbifold SCFT, a rank $r= 4$ SCFT with flavor symmetry of rank $f=3$. The group $D_{5,3}$ is generated by
    \begin{gather}
            D_{5,3}=\Bigg\langle
        \begin{pmatrix}
            \zeta_{6} & 0 & 0\\
            0 & \zeta_{6}^{-1} & 0\\
            0 & 0 & 1
        \end{pmatrix},
        \begin{pmatrix}
            0 & i & 0 \\
            i & 0 & 0 \\
            0 & 0 & 1
        \end{pmatrix}\cdot
        \begin{pmatrix}
            \zeta_{8} & 0 & 0\\
            0 & \zeta_{8} & 0\\
            0 & 0 & \zeta_{8}^{-2}
        \end{pmatrix}
    \Bigg\rangle.
\end{gather}
    We can now use the age grading of \cite{ItoReid} to understand the gauge and flavor ranks of the theory. Explicitly, we find that the number of age-1 (or `junior') classes is $7$, while the number of age-2 classes is $4$. This then gives us
    \begin{gather}
        r= b_4(X) =4, \quad f=b_2(X)-b_4(X)=3,
    \end{gather}
    where $X$ is the crepant resolution of $\bbC^3/D_{5,3}$, the $b_i(X)$ are its Betti numbers. As such, we are looking for a quiver of size $2r+f+1=12$ with an intersection pairing possessing a kernel of dimension 4. Indeed, we find the quiver in \cref{fig:D53} which possess these properties.

The $B$-matrix we obtain is
{\tiny
\begin{equation}
    \left(
\begin{array}{cccccccccccc}
 0 & 0 & 0 & 0 & 0 & 0 & 1 & -1 & 0 & 0 & 1 & -1 \\
 0 & 0 & 0 & 0 & 0 & 0 & -1 & 1 & 0 & 0 & 1 & -1 \\
 0 & 0 & 0 & -1 & 1 & 0 & 0 & 0 & -1 & 1 & 0 & 0 \\
 0 & 0 & 1 & 0 & 0 & -1 & 0 & 0 & 1 & -1 & 0 & 0 \\
 0 & 0 & -1 & 0 & 0 & 1 & 0 & 0 & 1 & -1 & 0 & 0 \\
 0 & 0 & 0 & 1 & -1 & 0 & 0 & 0 & -1 & 1 & 0 & 0 \\
 -1 & 1 & 0 & 0 & 0 & 0 & 0 & 0 & 0 & 0 & -1 & 1 \\
 1 & -1 & 0 & 0 & 0 & 0 & 0 & 0 & 0 & 0 & -1 & 1 \\
 0 & 0 & 1 & -1 & -1 & 1 & 0 & 0 & 0 & 0 & -1 & 1 \\
 0 & 0 & -1 & 1 & 1 & -1 & 0 & 0 & 0 & 0 & 1 & -1 \\
 -1 & -1 & 0 & 0 & 0 & 0 & 1 & 1 & 1 & -1 & 0 & 0 \\
 1 & 1 & 0 & 0 & 0 & 0 & -1 & -1 & -1 & 1 & 0 & 0 \\
\end{array}
\right)
\end{equation}}
From which it is easy to check the defect group for the 4d KK theory is indeed $\mathbb Z_4 \oplus \mathbb Z_4$ as expected.

\medskip

\begin{figure}
        \centering
        \includegraphics[scale=0.3]{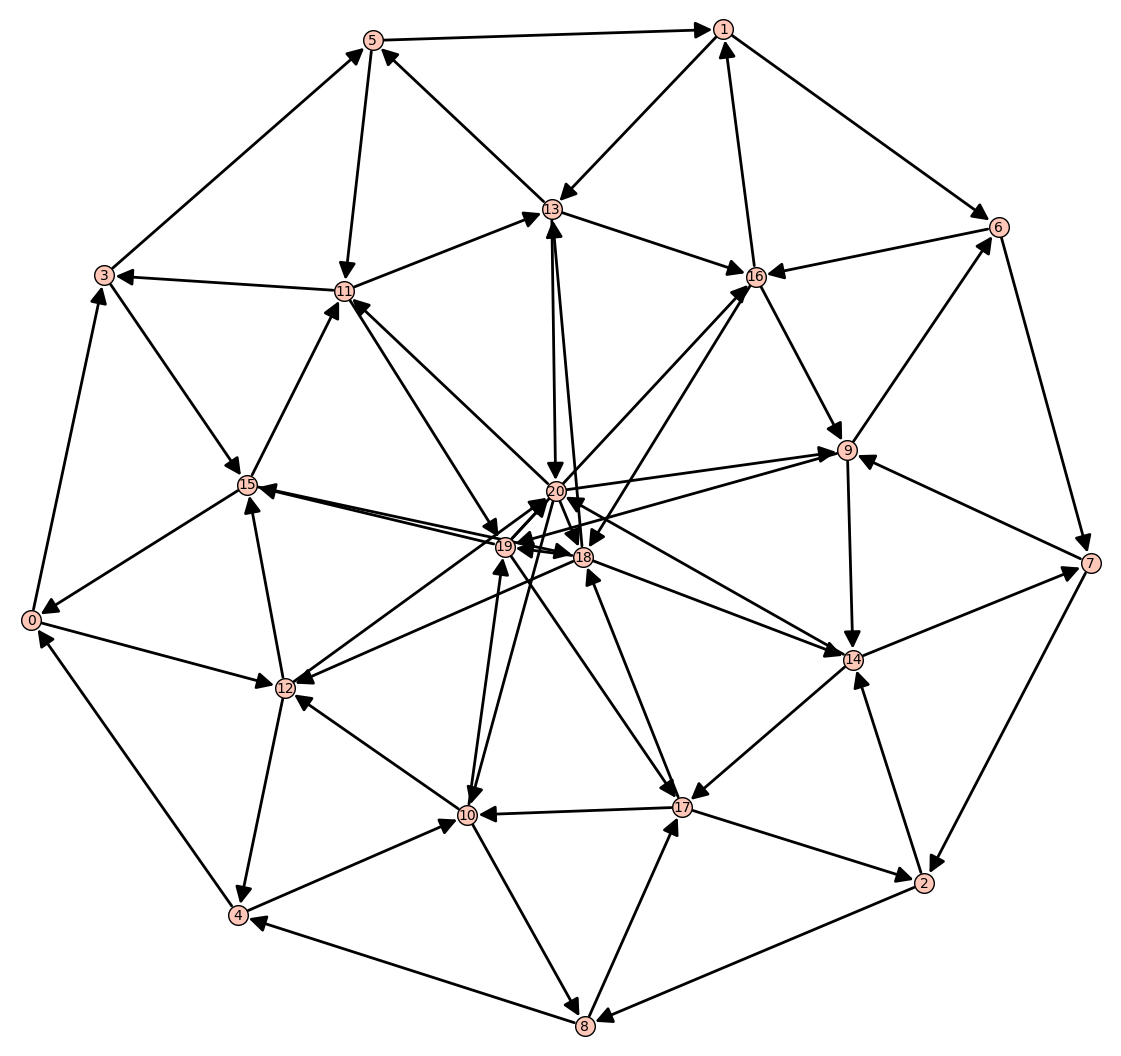}
        \caption{The quiver for the $\bbC^3/T_3$ orbifold theory.}
        \label{fig:t3}
    \end{figure}

\noindent\textbf{$T_3$ orbifold SCFT.} The $T_3$ group can be given as
    \begin{gather*}
        \Bigg\langle\begin{pmatrix}
            i & 0 & 0 \\
            0 & -i & 0 \\
            0 & 0 & 1
        \end{pmatrix},
        \begin{pmatrix}
            0 & 1 & 0 \\
            -1 & 0 & 0 \\
            0 & 0 & 1
        \end{pmatrix},
        \zeta_{18}\cdot\begin{pmatrix}
            (1+i)/2 & (-1+i)/2 & 0 \\
            (1+i)/2 & (1-i)/2 & 0 \\
            0 & 0 & \zeta_{18}^{-3}
        \end{pmatrix}\Bigg\rangle,
    \end{gather*}
    where $\zeta_{18}=e^{\pi i/9}$. From this we see that $|T_3|=72$ and that there are $21$ conjugacy classes\footnote{By the 3d McKay correspondence, this is equal to the Euler characteristic of the resolved $\bbC^3/T_3$ orbifold.}. Furthermore, the age grading gives us
    \begin{gather}
        r = b_4(X) = 9,\quad f =b_2(X)-b_4(X)= 2,
    \end{gather}
    where $X$ is the crepant resolution of $\bbC^3/T_3$, $r$ is the rank of the theory and $f$ is the rank of the flavor group. Note that these match with the constraint $2r+f+1=\chi(X)=21$. Interpreting this as quiver data, this means that we should find a quiver with $21$ nodes whose intersection pairing has a kernel of dimension $3$. Indeed, using our program to find the quiver, we obtain the quiver in \cref{fig:t3} which satisfies these conditions.

    The corresponding $B$-matrix is
    {\tiny
    \begin{equation}
        \left(
\begin{array}{ccccccccccccccccccccc}
 0 & 0 & 0 & 1 & -1 & 0 & 0 & 0 & 0 & 0 & 0 & 0 & 1 & 0 & 0 & -1 &
   0 & 0 & 0 & 0 & 0 \\
 0 & 0 & 0 & 0 & 0 & -1 & 1 & 0 & 0 & 0 & 0 & 0 & 0 & 1 & 0 & 0 &
   -1 & 0 & 0 & 0 & 0 \\
 0 & 0 & 0 & 0 & 0 & 0 & 0 & -1 & 1 & 0 & 0 & 0 & 0 & 0 & 1 & 0 & 0
   & -1 & 0 & 0 & 0 \\
 -1 & 0 & 0 & 0 & 0 & 1 & 0 & 0 & 0 & 0 & 0 & -1 & 0 & 0 & 0 & 1 &
   0 & 0 & 0 & 0 & 0 \\
 1 & 0 & 0 & 0 & 0 & 0 & 0 & 0 & -1 & 0 & 1 & 0 & -1 & 0 & 0 & 0 &
   0 & 0 & 0 & 0 & 0 \\
 0 & 1 & 0 & -1 & 0 & 0 & 0 & 0 & 0 & 0 & 0 & 1 & 0 & -1 & 0 & 0 &
   0 & 0 & 0 & 0 & 0 \\
 0 & -1 & 0 & 0 & 0 & 0 & 0 & 1 & 0 & -1 & 0 & 0 & 0 & 0 & 0 & 0 &
   1 & 0 & 0 & 0 & 0 \\
 0 & 0 & 1 & 0 & 0 & 0 & -1 & 0 & 0 & 1 & 0 & 0 & 0 & 0 & -1 & 0 &
   0 & 0 & 0 & 0 & 0 \\
 0 & 0 & -1 & 0 & 1 & 0 & 0 & 0 & 0 & 0 & -1 & 0 & 0 & 0 & 0 & 0 &
   0 & 1 & 0 & 0 & 0 \\
 0 & 0 & 0 & 0 & 0 & 0 & 1 & -1 & 0 & 0 & 0 & 0 & 0 & 0 & 1 & 0 &
   -1 & 0 & 0 & 1 & -1 \\
 0 & 0 & 0 & 0 & -1 & 0 & 0 & 0 & 1 & 0 & 0 & 0 & 1 & 0 & 0 & 0 & 0
   & -1 & 0 & 1 & -1 \\
 0 & 0 & 0 & 1 & 0 & -1 & 0 & 0 & 0 & 0 & 0 & 0 & 0 & 1 & 0 & -1 &
   0 & 0 & 0 & 1 & -1 \\
 -1 & 0 & 0 & 0 & 1 & 0 & 0 & 0 & 0 & 0 & -1 & 0 & 0 & 0 & 0 & 1 &
   0 & 0 & -1 & 0 & 1 \\
 0 & -1 & 0 & 0 & 0 & 1 & 0 & 0 & 0 & 0 & 0 & -1 & 0 & 0 & 0 & 0 &
   1 & 0 & -1 & 0 & 1 \\
 0 & 0 & -1 & 0 & 0 & 0 & 0 & 1 & 0 & -1 & 0 & 0 & 0 & 0 & 0 & 0 &
   0 & 1 & -1 & 0 & 1 \\
 1 & 0 & 0 & -1 & 0 & 0 & 0 & 0 & 0 & 0 & 0 & 1 & -1 & 0 & 0 & 0 &
   0 & 0 & 1 & -1 & 0 \\
 0 & 1 & 0 & 0 & 0 & 0 & -1 & 0 & 0 & 1 & 0 & 0 & 0 & -1 & 0 & 0 &
   0 & 0 & 1 & -1 & 0 \\
 0 & 0 & 1 & 0 & 0 & 0 & 0 & 0 & -1 & 0 & 1 & 0 & 0 & 0 & -1 & 0 &
   0 & 0 & 1 & -1 & 0 \\
 0 & 0 & 0 & 0 & 0 & 0 & 0 & 0 & 0 & 0 & 0 & 0 & 1 & 1 & 1 & -1 &
   -1 & -1 & 0 & 1 & -1 \\
 0 & 0 & 0 & 0 & 0 & 0 & 0 & 0 & 0 & -1 & -1 & -1 & 0 & 0 & 0 & 1 &
   1 & 1 & -1 & 0 & 1 \\
 0 & 0 & 0 & 0 & 0 & 0 & 0 & 0 & 0 & 1 & 1 & 1 & -1 & -1 & -1 & 0 &
   0 & 0 & 1 & -1 & 0 \\
\end{array}
\right)
    \end{equation}}

    Taking the Smith normal form we find that
    \begin{gather}
        \mathrm{coker}(B) = \bbZ^3\oplus \bbZ_9^{e}\oplus\bbZ_9^{m},
    \end{gather}
    giving us a $\bbZ_9$ electric one-form symmetry.

\begin{table}
\centering
$\begin{array}{|c|c|c|c|} \hline\hline
    \Gamma & |\Gamma| & \mathbb D^{(1)}_{M2} & \text{Ab}[\Gamma] \\ \hline
    \Delta(3n^2) & 3n^2      & 0 & \bbZ_3 \times \bbZ_3 (3|n);\ \ \bbZ_3 (3 \!\!\not | n)         \\ \hline
    \Delta(6n^2) & 6n^2   & 0   &  \bbZ_2    \\ \hline
    C^{(1)}_{3l, l}\ (3|l)& 9l^2  & 0   &      \bbZ_{3} \times \bbZ_{3}        \\ \hline
    C^{(2)}_{7l, l} & 21 l^2    & 0 &  \bbZ_3 \times \bbZ_3 (3|n);\ \ \bbZ_3 (3 \!\!\not | n) \\ \hline
    D^{(1)}_{3l, l}\ (2|l)& 18l^2\    & 0 &  \bbZ_2 \times \bbZ_3       \\ \hline\hline
    H_{36}  & 108      &            0       & \bbZ_{4}          \\ \hline
    H_{72}  & 216      &            0       & \bbZ_2 \times \bbZ_2 \\ \hline
    H_{216} & 648      &            0       & \bbZ_3          \\ \hline
    H_{60}  & 60       &            0       & \mathbbm{1}  \\ \hline
    H_{168} & 168      &            0       & \mathbbm{1}   \\ \hline
    H_{360} & 1080     &            0       & \mathbbm{1}   \\ \hline
    J       & 180      &            0       & \bbZ_3   \\ \hline
    K       & 504      &            0       & \bbZ_3   \\ \hline\hline
\end{array}$
\caption{Data for orbifold theories derived from transitive subgroups of $SU(3)$. We determined the higher-form symmetry by direct computation of $\mathrm{Ab}[\pi_1(S^5 / \Gamma)]$ and from an analysis of the corresponding 5d BPS quiver. In all cases, none of these theories exhibit any one-form symmetry. For completeness, we have also included the abelianization of all these groups. Here $J$ and $K$ follows the notation of Yau and Yu \cite{yau1993gorenstein}, while we have followed the notation of \cite{Tian:2021cif} in the remaining entries. See Appendix \ref{app:ABGAMMA} for the definitions of all of these groups.}\label{tbl:transitive}
\end{table}

\medskip

\begin{figure}
        \centering
        \begin{tabular}{cc}
        \includegraphics[scale=0.4]{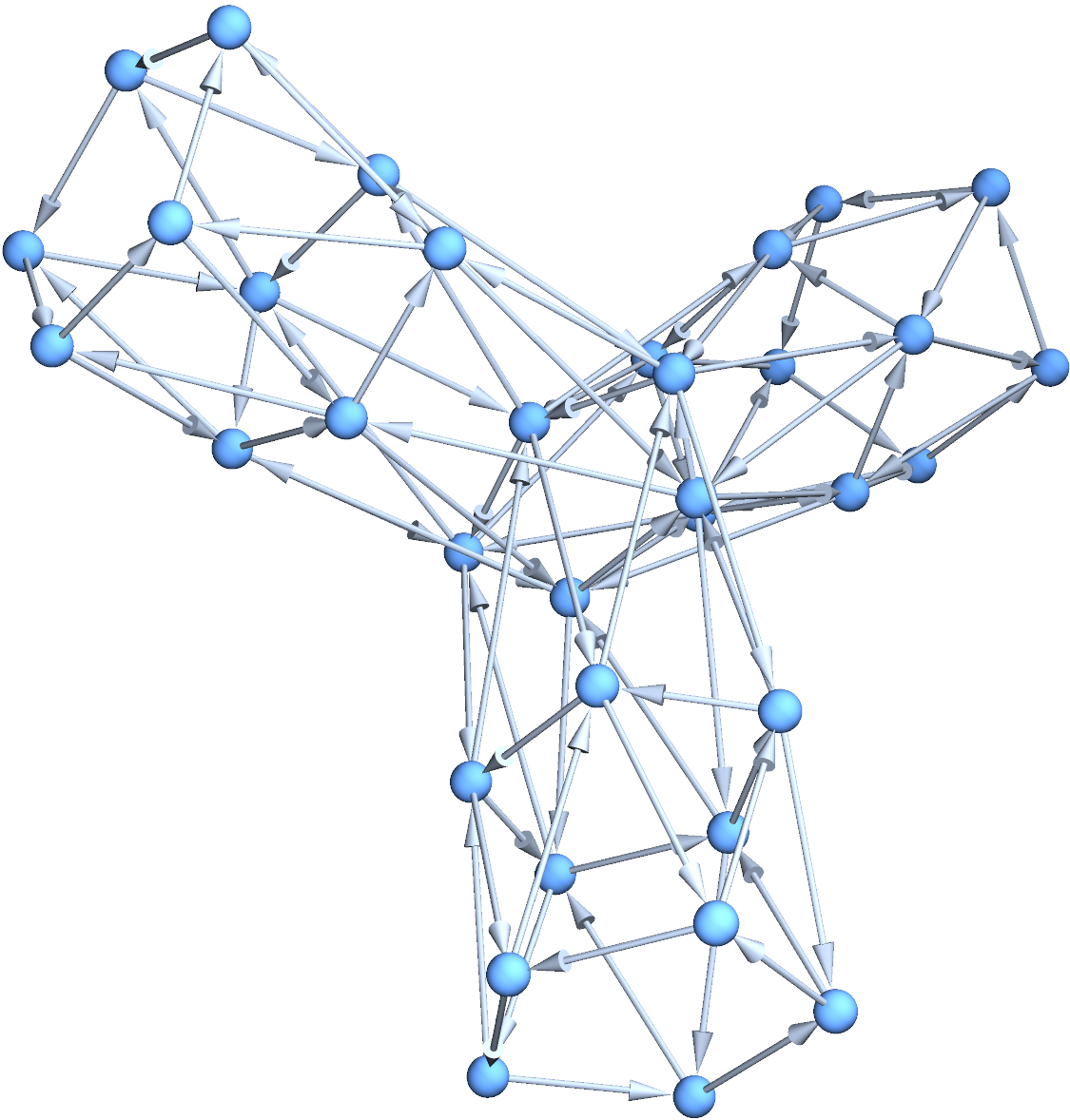}&\includegraphics[scale=0.4]{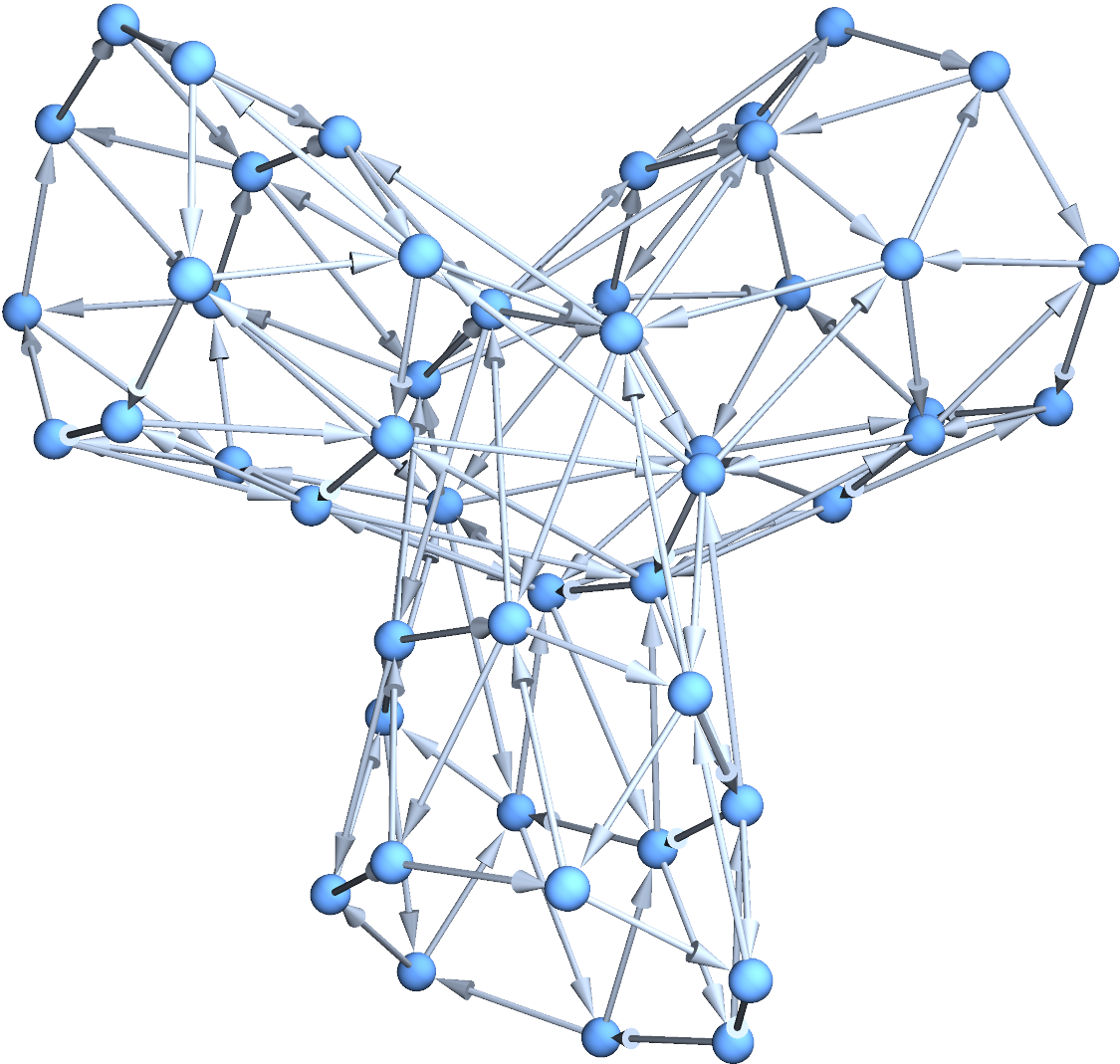}
        \end{tabular}
        \caption{\textsc{Left:} The BPS quiver corresponding to the orbifold group $T_5$; \textsc{Right:} The BPS quiver corresponding to the orbifold group $T_7$. Notice that we can recognize a box product-like structure in the quiver with an affine $\hat E_6$ structure.}\label{fig:t5}
    \end{figure}

\noindent\textbf{$T_5$ and $T_7$ orbifold SCFTs.} These cases can be analyzed in a similar way as above, and we again reproduce the defect groups $\mathbb Z_5^{(1)}$ for the $T_5$ orbifold SCFT and $\mathbb Z_7^{(1)}$ for the $T_7$ orbifold SCFT. We draw the corresponding quivers  in Figure \ref{fig:t5}, in a slightly different format to illustrate that these  have a box product form. We report the corresponding $B$ matrices in appendix \ref{app:Bmatrix}.

\medskip

\begin{figure}
        \centering
        \begin{tabular}{cc}
        \includegraphics[scale=0.2]{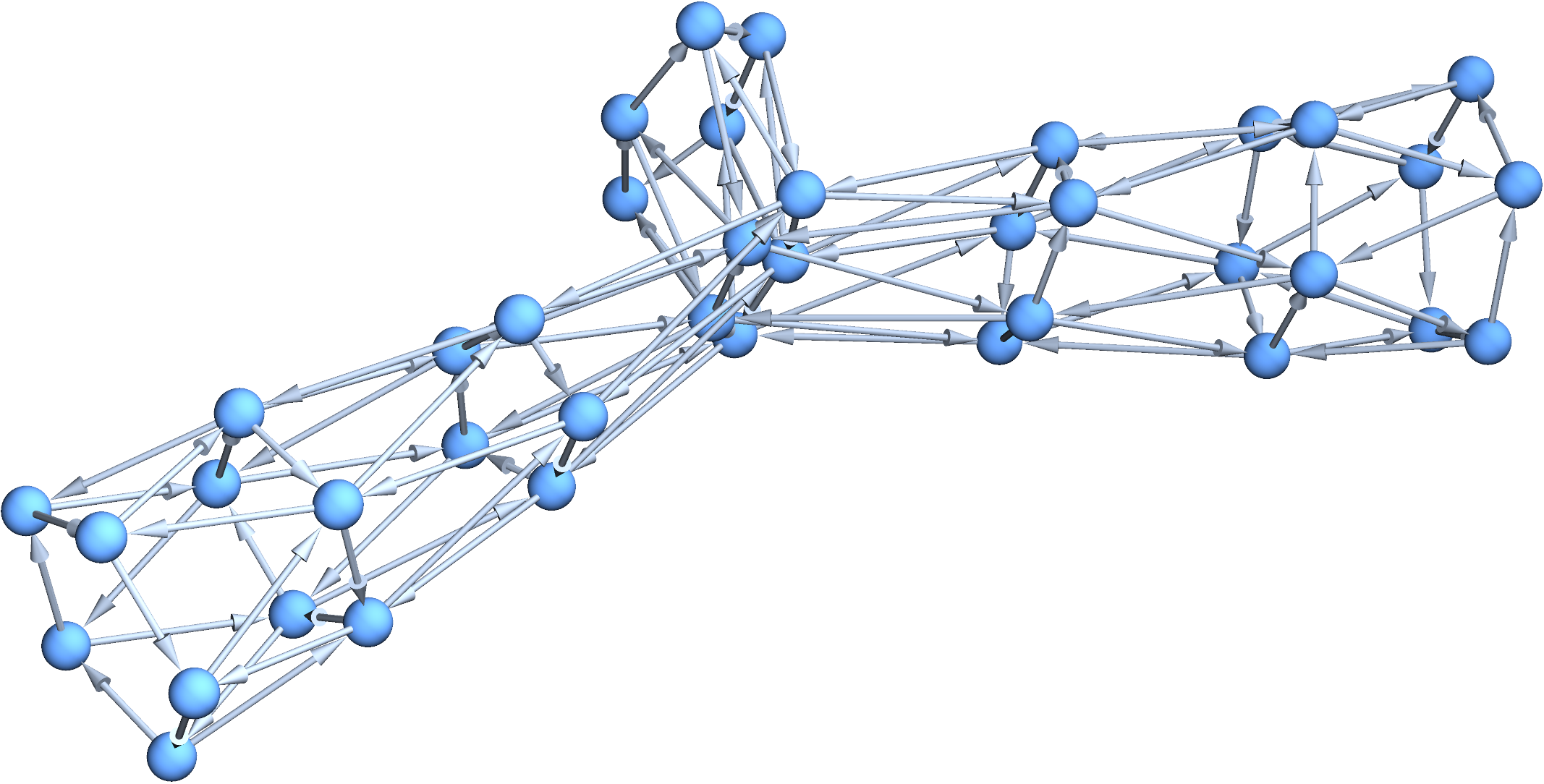}&\includegraphics[scale=0.4]{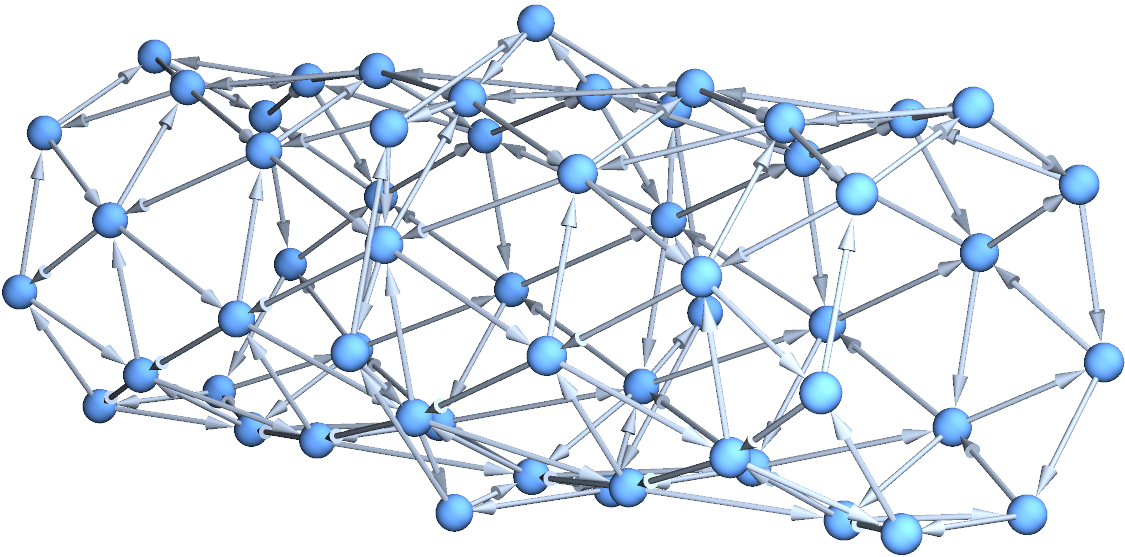}
        \end{tabular}
        \caption{\textsc{Left:} The BPS quiver corresponding to the orbifold group $O_5$; \textsc{Right:} The BPS quiver corresponding to the orbifold group $O_7$. Notice that we can recognize a box product-like structure in the quiver with an affine $\hat E_7$ structure.}\label{fig:o5}
    \end{figure}

\noindent\textbf{$O_5$ and $O_7$ orbifold SCFTs.} Also for these examples we report the relevant $B$-matrices in appendix \ref{app:Bmatrix}, which can be used to reproduce the results in table \ref{tbl:smallu2} also in these cases. We draw the corresponding quivers in Figure \ref{fig:o5} to illustrate that also these examples have quivers in a box product form.

\medskip

\noindent\textbf{General form of quivers for TOI orbifold theories.} All the other cases of orbifold 5D SCFTs with TOI orbifold groups can be analyzed similarly by extracting the corresponding quiver using the procedure of Appendix \ref{app:3dmckay}.  Based on these examples and further checks for the orbifold groups $T_m$, $O_m$ and $I_m$, we conjecture that, for suitable values of $m$, the quivers for many of the theories in this class can take the form of square tensor products too, leading to diagrams of the form
 \begin{equation}
     \hat{A}(m,0)\boxtimes \hat{E}_6\,, \qquad \hat{A}(m,0)\boxtimes \hat{E}_7\,, \qquad \hat{A}(m,0)\boxtimes \hat{E}_8,
 \end{equation}
    respectively for 5d orbifold theories of type $T_m$, $O_m$ and $I_m$. We stress this will not be the case for all values of $m$: for instance, in the case of $T_m$ theories with $3|m$, we expect to obtain quivers similar to \cref{fig:t3} consisting of an inner ring of $m$-many nodes surrounded by two rings of $3m$-many nodes connected appropriately. Moreover, these constructions will depend on choosing a suitable representative in the mutation class of the $\hat{E}_{6,7,8}$ type. Of course, for all the examples we checked, the 5d BPS quiver reproduces the result of Table \ref{tbl:smallu2}.

\subsection{Larger Subgroups}

Finally, let us briefly discuss the case of ``larger subgroups,'' namely
transitive finite subgroups $\Gamma \subset SU(3)$ which are not abelian, and which are also not induced from a subgroup of $U(2)$.
In these cases, we expect that the larger size of the group correlates with a larger fixed point set in terms of a group action on $S^5$. In fact, in all examples which we have checked, we find that the resulting defect group is trivial, simply because the normal subgroup of $\Gamma$ generated by the elements which have a fixed point is simply all of $\Gamma$! We have directly checked the adjacency matrix of the corresponding BPS quiver in these cases as well, and again confirm this result, which is in accord with the statements of \cite{Tian:2021cif}. See Table \ref{tbl:transitive} for an explicit list of these examples.

\section{$\mathrm{Ab}[\Gamma]$ and $2$-Group Symmetries}\label{sec:2group}

In the previous sections we saw that
the group $\mathrm{Ab}[\Gamma / H] = \mathbb D^{(1)}_{M2}$. Based on this, it is natural to ask whether
the abelianization of $\Gamma$ itself has any role to play in the 5d SCFT. Indeed, this structure
directly appears in the related context of 6d SCFTs. Recall that in the F-theory realization of 6d SCFTs, one
considers a canonical singularity of a non-compact elliptically fibered threefold $X \rightarrow B$. As shown in \cite{Heckman:2013pva},
the base $B$ is always of the form $\mathbb{C}^2 / \Gamma_{U(2)}$ for $\Gamma_{U(2)}$ a particular set of finite subgroups of $U(2)$, and
in all these cases, $\partial B = S^3 / \Gamma$. In this case, the corresponding
defect group is associated with a two-form symmetry, as specified by string-like defects
of the 6d SCFT \cite{DelZotto:2015isa} (see also \cite{GarciaEtxebarria:2019caf,Apruzzi:2020zot,Bhardwaj:2020phs,Apruzzi:2021mlh,Apruzzi:2021nmk,Bhardwaj:2021mzl}).
While we leave a more complete analysis for future work, in this section we observe that in
situations where the geometry faithfully reproduces the $0$-form symmetry of the system,
$\mathrm{Ab}[\Gamma]$ is closely correlated with the $2$-group symmetry of the 5d SCFT. Our plan in this section will be to first explain
some basic aspects of 2-group symmetries, following \cite{Benini:2018reh}, and then to turn to an analysis of a 5d SCFT where we can geometrically detect the $0$-form symmetry. For further discussion on aspects of $2$-group symmetries in 5d SCFTs, see e.g. \cite{Apruzzi:2021vcu}.

In order to investigate the potential role of  $\mathrm{Ab}[\Gamma]$ for 5d orbifold SCFTs,
we begin with the following two remarks:
\begin{enumerate}
\item Whenever $\Gamma$ acts without fixed points on $S^5$, $\mathbb D^{(1)}_{M2} \simeq \text{Ab}[\Gamma]$.
\item Whenever $\Gamma$ acts with fixed points on $S^5$, the theory $\TM$ typically has interesting 0-form symmetries, and  $\mathbb D^{(1)}_{M2}$ is a subgroup of $\text{Ab}[\Gamma]$.
\end{enumerate}
For a theory which has both 0-form symmetries and 1-form symmetries, the two can form a more interesting global categorical symmetry, that can organize into a 2-group. 2-groups are characterized by a 4-plet, consisting of a 0-form symmetry group $\mathbb F^{(0)}$, a 1-form symmetry group $\mathbb G^{(1)}$, a representation $\rho: \mathbb F^{(0)} \to \text{Aut}(\mathbb G^{(1)})$ and an element $\beta \in H^3(B\mathbb F^{(0)}, \mathbb G^{(1)})$, which characterizes the obstruction to switching on non-trivial backgrounds for $\mathbb F^{(0)}$ independently from backgrounds for $\mathbb G^{(1)}$. We can think of $\beta$ as if it is determining the 2-group structure constants, namely the extent to which the two higher symmetries mix with one another.

When $\Gamma$ acts with fixed points on $S^5$, the 5d SCFT typically has some non-trivial (non-abelian) 0-form symmetry, which can be characterized by the structure of the non-compact singularities in $\orb$. There is a ``naive'' answer dictated by lifting each simple Lie algebra factor to a simply connected Lie group, but this can in principle be quotiented to reach the true 0-form symmetry.
In such situations, $\mathbb D^{(1)}_{M2}$ is a strict subgroup of $\text{Ab}[\Gamma]$, and often the quotient can be understood in terms of the discrepancy between the naive 0-form symmetry and its quotiented counterpart. For example, in the case of the 5d $T_N$ theory we have an orbifold with structure $\mathbb C^3 / \mathbb Z_N \times \mathbb Z_N$. This geometry arises at the common intersection of three lines of singularities of the form $\mathbb C \times \mathbb C^2 / \mathbb Z_N$. In this case we have that $\mathbb D^{(1)}_{M2}$ is trivial and $\text{Ab}[\Gamma] = \mathbb Z_N \times \mathbb Z_N$ which in turn can be interpreted as the subgroup of the ``naive'' flavor symmetry $SU(N)^3$ by which we would quotient to reach the true global 0-form symmetry given by $SU(N)^3/\mathbb Z_N \times \mathbb Z_N$ (see e.g. \cite{Bhardwaj:2021ojs} for a discussion of the 
global symmetries in the closely related 4d $T_N$ theories). 
It is therefore tempting to claim that we have a nontrivial Postnikov class $\beta$ whenever the exact sequence
\begin{equation}\label{eq:thesequence}
1 \to \mathbb D^{(1)}_{M2} \to \text{Ab}[\Gamma] \to \text{Ab}[\Gamma]/\mathbb D^{(1)}_{M2} \to 1
\end{equation}
is non-split.

In order to check these general expectations, we seek a family of 5d orbifold SCFTs that have a Lagrangian interpretation and a non-trivial 1-form and 0-form symmetry. Since we also require that the geometry faithfully encodes the 0-form symmetry, an ideal class of examples in this case is provided by the abelian orbifolds $\mathbb C^3 / \mathbb Z_{2n}$ where $\mathbb Z_{2n}$ acts as $\frac{1}{2n}(1,1,2n-2)$. The latter give rise to 5d SCFTs with a gauge theory phase $SU(n)_{n}$. The 2-group structures have already been determined in \cite{Apruzzi:2021vcu}, and for $n$ even one finds that, choosing the electric global form of the theory, the 0-form symmetry of these models form a two group with the $\mathbb Z_n^{(1)}$ electric one-form symmetry. In our case since the group action is abelian we find that the sequence \eqref{eq:thesequence} reduces to
\begin{equation}
1 \to \mathbb Z_n^{(1)} \to \mathbb Z_{2n} \to \mathbb Z_2 \to 1\,.
\end{equation}
This sequence is non-split precisely when $n$ is even, which exactly reproduces the result of \cite{Apruzzi:2021vcu} obtained via other methods.

We find this remarkable, and it is natural to ask how this extends to other situations where geometry faithfully encodes the 0-form symmetry. As a further remark we stress here that if we were to choose the magnetic form of the theory, the 2-group structure disappears: this suggests that there is an interconnection between the Heisenberg algebra of non-commuting fluxes and the 2-group structure constants.

\section{Conclusions} \label{sec:CONC}

In this paper we have presented a general prescription for extracting the higher 1-form and 2-form symmetries
of 5d SCFTs obtained from M-theory on the orbifold $\mathbb{C}^3 / \Gamma$ with $\Gamma$ a
finite subgroup of $SU(3)$. We presented two complementary methods for extracting this data. First, building on \cite{GarciaEtxebarria:2019caf,Albertini:2020mdx,Morrison:2020ool}, we showed
how to extract it from the defining exact sequence for the defect group and the structure of $\pi_{1}(S^5 / \Gamma)$, generalizing the analysis to the case where $\Gamma$ has fixed points. Second, we showed that the same data can also be read off from the 5d BPS quiver of the corresponding SCFT, thus giving a nice consistency check to the method. We also provided some hints that the abelianization of $\Gamma$ detects the presence of a 2-group structure in such 5d SCFTs. We also remarked that the interplay with the Heisenberg algebra of non-commuting fluxes with the global form of the 5d SCFT must affect the 2-group structure in a non-trivial way. In the remainder of this section we discuss some avenues for further analysis.

Much of our analysis has focused on the computation of higher-form symmetries in these 5d SCFTs. We also saw hints of how the
2-group structure in these systems descends from the abelianization of $\Gamma$. It would be quite interesting to elucidate this structure. In particular, it would be desirable to extract the Postnikov class $\beta$ directly from the geometry of a string compactification. We think that in order to clarify this interplay it will be very fruitful to look at the symmetry TQFT \cite{Apruzzi:2021nmk} for these orbifold singularities, which arises from the reduction of the topological Chern-Simons term of M-theory on the horizon $S^5/\Gamma$.

Further compactification of these 5d SCFTs will give rise to a rich class of lower-dimensional systems. The same geometric methods
presented here can also be used to read off the corresponding higher-form symmetries of these systems. For example, compactification of our 5d SCFTs on a circle will give rise to 4d $\mathcal{N} = 2 $ SCFTs, and compactification on a $T^2$ will result in 3d $\mathcal{N} = 4$ SCFTs. Perhaps the most interesting case to study is the reduction of these 5d SCFTs on a Riemann surface $\Sigma_g$ as pioneered in \cite{Sacchi:2021wvg} to produce 3d $\mathcal N=2$ theories. In these cases the global structures we find in this paper can give rise to more interesting effects, along the lines explored for 6d (2,0) theories in \cite{Tachikawa:2013hya}.

Although it is notoriously difficult to engineer stable non-supersymmetric backgrounds in M-theory, it is nevertheless
natural to consider more general orbifold group actions $\Gamma \subset SU(4)$.\footnote{For some recent investigations
into 5d non-supersymmetric CFTs, see e.g. references \cite{BenettiGenolini:2019zth, Bertolini:2021cew,DeCesare:2021pfb}.}
In this case, we can still read off a corresponding
non-supersymmetric quiver gauge theory, though in this case the adjacency matrix for the bosonic and fermionic degrees of freedom will be different. This serves to define two separate notions of ``defect group'' depending on the boson/fermion number of the quantity in question. It would be interesting to see whether the presence of a non-trivial defect group could be used as a way of constraining the resulting non-supersymmetric dynamics.

\section*{Acknowledgments}

The work of MDZ and RM has received funding from the
European Research Council (ERC) under the European Union’s Horizon 2020 research
and innovation programme (grant agreement No. 851931) and from the Simons Foundation Grant \#888984 (Simons Collaboration on Global Categorical Symmetries). The work of JJH is supported by the DOE (HEP)
Award DE-SC0013528. The work of SM is funded by the MIUR PRIN Contract
2015 MP2CX4 ``Non-perturbative Aspects Of Gauge Theories And Strings”, by
INFN Iniziativa Specifica ST \& FI and by the scholarship granted by ``Fondazione Angelo Della Riccia".
The work of HYZ is supported by the Simons Foundation Collaboration Grant \#724069 (Simons Collaboration on Special Holonomy in Geometry, Analysis and Physics).

\newpage

\appendix

\section{Properties and Computation of \texorpdfstring{$\text{Ab}[\Gamma]$}{Ab[Gamma]} and \texorpdfstring{$\text{Ab}[\Gamma/H]$}{Ab[Gamma/H]}}\label{app:ABGAMMA}

In this Appendix, we present some computations used in the main body concerning the abelianization of $\Gamma \subset SU(3)$,
as well as its quotient $\Gamma / H$ by $H$ the normal subgroup generated by those elements of $\Gamma$ which have a fixed point locus on
$S^5 = \partial \mathbb{C}^3$.

The Appendix is organized as follows. First, in Appendix \ref{subapdx:ab_properties} we give the definition of the abelianization $\text{Ab}[\Gamma]$ of a finite group $\Gamma$ together with its properties. In Appendix \ref{subapdx:ab_algorithm} we give the algorithm of computing the abelianization via \texttt{Sage} \cite{sagemath}. The code requires as input the presentation of the finite group, which we give for all finite subgroups of $SU(3)$ in Appendix \ref{subapdx:presentations}. Finally in Appendix \ref{subapdx:ab_quotient_results}, we explain the algorithm for computing $\text{Ab}[\Gamma/H]$.

\subsection{General Aspects of Abelianization} \label{subapdx:ab_properties}
Recall that for a finite group $\Gamma$, the {\it abelianization} $\mathrm{Ab}[\Gamma]$ is defined as
\begin{gather}
    \mathrm{Ab}[\Gamma]=\frac{\Gamma}{[\Gamma,\Gamma]},
\end{gather}
where $[\Gamma,\Gamma]$ is the commutator subgroup, or derived subgroup, of $\Gamma$. A group that satisfies $\mathrm{Ab}[G]=\mathbbm{1}$ is called {\it perfect} while an abelian group clearly satisfies $\mathrm{Ab}[\Gamma]=\Gamma$.

A useful fact about $\mathrm{Ab}[\Gamma]$ is that any homomorphism $\varphi$ from $\Gamma$ to an abelian group $A$ factors through $\mathrm{Ab}[\Gamma]$, which is often referred to as the ``universal property of abelianizations". By this we mean that there exists a unique homomorphism $\psi:\mathrm{Ab}[\Gamma]\rightarrow A$ such that following diagram commutes
\begin{gather*}
    \xymatrix{\Gamma\ar[r]^{\varphi} \ar[d]_{\mathrm{quot.}} & A \\
    \mathrm{Ab}[\Gamma]\ar[ru]^{\psi}}
\end{gather*}
Using this property, we can set up the following diagram
\begin{gather*}
	\xymatrix{
		\Gamma \ar[r]^{q_1} \ar[d]^{p} & \Gamma/H \ar[r]^{q_2} & \mathrm{Ab}[\Gamma/H] \\
		 	\mathrm{Ab}[\Gamma]\ar[urr]_{u}
	}
\end{gather*}
where $N$ is a normal subgroup of $\Gamma$ and $p,q_1, q_2$ are all the natural quotient maps. From this it follows that $u$ must be surjective and, by the first isomorphism theorem, we get
\begin{gather}
    \mathrm{Ab}[\Gamma/H]\cong \frac{\mathrm{Ab}[\Gamma]}{\ker u}.
\end{gather}
Since quotients of abelian groups are isomorphic to subgroups of that same group, we get that $\mathrm{Ab}[\Gamma/H]$ is isomorphic to a subgroup of $\mathrm{Ab}[\Gamma]$.

\subsection{Algorithm} \label{subapdx:ab_algorithm}

If we can find a presentation of these groups, then \texttt{Sage} computes for us the abelianization (via \texttt{G.abelian\_invariants()}), where $\texttt{G}$ is a \texttt{FreeGroup} quotiented by equivalence relations. We can check that this works for $SU(2)$ subgroups where both the presentations and $\text{Ab}[\Gamma]$ are known.

For non-abelian subgroups of $SU(3)$, the presentation is not known explicitly. We take the following approach when finding them.
\begin{itemize}
    \item We start from the sets of matrix generators as given in \cite{Tian:2021cif}, and use them to construct a free group. We then find as many relations among them (as can be checked explicitly) as possible that we mod out by.

    \item If we do not yet have a complete set of defining relations, then the group we get $\hat{\Gamma}$ (either finite or infinite) would have $\Gamma$ as a non-trivial subgroup. In this case we go back to the previous step. If, on the other hand, we find a finite group with the correct order, then we have obtained a correct presentation of this $\Gamma \subset SU(3)$.
\end{itemize}
In the end, we use a \texttt{Sage} function to compute the abelianization of the group.

An alternative method of computing $\text{Ab}[\Gamma]$ is to input the groups as a matrix group and then use the {\tt as\_permutation\_group()} and {\tt as\_finitely\_presented\_group()} functions to convert the matrix groups into a form for which Sage can compute the abelianization readily. The advantage of this method is that we no longer need to find a presentation of the group. However, the running time of the \texttt{as\_permutation\_group()} function grows significantly as $|\Gamma|$ increases, so we have primarily used the previously listed method.

\subsection{Presentations of \texorpdfstring{$SU(3)$}{SU(3)} Subgroups} \label{subapdx:presentations}

In this Appendix we give the explicit presentations of various finite subgroups of $SU(3)$ used in the main text.

\subsubsection{Discrete Subgroups of \texorpdfstring{$U(2)$}{U(2)}}

This case is organized using rather different notation in \cite{Tian:2021cif} when compared to \cite{Carrasco2014}. We will start by doing most of the cases with the \cite{Tian:2021cif} notation, while shifting to \cite{Carrasco2014} notation when looking at ``sporadic subgroups".

\begin{itemize}

\item{$G_m$} $\quad a^{2m} = b^2 = c^2 = aba^{-1}b^{-1} = (ac)^2 = (bc)^2a^m = 1$, where $a = M_1, b = M_2, c = M_3$ as in (3.19) of \cite{Tian:2021cif}.

\item{$G_{p, q}$} $\quad a^{2pq} = b^{2q} = c^2 = aba^{-1}b^{-1} = (ac)^2 = (bc)^{4q} = (bc)^2a^p b^{-2} = 1$, where $a = M_1, b = M_2, c = M_3$ as in (3.26) of \cite{Tian:2021cif}.

\item{$G'_{m}$}
    \begin{itemize}
    \item{$m$ even} $\quad a^4 = b^{4m} = (ab)^2 = a^2 b^{2m}$, where $a = M_1, b = M_2$ as in (3.34) of \cite{Tian:2021cif}
    \item{$m$ odd} $\quad a^4 = b^{4m} = c^4 = (ab)^2 = a^2 b a^{-2} b^{-1} = a^2 b^{2m} = bcb^{-1}c^{-1} = (ac)^4 = (bc)^m$, with $a = M_1, b = M_2, c = M_3$ as in (3.37) of \cite{Tian:2021cif}
    \end{itemize}

\item{``Sporadic cases"}
    \begin{itemize}
        \item{$E^{(1)}$:} $a^3 b^{-3} = b^3 c^{-2} = abc^{-1} = d^3 a^4 = ada^{-1}d^{-1} = bdb^{-1}d^{-1} = cdc^{-1} d^{-1} = 1 $
        \item{$E^{(2)}$:} $a^4 = b^2 = aba^{-1}b^{-1} = c^3 =  cac^{-1}a^{-1} = cbc^{-1}b^{-1} = 1$
        \item{$E^{(3)}$:} $ a^4 b^{-3} = b^3 c^{-2} = abc^{-1} = d^2 a^4 = ada^{-1}d^{-1} = bdb^{-1}d^{-1} = cdc^{-1} d^{-1} = 1$
        \item{$E^{(4)}$:} $c^3 = bc^{-1}a^{-1}c = (da)^2 = c^{-1}bca^{-1}b = d^{-1}ba^{-2}d^{-1} = c^{-1}a^{-1}d^{-1}b^{-1}c^{-1}d^{-1} = 1$
        \item{$E^{(5)}$:} $ a^3 b^{-3} = b^3 c^{-2} = abc^{-1} = d^2 c^3 = ada^{-1}d^{-1} = bdb^{-1}d^{-1} = cdc^{-1} d^{-1} = 1 $
        \item{$E^{(6)}$:} $ a^3 b^{-3} = b^3 c^{-2} = abc^{-1} = d^2 a^3 = ada^{-1}d^{-1} = bdb^{-1}d^{-1} = cdc^{-1} d^{-1} = 1$
        \item{$E^{(7)}$:} $a^4 b^{-3} = b^3 c^{-2} = abc^{-1} = d^3 a^4 = ada^{-1}d^{-1} = bdb^{-1}d^{-1} = cdc^{-1} d^{-1} = 1$
        \item{$E^{(8)}$:} $ a^4 b^{-3} = b^3 c^{-2} = abc^{-1} = d^4 a^4 = ada^{-1}d^{-1} = bdb^{-1}d^{-1} = cdc^{-1} d^{-1} = 1$
        \item{$E^{(9)}$:} $ a^5 b^{-3} = b^3 c^{-2} = abc^{-1} = d^2 c^5 = = ada^{-1}d^{-1} = bdb^{-1}d^{-1} = cdc^{-1} d^{-1} = 1 $
        \item{$E^{(10)}$:} $ a^5 b^{-3} = b^3 c^{-2} = abc^{-1} = d^3 c^5 = = ada^{-1}d^{-1} = bdb^{-1}d^{-1} = cdc^{-1} d^{-1} = 1 $
        \item{$E^{(11)}$:} $ a^5 b^{-3} = b^3 c^{-2} = abc^{-1} = d^5 c^5 = = ada^{-1}d^{-1} = bdb^{-1}d^{-1} = cdc^{-1} d^{-1} = 1 $
    \end{itemize}

\item{$D_{n, q}$ cases}

    There are all small subgroups of $U(2)$ whose presentation and abelianization have been presented in (4.26) - (4.27) of \cite{DelZotto:2015isa}.)

\item{$T_m, O_m, I_m$ cases}

    We notice that $O_m = O \times \bbZ_m$ for all allowed $m$ (namely $(m, 6 = 1)$), $I_m = I \times \bbZ_m$ for all allowed $m$ (namely $(m, 30) = 1$), and $T_m = T \otimes \bbZ_m$ for $m \equiv 1, 5 \mod\ 6$, where $T, O, I \subset SU(2)$ are the tetrahedral, octahedral and isocahedral groups, respectively. So for these cases, the abelianization follows from those of $SU(2)$ finite subgroups, and for the latter we refer to \cite{DelZotto:2015isa}.

    For $T_m$ where $m = 6k+3$, the above direct product expression no longer holds. Instead, the presentation is given by $ bab^{-1}a = a^4 = ab^{-2}a = acbc^{-1} = bcabc^{-1} = c^{3m} = 1 $.
\end{itemize}

\subsubsection{Transitive subgroups of \texorpdfstring{$SU(3)$}{SU(3)}}

\begin{itemize}
    \item{$\Delta$ series}
    \begin{itemize}
    \item{$\Delta(3n^2)$} $\quad a^n = b^n = aba^{-1}b^{-1} = afb^{-1}f^{-1} = (af)^3$, where $a = L_n, f = E$ as in (3.56) of \cite{Tian:2021cif}, and $b = \text{diag}\{1, \omega_n, \omega_n^{-1}\}$.
    \item{$\Delta(6n^2)$} generated with $a, b, f$ as in $\Delta(3n^2)$ with the above relations, and $h = I$ as in (3.62) of \cite{Tian:2021cif} such that $h^2 = (hf)^2 = hahb = 1.$
    \end{itemize}
    \item{$C^{(k)}_{n, l}$}
    \begin{itemize}
        \item $(r, k, l) = (3, 1, l),\ 3 | l$  $\quad$ defined by $a^{3l} = b^l = f^3 = afa^{-1}f^{-1} = (af)^3 = faf^{-1}a^{-1}b^{-1} = 1$, where $a = B_{9, 1}, b = G_{7, 1}, f = E$ as in (3.83) of \cite{Tian:2021cif}.
        \item $(r, k, l) = (7, 2, l)$ $\quad$ defined by $a^{7l} = b^l = f^3 = afa^{-1}f^{-1} = (af)^3 = afa^{-1}f^{-1}a b^{-1} = 1$, where $a = B_{7, 2}, b = G_{7l, 7}, f = E$ as in (3.90) of \cite{Tian:2021cif}
    \end{itemize}
    \item{$D^{(1)}_{3l, l}, \ 2 | l$} $\quad$ generated with $a = B_{3l, l}, b = G_{7, 7}, f = E, h = I$ as in (3.62) and (3.83) such that $a^{3l} = b^l = f^3 = h^2 = aba^{-1}b^{-1} = (af)^3 = faf^{-1}a^{-1}b^{-1} = h a h b a^2 = (fh)^2 = afba^2f^{-1} = (ba^2f)^3 = 1$
    \item{Exceptional subgroups}
    \begin{itemize}
    \item{$H_{36}$:} $\quad a^3 = b^3 = f^3 = z^4 = aba^{-1}b^{-1} = afb^{-1}f^{-1}b^{-1}a = azafz^{-1}f^{-1}a^{-1}b = azfz^{-1} = 1$,  where $a = M_1, f = M_2, z = M_3, b = \text{diag}\{\omega_3, \omega_3^2, 1\}$
    \item{$H_{72}$:} $\quad a^3 = b^3 = f^3 = z^4 = w^4 = aba^{-1}b^{-1} = afb^{-1}f^{-1}b^{-1}a = azfz^{-1} = z^2aw^2 = z^2 f^2 z w^3 z f b^{-1} a w^{-1} = 1$, where $a, b, f, z$ as defined in $H_{36}$, and $w = M_4$ as in (3.125) of \cite{Tian:2021cif}.
    \item{$H_{216}$:} $\quad$ defined with $a, b, f, w, z$ as in $H_{72}$ satisfying all the above relations, and $y = M_4$ in (3.137) of \cite{Tian:2021cif} with the extra relation that $y^9 = yay^{-1}a^{-1} = (yz)^3 = wyz^{-1}y^{-1} = 1$
    \item{$H_{60}$:} $\quad a^5 = b^2 = c^2 = (ab)^3 = (ac)^2 = (abc)^5 = (bc)^2 = 1$, where $a = H_1, b = H_2, c = H_3$ according to section 3.4.1 of \cite{Carrasco2014}.
    \item{$H_{360}$} Defined with $a, b, c$ as in $H_{60}$ with the above relations, and $d = M_4$ in (3.145) of \cite{Tian:2021cif} with extra relations $d^2 = (ad)^3 = (cd)^2 = (adbdc)^3 = (aba^2cad)^3 = 1$
    \item{$H_{168}$} $a^7 = b^7 = aba^{-1}b^{-1} = E^3 = aEb^{-1}E^{-1} = Q^2 = (Qa)^4 = (Qb)^7 = (QE)^2 = (Qab)^3 = 1$, where $a = M_1, E = M_2, Q = M_3$ as in (3.132) of \cite{Tian:2021cif}, and $b = \text{diag}\{\omega_7^4, \omega_7, \omega_7^2\}$ is a suitable permutation of the diagonal entries in $a$.
    \item{$J$} This is a case with order 180 that was filled in by Yau and Yu \cite{yau1993gorenstein}, which is isomorphic to $H_{60} \times \bbZ_3$, so $\text{Ab}[J] = \text{Ab}[H_{60}] \times \bbZ_3$.
    \item{$K$} Similar to $J$, this is a case with order 504 such that $K = H_{168} \times \bbZ_3$, so $\text{Ab}[K] = \text{Ab}[H_{168}] \times \bbZ_3$.
    \end{itemize}

\end{itemize}

\subsection{Computation of $\text{Ab}[\Gamma/H]$} \label{subapdx:ab_quotient_results}

Given a explicit quotient singularity $\bbC^3/\Gamma$ specified by a discrete group $\Gamma$ and its explicit action on $\bbC^3$, we now give a \texttt{Sage} algorithm to compute the normal subgroup $H$ generated by all elements inside $\Gamma$ whose action has fixed points, and eventually $\text{Ab}[\Gamma/H]$.

An action of $\Gamma$ on $\bbC^3$ is determined by a representation $\rho: g \rightarrow GL(3, \bbC)$ that assign to an element $g \in \Gamma$ a 3-by-3 matrix $\rho(g)$ with complex entries. $\rho(g)$ will has a fixed element $\mathbf{v} \in \bbC^3$ if and only if
\begin{equation}
    \exists \mathbf{v}  \in \bbC^3\backslash\{0\} \ \text{s.t.}\ \rho(g) \mathbf{v} = \mathbf{v} \ \ \Leftrightarrow\ \  |\rho(g) - I| = 0,
\end{equation}
where $I$ is the 3-by-3 identity matrix. So our task is to ask \texttt{Sage} to compute $|\rho(g) - I|$ for all $g \in \Gamma$, and then determine the normal subgroup $H \trianglelefteq \Gamma$ that these elements generate.

\medskip

\noindent \textbf{Computing $H$.} Technically, one can simplify this by noticing that $\forall h \in \Gamma, |\rho(hgh^{-1} - I)| = |\rho(h)||\rho(g) - I||\rho^{-1}(h)| = |\rho(g) - I|$, so  $|\rho(g) - I|$ only depends on the conjugacy class that $g$ sits in. So we need these steps to compute $H$:
\begin{itemize}
    \item Determine the list of conjugacy classes of $\Gamma$
    \item Find an element $g$ for each such conjugacy class and compute its $|\rho(g) - I|$, and thus determining $|\rho(g) - I|$ for this entire conjugacy class
    \item Take the union of all conjugacy classes that has $|\rho(g) - I| = 0$, and compute the subgroup $H$ which they generate.
\end{itemize}

\medskip

\noindent \textbf{Computing $\text{Ab}|\Gamma/H|$.} Having both $\Gamma$ and $H$ explicitly, we can then use \texttt{Sage} to compute $\Gamma/H$ as well as $\mathrm{Ab}[\Gamma/H]$.

\section{Group Theory of \texorpdfstring{$D_{n,q}, T_m, O_m$ and $I_m$}{Dnq, Tm, Om and Im}}\label{app:conv}
In this Appendix we present the explicit generators for the finite subgroups of $SU(3)$ induced by finite subgroups of $U(2)$ specified
by $D_{n,q}, T_m, O_m$ and $I_m$. These are essentially just twists by an additional cyclic subgroup of the familiar $D$ and $E$-series finite subgroups of $SU(2)$. Our discussion follows that in reference \cite{yau1993gorenstein}.

\subsection{\texorpdfstring{$D_{n,q}$}{Dnq}}
This group is built from a small representation of an extension of the binary dihedral group $\mathbf{BD_n}$. To ensure the smallness of this representation, we require $1<q<n$ and $(n,q)=1$. Furthermore, we must split this case into two subclasses.

Taking $m=n-q$ to be odd, we can generate $D_{n,q}$ by
\begin{gather}
    D_{n,q}=\Bigg\langle
        \begin{pmatrix}
            \zeta_{2q} & 0 & 0\\
            0 & \zeta_{2q}^{-1} & 0\\
            0 & 0 & 1
        \end{pmatrix},
        \begin{pmatrix}
            0 & i & 0 \\
            i & 0 & 0 \\
            0 & 0 & 1
        \end{pmatrix},
        \begin{pmatrix}
            \zeta_{2m} & 0 & 0\\
            0 & \zeta_{2m} & 0\\
            0 & 0 & \zeta_{2m}^{-2}
        \end{pmatrix}
    \Bigg\rangle.
\end{gather}
where $\zeta_k=e^{2\pi i/k}$. Now taking $m=n-q$ even, we generate the group as
\begin{gather}
    D_{n,q}=\Bigg\langle
        \begin{pmatrix}
            \zeta_{2q} & 0 & 0\\
            0 & \zeta_{2q}^{-1} & 0\\
            0 & 0 & 1
        \end{pmatrix},
        \begin{pmatrix}
            0 & i & 0 \\
            i & 0 & 0 \\
            0 & 0 & 1
        \end{pmatrix}\cdot
        \begin{pmatrix}
            \zeta_{4m} & 0 & 0\\
            0 & \zeta_{4m} & 0\\
            0 & 0 & \zeta_{4m}^{-2}
        \end{pmatrix}
    \Bigg\rangle.
\end{gather}
Note that in the even case, there are only two generators.
\subsection{\texorpdfstring{$T_m$}{Tm}}
This group is built from a small representation of an extension of the binary tetrahedral group $\mathbf{BT}$. The smallness condition in this case is that $m$ must be odd. Again, we must split this into two further cases.

Taking $m=1$ or $5$ (mod $6)$, we generate the group as
\begin{gather}
    \Bigg\langle
    \begin{pmatrix}
        i & 0 & 0 \\
        0 & -i& 0 \\
        0 & 0 & 1
    \end{pmatrix},
    \begin{pmatrix}
        0 & i & 0 \\
        i & 0 & 0 \\
        0 & 0 & 1
    \end{pmatrix},
    \begin{pmatrix}
        (1+i)/2 & (-1+i)/2 & 0 \\
        (1+i)/2 & (1-i)/2 & 0 \\
        0 & 0 & 1
    \end{pmatrix},
    \begin{pmatrix}
        \zeta_{2m} & 0 & 0 \\
        0 & \zeta_{2m} & 0 \\
        0 & 0 & \zeta_{2m}^{-2}
    \end{pmatrix}\Bigg\rangle.
\end{gather}
Taking $m=3$ (mod $6)$, we generate the group by
\begin{gather}
    \Bigg\langle
    \begin{pmatrix}
        i & 0 & 0 \\
        0 & -i& 0 \\
        0 & 0 & 1
    \end{pmatrix},
    \begin{pmatrix}
        0 & i & 0 \\
        i & 0 & 0 \\
        0 & 0 & 1
    \end{pmatrix},
    \zeta_{6m}\cdot\begin{pmatrix}
        (1+i)/2 & (-1+i)/2 & 0 \\
        (1+i)/2 & (1-i)/2 & 0 \\
        0 & 0 & \zeta_{6m}^{-3}
    \end{pmatrix}\Bigg\rangle.
\end{gather}
\subsection{\texorpdfstring{$O_m$}{Om}}
This group is built from a small representation of an extension of the binary octahedral group $\mathbf{BO}$. To ensure smallness, we impose $(m,6)=1$. This is enough to give one set of generators for any valid $m$. The group is therefore always generated by
\begin{gather}
    \Bigg\langle
    \begin{pmatrix}
        \zeta_8 & 0 & 0\\
        0 & \zeta_8^{-1} & 0 \\
        0 & 0 & 1
    \end{pmatrix},
    \begin{pmatrix}
        0 & i & 0 \\
        i & 0 & 0 \\
        0 & 0 & 1
    \end{pmatrix},
    \begin{pmatrix}
        (1+i)/2 & (-1+i)/2 & 0 \\
        (1+i)/2 & (1-i)/2 & 0 \\
        0 & 0 & 1
    \end{pmatrix},
    \begin{pmatrix}
        \zeta_{2m} & 0 & 0 \\
        0 & \zeta_{2m} & 0 \\
        0 & 0 & \zeta_{2m}^{-2}
    \end{pmatrix}\Bigg\rangle.
\end{gather}
\subsection{\texorpdfstring{$I_m$}{Im}}
This group is built from a small representation of an extension of the binary icosahedral group $\mathbf{BI}$. The smallness condition is given by $(m,30)=1$. The group can always be generated as
\begin{gather}
    \Bigg\langle\begin{pmatrix}
        \zeta_{2m} & 0 & 0 \\
        0 & \zeta_{2m} & 0 \\
        0 & 0 & \zeta_{2m}^{-2}
    \end{pmatrix},
    \begin{pmatrix}
        0 & -1 & 0 \\
        1 & 0 & 0 \\
        0 & 0 & 1
    \end{pmatrix},
    \begin{pmatrix}
        \zeta_5^3 & 0 & 0 \\
        0 & \zeta_5^2 & 0 \\
        0 & 0 & 1
    \end{pmatrix},
    \frac{1}{\sqrt{5}}
    \begin{pmatrix}
        \zeta_5^4-\zeta_5 & \zeta_5^2-\zeta_5^3 & 0 \\
        \zeta_5^2-\zeta_5^3 & \zeta_5-\zeta_5^4 & 0\\
        0 & 0 & \sqrt{5}
    \end{pmatrix}\Bigg\rangle.
\end{gather}



\section{3d McKay Correspondence}\label{app:3dmckay}

In this Appendix we review the construction of the BPS quiver obtained from D-branes probing an orbifold singularity. In the main body of the text, this is used to compute the BPS quiver of the 5d SCFT, and is equivalent to extracting the worldvolume theory of a probe D0-brane of the orbifold singularity. With this in mind, it suffices to consider a T-dual description as obtained from 4d $\mathcal{N}=4$ Super-Yang Mills theory with gauge group $U(n)$. We then can apply the general orbifold prescription of \cite{Douglas:1996sw} as described in references \cite{Hanany:1998sd,Lawrence:1998ja}.

To begin, recall that the matter content of 4d $\mathcal{N}=4$ SYM is given by adjoint valued fields in the singlet, fundamental and two-index anti-symmetric representation of the R-symmetry $SU(4)$, respectively describing the vector bosons, fermions, and scalars of the theory:
\begin{center}
	\begin{tabular}{|c|c|c|}
		\hline
		&$U(n)$ gauge &  SU(4) R-symmetry \\
		\hline
		$A_{IJ}$ & Adj. & {\bf 1} \\ \hline
		$\psi^\alpha_{IJ}$ & Adj. & {\bf 4}  \\ \hline
		$\Phi^\alpha_{IJ}$ & Adj. & {\bf 6}  \\ \hline
	\end{tabular}
\end{center}

To construct the orbifolded theory, we can consider $\Gamma$, a finite subgroup of $SU(3)$, which is in turned embedded in the $SU(4)_R$ symmetry so that the fundamental decomposes as $\mathbf{4} \rightarrow \mathbf{3} \oplus \mathbf{1}$.
Under the orbifold action, the indices of the vector boson break into various representation of $\Gamma$, $\gamma_i$, such that
\begin{align}
	U(N) \to \prod_i U(N_i),
\end{align}
where $N_i=n \text{ dim}(\gamma_i)$.

Indeed, we can write adjoint fields of $U(N)$ as $ Hom(\mathbb{C}^N,\mathbb{C}^N)$.
When we take the orbifold quotient, we keep only the invariant homomorphisms
under the action of irreducible representations (irreps) of $\Gamma$
\begin{align}
	(\Hom(\mathbb{C}^N,\mathbb{C}^N))^\Gamma=\bigoplus_{i}\Hom(\mathbb{C}^{N_i},\mathbb{C}^{N_i}).
\end{align}
Since we are dealing with a brane probe theory which preserves 4d $\mathcal{N} = 1$ supersymmetry, it suffices to consider the fermions, which will be paired with scalar degrees of freedom.

The $\psi^\alpha_{IJ}$ transforms as ${\bf 4}_R \otimes \Hom(\mathbb{C}^N,\mathbb{C}^N)$. When we quotient we get:
\begin{align}
	({\bf 4}_R \otimes \Hom(\mathbb{C}^N,\mathbb{C}^N))^\Gamma=\bigoplus_{i,j} a^{\bf 4}_{i j }\Hom(\mathbb{C}^{N_i},\mathbb{C}^{N_j}).
\end{align}
The fermions are now bifundamentals charged under the various $U(N_i)$, and the matrix $a^{\bf 4}_{i j }$ gives the adjacency matrix for the quiver describing the theory. To be precise,  $a^{\bf 4}_{i j }$ gives the number of arrows from node $i$ to node $j$ in the quiver.
To compute $a^{\bf 4}_{i j }$, we use the following decomposition:
\begin{align}
	{\bf 4}_R \otimes \gamma_i = \oplus_{j} a^{\bf 4}_{i j } \gamma_j.
\end{align}
We now trace this relation to have a relation between characters of irreducible representations:
\begin{align}
	\chi({\bf 4}_R)^\alpha \chi(\gamma_i)^\alpha = \sum_j a^{\bf 4}_{i j } \chi(\gamma_j)^\alpha,
\end{align}
where $\alpha$ indicates the conjugacy class. Using the orthogonality of the characters we can express $a^{\bf 4}_{i j }$ as
\begin{align}\label{aijalpha}
	a^{\bf 4}_{i j }=\frac{1}{\vert \Gamma \vert} \sum_\alpha r_\alpha \chi({\bf 4}_R)^\alpha  \chi(\gamma_i)^\alpha \overline{\chi(\gamma_j)^\alpha},
\end{align}
where $r_\alpha$ counts the dimension of the $\alpha$ conjugacy class and the bar means complex conjugate.

What we need to specify now is $\chi({\bf 4}_R)^\alpha$. Using the decomposition ${\bf 4_R} \rightarrow \mathbf{3} \oplus \mathbf{1}$, for the fundamental of $SU(4)_R$ into $SU(3)$, the character becomes
\begin{align}
	\chi({\bf 4}_R)^\alpha=\chi({\bf 1})^\alpha+\chi({\bf 3})^\alpha=1+\chi({\bf 3})^\alpha.
\end{align}
This tells us that $\Gamma$ acts on fermions with a three dimensional representation, which needs not to be irreducible. We have the following possible decompositions for three dimensional representations:
\begin{align} \label{irrepdecomp}
	{\bf 1'}\oplus{\bf 1''}\oplus{\bf 1'''} &\to \chi({\bf 1'})^\alpha +\chi({\bf 1''})^\alpha+ \chi({\bf 1'''})^\alpha \nonumber  \\
	{\bf 1'}\oplus{\bf 2} &\to \chi({\bf 1'})^\alpha + \chi({\bf 2})^\alpha \nonumber \\
	{\bf 3} &\to \chi({\bf 3})^\alpha,
\end{align}
where ${\bf 1'}$ is a one-dimensional, possibly non-trivial, irreducible representation.

In order to choose a consistent decomposition of a three-dimensional representation in term of irreducible representations,
we require that the product of the determinant of all irreducible representations be unity.
This means that the one-dimensional representation must be chosen such that
\begin{align}
	\prod_\alpha \chi(\bf 1')^\alpha\chi(\bf 1'')^\alpha\chi(\bf 1''')^\alpha=1
\end{align}
since the character table of one dimensional irreps corresponds with the representations itself. For the ${\bf 1'}\oplus{\bf 2}$ we can also work out the determinant of the two-dimensional irreducible representations using the Adams Operations.

\subsection{Computing the Defect Group}

As briefly discussed in section \ref{sec:PRESCRIPTION},
from the matrix $a_{ij}^{\bf 4}$ we can compute the Dirac pairing
\begin{align}
    B_{ij}=a_{ij}^{\bf 4}-a_{ji}^{\bf 4},
\end{align}
and using the results of \cite{Caorsi:2017bnp} (see also \cite{Albertini:2020mdx,Hosseini:2021ged}), we can extract the defect group of the theory.

Since $B$ is a $n \times n$ matrix with entries in $\mathbb{Z}$, one can decompose it into Smith Normal Form (SNF). This amounts to finding
invertible matrices $S$ and $T$ over $\mathbb{Z}$ such that $B=SB_{SNF}T$, this is a change of base of $B$. In the new frame, $B_{SNF}=\text{diag}\{a_1,a_2,a_3,\hdots,a_m,0,\hdots,0\}$, such that $a_i$ are integers and $a_i$ divides $a_{i+1}$ for each $i < m$.

The matrices $B$ and $B_{SNF}$ have the same cokernel given by
\begin{equation}
    \text{coker} (B)=\text{coker} (B_{SNF})=\mathbb{Z}^l\oplus \mathbb{Z}^m / (a_1\mathbb{Z}\oplus  a_2\mathbb{Z}\oplus\hdots\oplus a_m\mathbb{Z})\, ,
\end{equation}
with $l$ the number of zero diagonal elements of $B_{SNF}$, corresponding to vectors which lie in $\mathrm{ker}(Q)$. We also have that the $a_i$ comes in equal pair, corresponding to electric and magnetic defect charges.

As discussed in \cite{Albertini:2020mdx}, the cokernel of $B$ gives the defect group of the theory. In particular, we have that
\begin{equation}
   \text{Tor } \bbD^{(1)} = \text{Tor}(\text{coker}(B))=(\bbZ_{a_1}\oplus \bbZ_{a_1}) \oplus \hdots \oplus (\bbZ_{a_{m/2}} \oplus \bbZ_{a_{m/2}}),
\end{equation}
where the each pair in parenthesis represent a couple of non local defect charges. This information is not enough to fully determine the Heisenberg algebra of non-commuting fluxes. The latter can be reconstructed exploiting the prescription on the Weyl pairing discussed in \cite{Caorsi:2017bnp}, to which we refer our readers.

\section{Selected B-matrices}\label{app:Bmatrix}

\subsection{$T_5$ orbifold SCFT}

\scalebox{0.4}{$
    \left(
\begin{array}{ccccccccccccccccccccccccccccccccccc}
 0 & 1 & 0 & 0 & -1 & 0 & 0 & 0 & 0 & 0 & 0 & 0 & 0 & 0 & 0 & 0 & 0
   & 0 & 0 & 0 & 0 & 0 & 0 & 0 & 0 & 0 & 0 & 1 & -1 & 0 & 0 & 0 & 0
   & 0 & 0 \\
 -1 & 0 & 1 & 0 & 0 & 0 & 0 & 0 & 0 & 0 & 0 & 0 & 0 & 0 & 0 & 0 & 0
   & 0 & 0 & 0 & 0 & 0 & 0 & 0 & 0 & 0 & 0 & 0 & 1 & -1 & 0 & 0 & 0
   & 0 & 0 \\
 0 & -1 & 0 & 1 & 0 & 0 & 0 & 0 & 0 & 0 & 0 & 0 & 0 & 0 & 0 & -1 &
   0 & 0 & 0 & 0 & 0 & 0 & 0 & 0 & 0 & 0 & 0 & 0 & 0 & 1 & 0 & 0 &
   0 & 0 & 0 \\
 0 & 0 & -1 & 0 & 1 & 0 & 0 & 0 & 0 & 0 & 0 & 0 & 0 & 0 & 0 & 1 & 0
   & 0 & 0 & 0 & 0 & 0 & 0 & 0 & 0 & 0 & -1 & 0 & 0 & 0 & 0 & 0 & 0
   & 0 & 0 \\
 1 & 0 & 0 & -1 & 0 & 0 & 0 & 0 & 0 & 0 & 0 & 0 & 0 & 0 & 0 & 0 & 0
   & 0 & 0 & 0 & 0 & 0 & 0 & 0 & 0 & 0 & 1 & -1 & 0 & 0 & 0 & 0 & 0
   & 0 & 0 \\
 0 & 0 & 0 & 0 & 0 & 0 & 0 & 1 & 0 & 0 & -1 & 0 & 0 & 0 & 0 & 0 & 0
   & 0 & 0 & 1 & -1 & 0 & 0 & 0 & 0 & 0 & 0 & 0 & 0 & 0 & 0 & 0 & 0
   & 0 & 0 \\
 0 & 0 & 0 & 0 & 0 & 0 & 0 & 0 & 0 & 0 & 0 & 1 & 0 & 0 & -1 & 0 & 0
   & 0 & 0 & 0 & 0 & 0 & 0 & 1 & -1 & 0 & 0 & 0 & 0 & 0 & 0 & 0 & 0
   & 0 & 0 \\
 0 & 0 & 0 & 0 & 0 & -1 & 0 & 0 & 1 & 0 & 0 & 0 & 0 & 0 & 0 & 0 & 0
   & 0 & 0 & 0 & 1 & -1 & 0 & 0 & 0 & 0 & 0 & 0 & 0 & 0 & 0 & 0 & 0
   & 0 & 0 \\
 0 & 0 & 0 & 0 & 0 & 0 & 0 & -1 & 0 & 1 & 0 & 0 & 0 & 0 & 0 & 0 &
   -1 & 0 & 0 & 0 & 0 & 1 & 0 & 0 & 0 & 0 & 0 & 0 & 0 & 0 & 0 & 0 &
   0 & 0 & 0 \\
 0 & 0 & 0 & 0 & 0 & 0 & 0 & 0 & -1 & 0 & 1 & 0 & 0 & 0 & 0 & 0 & 1
   & 0 & -1 & 0 & 0 & 0 & 0 & 0 & 0 & 0 & 0 & 0 & 0 & 0 & 0 & 0 & 0
   & 0 & 0 \\
 0 & 0 & 0 & 0 & 0 & 1 & 0 & 0 & 0 & -1 & 0 & 0 & 0 & 0 & 0 & 0 & 0
   & 0 & 1 & -1 & 0 & 0 & 0 & 0 & 0 & 0 & 0 & 0 & 0 & 0 & 0 & 0 & 0
   & 0 & 0 \\
 0 & 0 & 0 & 0 & 0 & 0 & -1 & 0 & 0 & 0 & 0 & 0 & 1 & 0 & 0 & 0 & 0
   & 0 & 0 & 0 & 0 & 0 & 0 & 0 & 1 & -1 & 0 & 0 & 0 & 0 & 0 & 0 & 0
   & 0 & 0 \\
 0 & 0 & 0 & 0 & 0 & 0 & 0 & 0 & 0 & 0 & 0 & -1 & 0 & 1 & 0 & 0 & 0
   & -1 & 0 & 0 & 0 & 0 & 0 & 0 & 0 & 1 & 0 & 0 & 0 & 0 & 0 & 0 & 0
   & 0 & 0 \\
 0 & 0 & 0 & 0 & 0 & 0 & 0 & 0 & 0 & 0 & 0 & 0 & -1 & 0 & 1 & 0 & 0
   & 1 & 0 & 0 & 0 & 0 & -1 & 0 & 0 & 0 & 0 & 0 & 0 & 0 & 0 & 0 & 0
   & 0 & 0 \\
 0 & 0 & 0 & 0 & 0 & 0 & 1 & 0 & 0 & 0 & 0 & 0 & 0 & -1 & 0 & 0 & 0
   & 0 & 0 & 0 & 0 & 0 & 1 & -1 & 0 & 0 & 0 & 0 & 0 & 0 & 0 & 0 & 0
   & 0 & 0 \\
 0 & 0 & 1 & -1 & 0 & 0 & 0 & 0 & 0 & 0 & 0 & 0 & 0 & 0 & 0 & 0 & 0
   & 0 & 0 & 0 & 0 & 0 & 0 & 0 & 0 & 0 & 1 & 0 & 0 & -1 & 0 & 0 & 1
   & -1 & 0 \\
 0 & 0 & 0 & 0 & 0 & 0 & 0 & 0 & 1 & -1 & 0 & 0 & 0 & 0 & 0 & 0 & 0
   & 0 & 1 & 0 & 0 & -1 & 0 & 0 & 0 & 0 & 0 & 0 & 0 & 0 & 0 & 0 & 1
   & -1 & 0 \\
 0 & 0 & 0 & 0 & 0 & 0 & 0 & 0 & 0 & 0 & 0 & 0 & 1 & -1 & 0 & 0 & 0
   & 0 & 0 & 0 & 0 & 0 & 1 & 0 & 0 & -1 & 0 & 0 & 0 & 0 & 0 & 0 & 1
   & -1 & 0 \\
 0 & 0 & 0 & 0 & 0 & 0 & 0 & 0 & 0 & 1 & -1 & 0 & 0 & 0 & 0 & 0 &
   -1 & 0 & 0 & 1 & 0 & 0 & 0 & 0 & 0 & 0 & 0 & 0 & 0 & 0 & 0 & 0 &
   0 & 1 & -1 \\
 0 & 0 & 0 & 0 & 0 & -1 & 0 & 0 & 0 & 0 & 1 & 0 & 0 & 0 & 0 & 0 & 0
   & 0 & -1 & 0 & 1 & 0 & 0 & 0 & 0 & 0 & 0 & 0 & 0 & 0 & -1 & 0 &
   0 & 0 & 1 \\
 0 & 0 & 0 & 0 & 0 & 1 & 0 & -1 & 0 & 0 & 0 & 0 & 0 & 0 & 0 & 0 & 0
   & 0 & 0 & -1 & 0 & 1 & 0 & 0 & 0 & 0 & 0 & 0 & 0 & 0 & 1 & -1 &
   0 & 0 & 0 \\
 0 & 0 & 0 & 0 & 0 & 0 & 0 & 1 & -1 & 0 & 0 & 0 & 0 & 0 & 0 & 0 & 1
   & 0 & 0 & 0 & -1 & 0 & 0 & 0 & 0 & 0 & 0 & 0 & 0 & 0 & 0 & 1 &
   -1 & 0 & 0 \\
 0 & 0 & 0 & 0 & 0 & 0 & 0 & 0 & 0 & 0 & 0 & 0 & 0 & 1 & -1 & 0 & 0
   & -1 & 0 & 0 & 0 & 0 & 0 & 1 & 0 & 0 & 0 & 0 & 0 & 0 & 0 & 0 & 0
   & 1 & -1 \\
 0 & 0 & 0 & 0 & 0 & 0 & -1 & 0 & 0 & 0 & 0 & 0 & 0 & 0 & 1 & 0 & 0
   & 0 & 0 & 0 & 0 & 0 & -1 & 0 & 1 & 0 & 0 & 0 & 0 & 0 & -1 & 0 &
   0 & 0 & 1 \\
 0 & 0 & 0 & 0 & 0 & 0 & 1 & 0 & 0 & 0 & 0 & -1 & 0 & 0 & 0 & 0 & 0
   & 0 & 0 & 0 & 0 & 0 & 0 & -1 & 0 & 1 & 0 & 0 & 0 & 0 & 1 & -1 &
   0 & 0 & 0 \\
 0 & 0 & 0 & 0 & 0 & 0 & 0 & 0 & 0 & 0 & 0 & 1 & -1 & 0 & 0 & 0 & 0
   & 1 & 0 & 0 & 0 & 0 & 0 & 0 & -1 & 0 & 0 & 0 & 0 & 0 & 0 & 1 &
   -1 & 0 & 0 \\
 0 & 0 & 0 & 1 & -1 & 0 & 0 & 0 & 0 & 0 & 0 & 0 & 0 & 0 & 0 & -1 &
   0 & 0 & 0 & 0 & 0 & 0 & 0 & 0 & 0 & 0 & 0 & 1 & 0 & 0 & 0 & 0 &
   0 & 1 & -1 \\
 -1 & 0 & 0 & 0 & 1 & 0 & 0 & 0 & 0 & 0 & 0 & 0 & 0 & 0 & 0 & 0 & 0
   & 0 & 0 & 0 & 0 & 0 & 0 & 0 & 0 & 0 & -1 & 0 & 1 & 0 & -1 & 0 &
   0 & 0 & 1 \\
 1 & -1 & 0 & 0 & 0 & 0 & 0 & 0 & 0 & 0 & 0 & 0 & 0 & 0 & 0 & 0 & 0
   & 0 & 0 & 0 & 0 & 0 & 0 & 0 & 0 & 0 & 0 & -1 & 0 & 1 & 1 & -1 &
   0 & 0 & 0 \\
 0 & 1 & -1 & 0 & 0 & 0 & 0 & 0 & 0 & 0 & 0 & 0 & 0 & 0 & 0 & 1 & 0
   & 0 & 0 & 0 & 0 & 0 & 0 & 0 & 0 & 0 & 0 & 0 & -1 & 0 & 0 & 1 &
   -1 & 0 & 0 \\
 0 & 0 & 0 & 0 & 0 & 0 & 0 & 0 & 0 & 0 & 0 & 0 & 0 & 0 & 0 & 0 & 0
   & 0 & 0 & 1 & -1 & 0 & 0 & 1 & -1 & 0 & 0 & 1 & -1 & 0 & 0 & 1 &
   0 & 0 & -1 \\
 0 & 0 & 0 & 0 & 0 & 0 & 0 & 0 & 0 & 0 & 0 & 0 & 0 & 0 & 0 & 0 & 0
   & 0 & 0 & 0 & 1 & -1 & 0 & 0 & 1 & -1 & 0 & 0 & 1 & -1 & -1 & 0
   & 1 & 0 & 0 \\
 0 & 0 & 0 & 0 & 0 & 0 & 0 & 0 & 0 & 0 & 0 & 0 & 0 & 0 & 0 & -1 &
   -1 & -1 & 0 & 0 & 0 & 1 & 0 & 0 & 0 & 1 & 0 & 0 & 0 & 1 & 0 & -1
   & 0 & 1 & 0 \\
 0 & 0 & 0 & 0 & 0 & 0 & 0 & 0 & 0 & 0 & 0 & 0 & 0 & 0 & 0 & 1 & 1
   & 1 & -1 & 0 & 0 & 0 & -1 & 0 & 0 & 0 & -1 & 0 & 0 & 0 & 0 & 0 &
   -1 & 0 & 1 \\
 0 & 0 & 0 & 0 & 0 & 0 & 0 & 0 & 0 & 0 & 0 & 0 & 0 & 0 & 0 & 0 & 0
   & 0 & 1 & -1 & 0 & 0 & 1 & -1 & 0 & 0 & 1 & -1 & 0 & 0 & 1 & 0 &
   0 & -1 & 0 \\
\end{array}
\right)
$}

\subsection{$T_7$ orbifold SCFT}
\scalebox{0.4}{$\left(
\begin{array}{ccccccccccccccccccccccccccccccccccccccccccccccccc}
 0 & 1 & 0 & 0 & 0 & 0 & -1 & 0 & 0 & 0 & 0 & 0 & 0 & 0 & 0 & 0 & 0
   & 0 & 0 & 0 & 0 & 0 & 0 & 0 & 0 & 0 & 0 & 0 & 0 & 0 & 0 & 0 & 0
   & 0 & 0 & 0 & 0 & 0 & 1 & -1 & 0 & 0 & 0 & 0 & 0 & 0 & 0 & 0 & 0
   \\
 -1 & 0 & 1 & 0 & 0 & 0 & 0 & 0 & 0 & 0 & 0 & 0 & 0 & 0 & 0 & 0 & 0
   & 0 & 0 & 0 & 0 & 0 & 0 & 0 & 0 & 0 & 0 & 0 & 0 & 0 & 0 & 0 & 0
   & 0 & 0 & 0 & 0 & 0 & 0 & 1 & -1 & 0 & 0 & 0 & 0 & 0 & 0 & 0 & 0
   \\
 0 & -1 & 0 & 1 & 0 & 0 & 0 & 0 & 0 & 0 & 0 & 0 & 0 & 0 & 0 & 0 & 0
   & 0 & 0 & 0 & 0 & 0 & 0 & 0 & 0 & 0 & 0 & 0 & 0 & 0 & 0 & 0 & 0
   & 0 & 0 & 0 & 0 & 0 & 0 & 0 & 1 & -1 & 0 & 0 & 0 & 0 & 0 & 0 & 0
   \\
 0 & 0 & -1 & 0 & 1 & 0 & 0 & 0 & 0 & 0 & 0 & 0 & 0 & 0 & 0 & 0 & 0
   & 0 & 0 & 0 & 0 & -1 & 0 & 0 & 0 & 0 & 0 & 0 & 0 & 0 & 0 & 0 & 0
   & 0 & 0 & 0 & 0 & 0 & 0 & 0 & 0 & 1 & 0 & 0 & 0 & 0 & 0 & 0 & 0
   \\
 0 & 0 & 0 & -1 & 0 & 1 & 0 & 0 & 0 & 0 & 0 & 0 & 0 & 0 & 0 & 0 & 0
   & 0 & 0 & 0 & 0 & 1 & 0 & 0 & 0 & 0 & 0 & 0 & 0 & 0 & 0 & 0 & 0
   & 0 & 0 & 0 & -1 & 0 & 0 & 0 & 0 & 0 & 0 & 0 & 0 & 0 & 0 & 0 & 0
   \\
 0 & 0 & 0 & 0 & -1 & 0 & 1 & 0 & 0 & 0 & 0 & 0 & 0 & 0 & 0 & 0 & 0
   & 0 & 0 & 0 & 0 & 0 & 0 & 0 & 0 & 0 & 0 & 0 & 0 & 0 & 0 & 0 & 0
   & 0 & 0 & 0 & 1 & -1 & 0 & 0 & 0 & 0 & 0 & 0 & 0 & 0 & 0 & 0 & 0
   \\
 1 & 0 & 0 & 0 & 0 & -1 & 0 & 0 & 0 & 0 & 0 & 0 & 0 & 0 & 0 & 0 & 0
   & 0 & 0 & 0 & 0 & 0 & 0 & 0 & 0 & 0 & 0 & 0 & 0 & 0 & 0 & 0 & 0
   & 0 & 0 & 0 & 0 & 1 & -1 & 0 & 0 & 0 & 0 & 0 & 0 & 0 & 0 & 0 & 0
   \\
 0 & 0 & 0 & 0 & 0 & 0 & 0 & 0 & 0 & 1 & 0 & 0 & 0 & 0 & -1 & 0 & 0
   & 0 & 0 & 0 & 0 & 0 & 0 & 0 & 0 & 0 & 0 & 0 & 0 & 0 & 0 & 0 & 1
   & -1 & 0 & 0 & 0 & 0 & 0 & 0 & 0 & 0 & 0 & 0 & 0 & 0 & 0 & 0 & 0
   \\
 0 & 0 & 0 & 0 & 0 & 0 & 0 & 0 & 0 & 0 & 0 & 0 & 0 & 0 & 0 & 1 & 0
   & 0 & 0 & 0 & -1 & 0 & 0 & 0 & 0 & 0 & 1 & -1 & 0 & 0 & 0 & 0 &
   0 & 0 & 0 & 0 & 0 & 0 & 0 & 0 & 0 & 0 & 0 & 0 & 0 & 0 & 0 & 0 &
   0 \\
 0 & 0 & 0 & 0 & 0 & 0 & 0 & -1 & 0 & 0 & 1 & 0 & 0 & 0 & 0 & 0 & 0
   & 0 & 0 & 0 & 0 & 0 & 0 & 0 & 0 & 0 & 0 & 0 & 0 & 0 & 0 & 0 & 0
   & 1 & -1 & 0 & 0 & 0 & 0 & 0 & 0 & 0 & 0 & 0 & 0 & 0 & 0 & 0 & 0
   \\
 0 & 0 & 0 & 0 & 0 & 0 & 0 & 0 & 0 & -1 & 0 & 1 & 0 & 0 & 0 & 0 & 0
   & 0 & 0 & 0 & 0 & 0 & 0 & 0 & 0 & 0 & 0 & 0 & 0 & 0 & 0 & 0 & 0
   & 0 & 1 & -1 & 0 & 0 & 0 & 0 & 0 & 0 & 0 & 0 & 0 & 0 & 0 & 0 & 0
   \\
 0 & 0 & 0 & 0 & 0 & 0 & 0 & 0 & 0 & 0 & -1 & 0 & 1 & 0 & 0 & 0 & 0
   & 0 & 0 & 0 & 0 & 0 & 0 & -1 & 0 & 0 & 0 & 0 & 0 & 0 & 0 & 0 & 0
   & 0 & 0 & 1 & 0 & 0 & 0 & 0 & 0 & 0 & 0 & 0 & 0 & 0 & 0 & 0 & 0
   \\
 0 & 0 & 0 & 0 & 0 & 0 & 0 & 0 & 0 & 0 & 0 & -1 & 0 & 1 & 0 & 0 & 0
   & 0 & 0 & 0 & 0 & 0 & 0 & 1 & 0 & 0 & 0 & 0 & 0 & 0 & -1 & 0 & 0
   & 0 & 0 & 0 & 0 & 0 & 0 & 0 & 0 & 0 & 0 & 0 & 0 & 0 & 0 & 0 & 0
   \\
 0 & 0 & 0 & 0 & 0 & 0 & 0 & 0 & 0 & 0 & 0 & 0 & -1 & 0 & 1 & 0 & 0
   & 0 & 0 & 0 & 0 & 0 & 0 & 0 & 0 & 0 & 0 & 0 & 0 & 0 & 1 & -1 & 0
   & 0 & 0 & 0 & 0 & 0 & 0 & 0 & 0 & 0 & 0 & 0 & 0 & 0 & 0 & 0 & 0
   \\
 0 & 0 & 0 & 0 & 0 & 0 & 0 & 1 & 0 & 0 & 0 & 0 & 0 & -1 & 0 & 0 & 0
   & 0 & 0 & 0 & 0 & 0 & 0 & 0 & 0 & 0 & 0 & 0 & 0 & 0 & 0 & 1 & -1
   & 0 & 0 & 0 & 0 & 0 & 0 & 0 & 0 & 0 & 0 & 0 & 0 & 0 & 0 & 0 & 0
   \\
 0 & 0 & 0 & 0 & 0 & 0 & 0 & 0 & -1 & 0 & 0 & 0 & 0 & 0 & 0 & 0 & 1
   & 0 & 0 & 0 & 0 & 0 & 0 & 0 & 0 & 0 & 0 & 1 & -1 & 0 & 0 & 0 & 0
   & 0 & 0 & 0 & 0 & 0 & 0 & 0 & 0 & 0 & 0 & 0 & 0 & 0 & 0 & 0 & 0
   \\
 0 & 0 & 0 & 0 & 0 & 0 & 0 & 0 & 0 & 0 & 0 & 0 & 0 & 0 & 0 & -1 & 0
   & 1 & 0 & 0 & 0 & 0 & 0 & 0 & 0 & 0 & 0 & 0 & 1 & -1 & 0 & 0 & 0
   & 0 & 0 & 0 & 0 & 0 & 0 & 0 & 0 & 0 & 0 & 0 & 0 & 0 & 0 & 0 & 0
   \\
 0 & 0 & 0 & 0 & 0 & 0 & 0 & 0 & 0 & 0 & 0 & 0 & 0 & 0 & 0 & 0 & -1
   & 0 & 1 & 0 & 0 & 0 & -1 & 0 & 0 & 0 & 0 & 0 & 0 & 1 & 0 & 0 & 0
   & 0 & 0 & 0 & 0 & 0 & 0 & 0 & 0 & 0 & 0 & 0 & 0 & 0 & 0 & 0 & 0
   \\
 0 & 0 & 0 & 0 & 0 & 0 & 0 & 0 & 0 & 0 & 0 & 0 & 0 & 0 & 0 & 0 & 0
   & -1 & 0 & 1 & 0 & 0 & 1 & 0 & -1 & 0 & 0 & 0 & 0 & 0 & 0 & 0 &
   0 & 0 & 0 & 0 & 0 & 0 & 0 & 0 & 0 & 0 & 0 & 0 & 0 & 0 & 0 & 0 &
   0 \\
 0 & 0 & 0 & 0 & 0 & 0 & 0 & 0 & 0 & 0 & 0 & 0 & 0 & 0 & 0 & 0 & 0
   & 0 & -1 & 0 & 1 & 0 & 0 & 0 & 1 & -1 & 0 & 0 & 0 & 0 & 0 & 0 &
   0 & 0 & 0 & 0 & 0 & 0 & 0 & 0 & 0 & 0 & 0 & 0 & 0 & 0 & 0 & 0 &
   0 \\
 0 & 0 & 0 & 0 & 0 & 0 & 0 & 0 & 1 & 0 & 0 & 0 & 0 & 0 & 0 & 0 & 0
   & 0 & 0 & -1 & 0 & 0 & 0 & 0 & 0 & 1 & -1 & 0 & 0 & 0 & 0 & 0 &
   0 & 0 & 0 & 0 & 0 & 0 & 0 & 0 & 0 & 0 & 0 & 0 & 0 & 0 & 0 & 0 &
   0 \\
 0 & 0 & 0 & 1 & -1 & 0 & 0 & 0 & 0 & 0 & 0 & 0 & 0 & 0 & 0 & 0 & 0
   & 0 & 0 & 0 & 0 & 0 & 0 & 0 & 0 & 0 & 0 & 0 & 0 & 0 & 0 & 0 & 0
   & 0 & 0 & 0 & 1 & 0 & 0 & 0 & 0 & -1 & 0 & 0 & 0 & 1 & -1 & 0 &
   0 \\
 0 & 0 & 0 & 0 & 0 & 0 & 0 & 0 & 0 & 0 & 0 & 0 & 0 & 0 & 0 & 0 & 0
   & 1 & -1 & 0 & 0 & 0 & 0 & 0 & 1 & 0 & 0 & 0 & 0 & -1 & 0 & 0 &
   0 & 0 & 0 & 0 & 0 & 0 & 0 & 0 & 0 & 0 & 0 & 0 & 0 & 1 & -1 & 0 &
   0 \\
 0 & 0 & 0 & 0 & 0 & 0 & 0 & 0 & 0 & 0 & 0 & 1 & -1 & 0 & 0 & 0 & 0
   & 0 & 0 & 0 & 0 & 0 & 0 & 0 & 0 & 0 & 0 & 0 & 0 & 0 & 1 & 0 & 0
   & 0 & 0 & -1 & 0 & 0 & 0 & 0 & 0 & 0 & 0 & 0 & 0 & 1 & -1 & 0 &
   0 \\
 0 & 0 & 0 & 0 & 0 & 0 & 0 & 0 & 0 & 0 & 0 & 0 & 0 & 0 & 0 & 0 & 0
   & 0 & 1 & -1 & 0 & 0 & -1 & 0 & 0 & 1 & 0 & 0 & 0 & 0 & 0 & 0 &
   0 & 0 & 0 & 0 & 0 & 0 & 0 & 0 & 0 & 0 & 0 & 0 & 0 & 0 & 1 & -1 &
   0 \\
 0 & 0 & 0 & 0 & 0 & 0 & 0 & 0 & 0 & 0 & 0 & 0 & 0 & 0 & 0 & 0 & 0
   & 0 & 0 & 1 & -1 & 0 & 0 & 0 & -1 & 0 & 1 & 0 & 0 & 0 & 0 & 0 &
   0 & 0 & 0 & 0 & 0 & 0 & 0 & 0 & 0 & 0 & 0 & 0 & 0 & 0 & 0 & 1 &
   -1 \\
 0 & 0 & 0 & 0 & 0 & 0 & 0 & 0 & -1 & 0 & 0 & 0 & 0 & 0 & 0 & 0 & 0
   & 0 & 0 & 0 & 1 & 0 & 0 & 0 & 0 & -1 & 0 & 1 & 0 & 0 & 0 & 0 & 0
   & 0 & 0 & 0 & 0 & 0 & 0 & 0 & 0 & 0 & -1 & 0 & 0 & 0 & 0 & 0 & 1
   \\
 0 & 0 & 0 & 0 & 0 & 0 & 0 & 0 & 1 & 0 & 0 & 0 & 0 & 0 & 0 & -1 & 0
   & 0 & 0 & 0 & 0 & 0 & 0 & 0 & 0 & 0 & -1 & 0 & 1 & 0 & 0 & 0 & 0
   & 0 & 0 & 0 & 0 & 0 & 0 & 0 & 0 & 0 & 1 & -1 & 0 & 0 & 0 & 0 & 0
   \\
 0 & 0 & 0 & 0 & 0 & 0 & 0 & 0 & 0 & 0 & 0 & 0 & 0 & 0 & 0 & 1 & -1
   & 0 & 0 & 0 & 0 & 0 & 0 & 0 & 0 & 0 & 0 & -1 & 0 & 1 & 0 & 0 & 0
   & 0 & 0 & 0 & 0 & 0 & 0 & 0 & 0 & 0 & 0 & 1 & -1 & 0 & 0 & 0 & 0
   \\
 0 & 0 & 0 & 0 & 0 & 0 & 0 & 0 & 0 & 0 & 0 & 0 & 0 & 0 & 0 & 0 & 1
   & -1 & 0 & 0 & 0 & 0 & 1 & 0 & 0 & 0 & 0 & 0 & -1 & 0 & 0 & 0 &
   0 & 0 & 0 & 0 & 0 & 0 & 0 & 0 & 0 & 0 & 0 & 0 & 1 & -1 & 0 & 0 &
   0 \\
 0 & 0 & 0 & 0 & 0 & 0 & 0 & 0 & 0 & 0 & 0 & 0 & 1 & -1 & 0 & 0 & 0
   & 0 & 0 & 0 & 0 & 0 & 0 & -1 & 0 & 0 & 0 & 0 & 0 & 0 & 0 & 1 & 0
   & 0 & 0 & 0 & 0 & 0 & 0 & 0 & 0 & 0 & 0 & 0 & 0 & 0 & 1 & -1 & 0
   \\
 0 & 0 & 0 & 0 & 0 & 0 & 0 & 0 & 0 & 0 & 0 & 0 & 0 & 1 & -1 & 0 & 0
   & 0 & 0 & 0 & 0 & 0 & 0 & 0 & 0 & 0 & 0 & 0 & 0 & 0 & -1 & 0 & 1
   & 0 & 0 & 0 & 0 & 0 & 0 & 0 & 0 & 0 & 0 & 0 & 0 & 0 & 0 & 1 & -1
   \\
 0 & 0 & 0 & 0 & 0 & 0 & 0 & -1 & 0 & 0 & 0 & 0 & 0 & 0 & 1 & 0 & 0
   & 0 & 0 & 0 & 0 & 0 & 0 & 0 & 0 & 0 & 0 & 0 & 0 & 0 & 0 & -1 & 0
   & 1 & 0 & 0 & 0 & 0 & 0 & 0 & 0 & 0 & -1 & 0 & 0 & 0 & 0 & 0 & 1
   \\
 0 & 0 & 0 & 0 & 0 & 0 & 0 & 1 & 0 & -1 & 0 & 0 & 0 & 0 & 0 & 0 & 0
   & 0 & 0 & 0 & 0 & 0 & 0 & 0 & 0 & 0 & 0 & 0 & 0 & 0 & 0 & 0 & -1
   & 0 & 1 & 0 & 0 & 0 & 0 & 0 & 0 & 0 & 1 & -1 & 0 & 0 & 0 & 0 & 0
   \\
 0 & 0 & 0 & 0 & 0 & 0 & 0 & 0 & 0 & 1 & -1 & 0 & 0 & 0 & 0 & 0 & 0
   & 0 & 0 & 0 & 0 & 0 & 0 & 0 & 0 & 0 & 0 & 0 & 0 & 0 & 0 & 0 & 0
   & -1 & 0 & 1 & 0 & 0 & 0 & 0 & 0 & 0 & 0 & 1 & -1 & 0 & 0 & 0 &
   0 \\
 0 & 0 & 0 & 0 & 0 & 0 & 0 & 0 & 0 & 0 & 1 & -1 & 0 & 0 & 0 & 0 & 0
   & 0 & 0 & 0 & 0 & 0 & 0 & 1 & 0 & 0 & 0 & 0 & 0 & 0 & 0 & 0 & 0
   & 0 & -1 & 0 & 0 & 0 & 0 & 0 & 0 & 0 & 0 & 0 & 1 & -1 & 0 & 0 &
   0 \\
 0 & 0 & 0 & 0 & 1 & -1 & 0 & 0 & 0 & 0 & 0 & 0 & 0 & 0 & 0 & 0 & 0
   & 0 & 0 & 0 & 0 & -1 & 0 & 0 & 0 & 0 & 0 & 0 & 0 & 0 & 0 & 0 & 0
   & 0 & 0 & 0 & 0 & 1 & 0 & 0 & 0 & 0 & 0 & 0 & 0 & 0 & 1 & -1 & 0
   \\
 0 & 0 & 0 & 0 & 0 & 1 & -1 & 0 & 0 & 0 & 0 & 0 & 0 & 0 & 0 & 0 & 0
   & 0 & 0 & 0 & 0 & 0 & 0 & 0 & 0 & 0 & 0 & 0 & 0 & 0 & 0 & 0 & 0
   & 0 & 0 & 0 & -1 & 0 & 1 & 0 & 0 & 0 & 0 & 0 & 0 & 0 & 0 & 1 &
   -1 \\
 -1 & 0 & 0 & 0 & 0 & 0 & 1 & 0 & 0 & 0 & 0 & 0 & 0 & 0 & 0 & 0 & 0
   & 0 & 0 & 0 & 0 & 0 & 0 & 0 & 0 & 0 & 0 & 0 & 0 & 0 & 0 & 0 & 0
   & 0 & 0 & 0 & 0 & -1 & 0 & 1 & 0 & 0 & -1 & 0 & 0 & 0 & 0 & 0 &
   1 \\
 1 & -1 & 0 & 0 & 0 & 0 & 0 & 0 & 0 & 0 & 0 & 0 & 0 & 0 & 0 & 0 & 0
   & 0 & 0 & 0 & 0 & 0 & 0 & 0 & 0 & 0 & 0 & 0 & 0 & 0 & 0 & 0 & 0
   & 0 & 0 & 0 & 0 & 0 & -1 & 0 & 1 & 0 & 1 & -1 & 0 & 0 & 0 & 0 &
   0 \\
 0 & 1 & -1 & 0 & 0 & 0 & 0 & 0 & 0 & 0 & 0 & 0 & 0 & 0 & 0 & 0 & 0
   & 0 & 0 & 0 & 0 & 0 & 0 & 0 & 0 & 0 & 0 & 0 & 0 & 0 & 0 & 0 & 0
   & 0 & 0 & 0 & 0 & 0 & 0 & -1 & 0 & 1 & 0 & 1 & -1 & 0 & 0 & 0 &
   0 \\
 0 & 0 & 1 & -1 & 0 & 0 & 0 & 0 & 0 & 0 & 0 & 0 & 0 & 0 & 0 & 0 & 0
   & 0 & 0 & 0 & 0 & 1 & 0 & 0 & 0 & 0 & 0 & 0 & 0 & 0 & 0 & 0 & 0
   & 0 & 0 & 0 & 0 & 0 & 0 & 0 & -1 & 0 & 0 & 0 & 1 & -1 & 0 & 0 &
   0 \\
 0 & 0 & 0 & 0 & 0 & 0 & 0 & 0 & 0 & 0 & 0 & 0 & 0 & 0 & 0 & 0 & 0
   & 0 & 0 & 0 & 0 & 0 & 0 & 0 & 0 & 0 & 1 & -1 & 0 & 0 & 0 & 0 & 1
   & -1 & 0 & 0 & 0 & 0 & 1 & -1 & 0 & 0 & 0 & 1 & 0 & 0 & 0 & 0 &
   -1 \\
 0 & 0 & 0 & 0 & 0 & 0 & 0 & 0 & 0 & 0 & 0 & 0 & 0 & 0 & 0 & 0 & 0
   & 0 & 0 & 0 & 0 & 0 & 0 & 0 & 0 & 0 & 0 & 1 & -1 & 0 & 0 & 0 & 0
   & 1 & -1 & 0 & 0 & 0 & 0 & 1 & -1 & 0 & -1 & 0 & 1 & 0 & 0 & 0 &
   0 \\
 0 & 0 & 0 & 0 & 0 & 0 & 0 & 0 & 0 & 0 & 0 & 0 & 0 & 0 & 0 & 0 & 0
   & 0 & 0 & 0 & 0 & 0 & 0 & 0 & 0 & 0 & 0 & 0 & 1 & -1 & 0 & 0 & 0
   & 0 & 1 & -1 & 0 & 0 & 0 & 0 & 1 & -1 & 0 & -1 & 0 & 1 & 0 & 0 &
   0 \\
 0 & 0 & 0 & 0 & 0 & 0 & 0 & 0 & 0 & 0 & 0 & 0 & 0 & 0 & 0 & 0 & 0
   & 0 & 0 & 0 & 0 & -1 & -1 & -1 & 0 & 0 & 0 & 0 & 0 & 1 & 0 & 0 &
   0 & 0 & 0 & 1 & 0 & 0 & 0 & 0 & 0 & 1 & 0 & 0 & -1 & 0 & 1 & 0 &
   0 \\
 0 & 0 & 0 & 0 & 0 & 0 & 0 & 0 & 0 & 0 & 0 & 0 & 0 & 0 & 0 & 0 & 0
   & 0 & 0 & 0 & 0 & 1 & 1 & 1 & -1 & 0 & 0 & 0 & 0 & 0 & -1 & 0 &
   0 & 0 & 0 & 0 & -1 & 0 & 0 & 0 & 0 & 0 & 0 & 0 & 0 & -1 & 0 & 1
   & 0 \\
 0 & 0 & 0 & 0 & 0 & 0 & 0 & 0 & 0 & 0 & 0 & 0 & 0 & 0 & 0 & 0 & 0
   & 0 & 0 & 0 & 0 & 0 & 0 & 0 & 1 & -1 & 0 & 0 & 0 & 0 & 1 & -1 &
   0 & 0 & 0 & 0 & 1 & -1 & 0 & 0 & 0 & 0 & 0 & 0 & 0 & 0 & -1 & 0
   & 1 \\
 0 & 0 & 0 & 0 & 0 & 0 & 0 & 0 & 0 & 0 & 0 & 0 & 0 & 0 & 0 & 0 & 0
   & 0 & 0 & 0 & 0 & 0 & 0 & 0 & 0 & 1 & -1 & 0 & 0 & 0 & 0 & 1 &
   -1 & 0 & 0 & 0 & 0 & 1 & -1 & 0 & 0 & 0 & 1 & 0 & 0 & 0 & 0 & -1
   & 0 \\
\end{array}
\right)$}

\subsection{$O_5$ orbifold SCFT}
\scalebox{0.35}{$\left(
\begin{array}{cccccccccccccccccccccccccccccccccccccccc}
 0 & 0 & 0 & 0 & 0 & 0 & 1 & 0 & 0 & -1 & 0 & 0 & 0 & 0 & 0 & 0 & 0
   & 0 & 0 & 0 & 1 & 0 & -1 & 0 & 0 & 0 & 0 & 0 & 0 & 0 & 0 & 0 & 0
   & 0 & 0 & 0 & 0 & 0 & 0 & 0 \\
 0 & 0 & 1 & 0 & 0 & -1 & 0 & 0 & 0 & 0 & 0 & 0 & 0 & 0 & 0 & 0 & 0
   & 0 & 0 & 1 & 0 & -1 & 0 & 0 & 0 & 0 & 0 & 0 & 0 & 0 & 0 & 0 & 0
   & 0 & 0 & 0 & 0 & 0 & 0 & 0 \\
 0 & -1 & 0 & 1 & 0 & 0 & 0 & 0 & 0 & 0 & 0 & 0 & 0 & 0 & 0 & 0 & 0
   & 0 & 0 & 0 & 0 & 1 & 0 & -1 & 0 & 0 & 0 & 0 & 0 & 0 & 0 & 0 & 0
   & 0 & 0 & 0 & 0 & 0 & 0 & 0 \\
 0 & 0 & -1 & 0 & 1 & 0 & 0 & 0 & 0 & 0 & 0 & 0 & 0 & 0 & 0 & -1 &
   0 & 0 & 0 & 0 & 0 & 0 & 0 & 1 & 0 & 0 & 0 & 0 & 0 & 0 & 0 & 0 &
   0 & 0 & 0 & 0 & 0 & 0 & 0 & 0 \\
 0 & 0 & 0 & -1 & 0 & 1 & 0 & 0 & 0 & 0 & 0 & 0 & 0 & 0 & 0 & 1 & 0
   & -1 & 0 & 0 & 0 & 0 & 0 & 0 & 0 & 0 & 0 & 0 & 0 & 0 & 0 & 0 & 0
   & 0 & 0 & 0 & 0 & 0 & 0 & 0 \\
 0 & 1 & 0 & 0 & -1 & 0 & 0 & 0 & 0 & 0 & 0 & 0 & 0 & 0 & 0 & 0 & 0
   & 1 & 0 & -1 & 0 & 0 & 0 & 0 & 0 & 0 & 0 & 0 & 0 & 0 & 0 & 0 & 0
   & 0 & 0 & 0 & 0 & 0 & 0 & 0 \\
 -1 & 0 & 0 & 0 & 0 & 0 & 0 & 1 & 0 & 0 & 0 & 0 & 0 & 0 & 0 & 0 & 0
   & 0 & 0 & 0 & 0 & 0 & 1 & 0 & -1 & 0 & 0 & 0 & 0 & 0 & 0 & 0 & 0
   & 0 & 0 & 0 & 0 & 0 & 0 & 0 \\
 0 & 0 & 0 & 0 & 0 & 0 & -1 & 0 & 1 & 0 & 0 & 0 & 0 & 0 & 0 & 0 &
   -1 & 0 & 0 & 0 & 0 & 0 & 0 & 0 & 1 & 0 & 0 & 0 & 0 & 0 & 0 & 0 &
   0 & 0 & 0 & 0 & 0 & 0 & 0 & 0 \\
 0 & 0 & 0 & 0 & 0 & 0 & 0 & -1 & 0 & 1 & 0 & 0 & 0 & 0 & 0 & 0 & 1
   & 0 & -1 & 0 & 0 & 0 & 0 & 0 & 0 & 0 & 0 & 0 & 0 & 0 & 0 & 0 & 0
   & 0 & 0 & 0 & 0 & 0 & 0 & 0 \\
 1 & 0 & 0 & 0 & 0 & 0 & 0 & 0 & -1 & 0 & 0 & 0 & 0 & 0 & 0 & 0 & 0
   & 0 & 1 & 0 & -1 & 0 & 0 & 0 & 0 & 0 & 0 & 0 & 0 & 0 & 0 & 0 & 0
   & 0 & 0 & 0 & 0 & 0 & 0 & 0 \\
 0 & 0 & 0 & 0 & 0 & 0 & 0 & 0 & 0 & 0 & 0 & 1 & 0 & 0 & -1 & 0 & 0
   & 0 & 0 & 0 & 0 & 0 & 0 & 0 & 0 & 0 & 0 & 0 & 0 & 0 & 0 & 0 & 0
   & 0 & 0 & 0 & 0 & 1 & -1 & 0 \\
 0 & 0 & 0 & 0 & 0 & 0 & 0 & 0 & 0 & 0 & -1 & 0 & 1 & 0 & 0 & 0 & 0
   & 0 & 0 & 0 & 0 & 0 & 0 & 0 & 0 & 0 & 0 & 0 & 0 & 0 & 0 & 0 & 0
   & 0 & 0 & 0 & 0 & 0 & 1 & -1 \\
 0 & 0 & 0 & 0 & 0 & 0 & 0 & 0 & 0 & 0 & 0 & -1 & 0 & 1 & 0 & 0 & 0
   & 0 & 0 & 0 & 0 & 0 & 0 & 0 & 0 & 0 & 0 & 0 & 0 & 0 & 0 & 0 & 0
   & 0 & 0 & -1 & 0 & 0 & 0 & 1 \\
 0 & 0 & 0 & 0 & 0 & 0 & 0 & 0 & 0 & 0 & 0 & 0 & -1 & 0 & 1 & 0 & 0
   & 0 & 0 & 0 & 0 & 0 & 0 & 0 & 0 & 0 & 0 & 0 & 0 & 0 & 0 & 0 & 0
   & 0 & 0 & 1 & -1 & 0 & 0 & 0 \\
 0 & 0 & 0 & 0 & 0 & 0 & 0 & 0 & 0 & 0 & 1 & 0 & 0 & -1 & 0 & 0 & 0
   & 0 & 0 & 0 & 0 & 0 & 0 & 0 & 0 & 0 & 0 & 0 & 0 & 0 & 0 & 0 & 0
   & 0 & 0 & 0 & 1 & -1 & 0 & 0 \\
 0 & 0 & 0 & 1 & -1 & 0 & 0 & 0 & 0 & 0 & 0 & 0 & 0 & 0 & 0 & 0 & 0
   & 1 & 0 & 0 & 0 & 0 & 0 & -1 & 0 & 0 & 0 & 0 & 0 & 0 & 0 & 0 & 1
   & -1 & 0 & 0 & 0 & 0 & 0 & 0 \\
 0 & 0 & 0 & 0 & 0 & 0 & 0 & 1 & -1 & 0 & 0 & 0 & 0 & 0 & 0 & 0 & 0
   & 0 & 1 & 0 & 0 & 0 & 0 & 0 & -1 & 0 & 0 & 0 & 1 & -1 & 0 & 0 &
   0 & 0 & 0 & 0 & 0 & 0 & 0 & 0 \\
 0 & 0 & 0 & 0 & 1 & -1 & 0 & 0 & 0 & 0 & 0 & 0 & 0 & 0 & 0 & -1 &
   0 & 0 & 0 & 1 & 0 & 0 & 0 & 0 & 0 & 0 & 0 & 0 & 0 & 0 & 0 & 0 &
   0 & 1 & -1 & 0 & 0 & 0 & 0 & 0 \\
 0 & 0 & 0 & 0 & 0 & 0 & 0 & 0 & 1 & -1 & 0 & 0 & 0 & 0 & 0 & 0 &
   -1 & 0 & 0 & 0 & 1 & 0 & 0 & 0 & 0 & 0 & 0 & 0 & 0 & 1 & -1 & 0
   & 0 & 0 & 0 & 0 & 0 & 0 & 0 & 0 \\
 0 & -1 & 0 & 0 & 0 & 1 & 0 & 0 & 0 & 0 & 0 & 0 & 0 & 0 & 0 & 0 & 0
   & -1 & 0 & 0 & 0 & 1 & 0 & 0 & 0 & 0 & -1 & 0 & 0 & 0 & 0 & 0 &
   0 & 0 & 1 & 0 & 0 & 0 & 0 & 0 \\
 -1 & 0 & 0 & 0 & 0 & 0 & 0 & 0 & 0 & 1 & 0 & 0 & 0 & 0 & 0 & 0 & 0
   & 0 & -1 & 0 & 0 & 0 & 1 & 0 & 0 & -1 & 0 & 0 & 0 & 0 & 1 & 0 &
   0 & 0 & 0 & 0 & 0 & 0 & 0 & 0 \\
 0 & 1 & -1 & 0 & 0 & 0 & 0 & 0 & 0 & 0 & 0 & 0 & 0 & 0 & 0 & 0 & 0
   & 0 & 0 & -1 & 0 & 0 & 0 & 1 & 0 & 0 & 1 & 0 & 0 & 0 & 0 & -1 &
   0 & 0 & 0 & 0 & 0 & 0 & 0 & 0 \\
 1 & 0 & 0 & 0 & 0 & 0 & -1 & 0 & 0 & 0 & 0 & 0 & 0 & 0 & 0 & 0 & 0
   & 0 & 0 & 0 & -1 & 0 & 0 & 0 & 1 & 1 & 0 & -1 & 0 & 0 & 0 & 0 &
   0 & 0 & 0 & 0 & 0 & 0 & 0 & 0 \\
 0 & 0 & 1 & -1 & 0 & 0 & 0 & 0 & 0 & 0 & 0 & 0 & 0 & 0 & 0 & 1 & 0
   & 0 & 0 & 0 & 0 & -1 & 0 & 0 & 0 & 0 & 0 & 0 & 0 & 0 & 0 & 1 &
   -1 & 0 & 0 & 0 & 0 & 0 & 0 & 0 \\
 0 & 0 & 0 & 0 & 0 & 0 & 1 & -1 & 0 & 0 & 0 & 0 & 0 & 0 & 0 & 0 & 1
   & 0 & 0 & 0 & 0 & 0 & -1 & 0 & 0 & 0 & 0 & 1 & -1 & 0 & 0 & 0 &
   0 & 0 & 0 & 0 & 0 & 0 & 0 & 0 \\
 0 & 0 & 0 & 0 & 0 & 0 & 0 & 0 & 0 & 0 & 0 & 0 & 0 & 0 & 0 & 0 & 0
   & 0 & 0 & 0 & 1 & 0 & -1 & 0 & 0 & 0 & 0 & 1 & 0 & 0 & -1 & 0 &
   0 & 0 & 0 & 0 & 0 & 1 & -1 & 0 \\
 0 & 0 & 0 & 0 & 0 & 0 & 0 & 0 & 0 & 0 & 0 & 0 & 0 & 0 & 0 & 0 & 0
   & 0 & 0 & 1 & 0 & -1 & 0 & 0 & 0 & 0 & 0 & 0 & 0 & 0 & 0 & 1 & 0
   & 0 & -1 & 0 & 0 & 1 & -1 & 0 \\
 0 & 0 & 0 & 0 & 0 & 0 & 0 & 0 & 0 & 0 & 0 & 0 & 0 & 0 & 0 & 0 & 0
   & 0 & 0 & 0 & 0 & 0 & 1 & 0 & -1 & -1 & 0 & 0 & 1 & 0 & 0 & 0 &
   0 & 0 & 0 & 0 & 0 & 0 & 1 & -1 \\
 0 & 0 & 0 & 0 & 0 & 0 & 0 & 0 & 0 & 0 & 0 & 0 & 0 & 0 & 0 & 0 & -1
   & 0 & 0 & 0 & 0 & 0 & 0 & 0 & 1 & 0 & 0 & -1 & 0 & 1 & 0 & 0 & 0
   & 0 & 0 & -1 & 0 & 0 & 0 & 1 \\
 0 & 0 & 0 & 0 & 0 & 0 & 0 & 0 & 0 & 0 & 0 & 0 & 0 & 0 & 0 & 0 & 1
   & 0 & -1 & 0 & 0 & 0 & 0 & 0 & 0 & 0 & 0 & 0 & -1 & 0 & 1 & 0 &
   0 & 0 & 0 & 1 & -1 & 0 & 0 & 0 \\
 0 & 0 & 0 & 0 & 0 & 0 & 0 & 0 & 0 & 0 & 0 & 0 & 0 & 0 & 0 & 0 & 0
   & 0 & 1 & 0 & -1 & 0 & 0 & 0 & 0 & 1 & 0 & 0 & 0 & -1 & 0 & 0 &
   0 & 0 & 0 & 0 & 1 & -1 & 0 & 0 \\
 0 & 0 & 0 & 0 & 0 & 0 & 0 & 0 & 0 & 0 & 0 & 0 & 0 & 0 & 0 & 0 & 0
   & 0 & 0 & 0 & 0 & 1 & 0 & -1 & 0 & 0 & -1 & 0 & 0 & 0 & 0 & 0 &
   1 & 0 & 0 & 0 & 0 & 0 & 1 & -1 \\
 0 & 0 & 0 & 0 & 0 & 0 & 0 & 0 & 0 & 0 & 0 & 0 & 0 & 0 & 0 & -1 & 0
   & 0 & 0 & 0 & 0 & 0 & 0 & 1 & 0 & 0 & 0 & 0 & 0 & 0 & 0 & -1 & 0
   & 1 & 0 & -1 & 0 & 0 & 0 & 1 \\
 0 & 0 & 0 & 0 & 0 & 0 & 0 & 0 & 0 & 0 & 0 & 0 & 0 & 0 & 0 & 1 & 0
   & -1 & 0 & 0 & 0 & 0 & 0 & 0 & 0 & 0 & 0 & 0 & 0 & 0 & 0 & 0 &
   -1 & 0 & 1 & 1 & -1 & 0 & 0 & 0 \\
 0 & 0 & 0 & 0 & 0 & 0 & 0 & 0 & 0 & 0 & 0 & 0 & 0 & 0 & 0 & 0 & 0
   & 1 & 0 & -1 & 0 & 0 & 0 & 0 & 0 & 0 & 1 & 0 & 0 & 0 & 0 & 0 & 0
   & -1 & 0 & 0 & 1 & -1 & 0 & 0 \\
 0 & 0 & 0 & 0 & 0 & 0 & 0 & 0 & 0 & 0 & 0 & 0 & 1 & -1 & 0 & 0 & 0
   & 0 & 0 & 0 & 0 & 0 & 0 & 0 & 0 & 0 & 0 & 0 & 1 & -1 & 0 & 0 & 1
   & -1 & 0 & 0 & 1 & 0 & 0 & -1 \\
 0 & 0 & 0 & 0 & 0 & 0 & 0 & 0 & 0 & 0 & 0 & 0 & 0 & 1 & -1 & 0 & 0
   & 0 & 0 & 0 & 0 & 0 & 0 & 0 & 0 & 0 & 0 & 0 & 0 & 1 & -1 & 0 & 0
   & 1 & -1 & -1 & 0 & 1 & 0 & 0 \\
 0 & 0 & 0 & 0 & 0 & 0 & 0 & 0 & 0 & 0 & -1 & 0 & 0 & 0 & 1 & 0 & 0
   & 0 & 0 & 0 & 0 & 0 & 0 & 0 & 0 & -1 & -1 & 0 & 0 & 0 & 1 & 0 &
   0 & 0 & 1 & 0 & -1 & 0 & 1 & 0 \\
 0 & 0 & 0 & 0 & 0 & 0 & 0 & 0 & 0 & 0 & 1 & -1 & 0 & 0 & 0 & 0 & 0
   & 0 & 0 & 0 & 0 & 0 & 0 & 0 & 0 & 1 & 1 & -1 & 0 & 0 & 0 & -1 &
   0 & 0 & 0 & 0 & 0 & -1 & 0 & 1 \\
 0 & 0 & 0 & 0 & 0 & 0 & 0 & 0 & 0 & 0 & 0 & 1 & -1 & 0 & 0 & 0 & 0
   & 0 & 0 & 0 & 0 & 0 & 0 & 0 & 0 & 0 & 0 & 1 & -1 & 0 & 0 & 1 &
   -1 & 0 & 0 & 1 & 0 & 0 & -1 & 0 \\
\end{array}
\right)$}

\subsection{$O_7$ orbifold SCFT}

\scalebox{0.35}{$\left(
\begin{array}{ccccccccccccccccccccccccccccccccccccccccccccccccccccc
   ccc}
 0 & 0 & 0 & 0 & 0 & 0 & 0 & 0 & 1 & 0 & 0 & 0 & 0 & -1 & 0 & 0 & 0
   & 0 & 0 & 0 & 0 & 0 & 0 & 0 & 0 & 0 & 0 & 0 & 1 & 0 & -1 & 0 & 0
   & 0 & 0 & 0 & 0 & 0 & 0 & 0 & 0 & 0 & 0 & 0 & 0 & 0 & 0 & 0 & 0
   & 0 & 0 & 0 & 0 & 0 & 0 & 0 \\
 0 & 0 & 1 & 0 & 0 & 0 & 0 & -1 & 0 & 0 & 0 & 0 & 0 & 0 & 0 & 0 & 0
   & 0 & 0 & 0 & 0 & 0 & 0 & 0 & 0 & 0 & 0 & 1 & 0 & -1 & 0 & 0 & 0
   & 0 & 0 & 0 & 0 & 0 & 0 & 0 & 0 & 0 & 0 & 0 & 0 & 0 & 0 & 0 & 0
   & 0 & 0 & 0 & 0 & 0 & 0 & 0 \\
 0 & -1 & 0 & 1 & 0 & 0 & 0 & 0 & 0 & 0 & 0 & 0 & 0 & 0 & 0 & 0 & 0
   & 0 & 0 & 0 & 0 & 0 & 0 & 0 & 0 & 0 & 0 & 0 & 0 & 1 & 0 & -1 & 0
   & 0 & 0 & 0 & 0 & 0 & 0 & 0 & 0 & 0 & 0 & 0 & 0 & 0 & 0 & 0 & 0
   & 0 & 0 & 0 & 0 & 0 & 0 & 0 \\
 0 & 0 & -1 & 0 & 1 & 0 & 0 & 0 & 0 & 0 & 0 & 0 & 0 & 0 & 0 & 0 & 0
   & 0 & 0 & 0 & 0 & 0 & 0 & 0 & 0 & 0 & 0 & 0 & 0 & 0 & 0 & 1 & 0
   & -1 & 0 & 0 & 0 & 0 & 0 & 0 & 0 & 0 & 0 & 0 & 0 & 0 & 0 & 0 & 0
   & 0 & 0 & 0 & 0 & 0 & 0 & 0 \\
 0 & 0 & 0 & -1 & 0 & 1 & 0 & 0 & 0 & 0 & 0 & 0 & 0 & 0 & 0 & 0 & 0
   & 0 & 0 & 0 & 0 & -1 & 0 & 0 & 0 & 0 & 0 & 0 & 0 & 0 & 0 & 0 & 0
   & 1 & 0 & 0 & 0 & 0 & 0 & 0 & 0 & 0 & 0 & 0 & 0 & 0 & 0 & 0 & 0
   & 0 & 0 & 0 & 0 & 0 & 0 & 0 \\
 0 & 0 & 0 & 0 & -1 & 0 & 1 & 0 & 0 & 0 & 0 & 0 & 0 & 0 & 0 & 0 & 0
   & 0 & 0 & 0 & 0 & 1 & 0 & -1 & 0 & 0 & 0 & 0 & 0 & 0 & 0 & 0 & 0
   & 0 & 0 & 0 & 0 & 0 & 0 & 0 & 0 & 0 & 0 & 0 & 0 & 0 & 0 & 0 & 0
   & 0 & 0 & 0 & 0 & 0 & 0 & 0 \\
 0 & 0 & 0 & 0 & 0 & -1 & 0 & 1 & 0 & 0 & 0 & 0 & 0 & 0 & 0 & 0 & 0
   & 0 & 0 & 0 & 0 & 0 & 0 & 1 & 0 & -1 & 0 & 0 & 0 & 0 & 0 & 0 & 0
   & 0 & 0 & 0 & 0 & 0 & 0 & 0 & 0 & 0 & 0 & 0 & 0 & 0 & 0 & 0 & 0
   & 0 & 0 & 0 & 0 & 0 & 0 & 0 \\
 0 & 1 & 0 & 0 & 0 & 0 & -1 & 0 & 0 & 0 & 0 & 0 & 0 & 0 & 0 & 0 & 0
   & 0 & 0 & 0 & 0 & 0 & 0 & 0 & 0 & 1 & 0 & -1 & 0 & 0 & 0 & 0 & 0
   & 0 & 0 & 0 & 0 & 0 & 0 & 0 & 0 & 0 & 0 & 0 & 0 & 0 & 0 & 0 & 0
   & 0 & 0 & 0 & 0 & 0 & 0 & 0 \\
 -1 & 0 & 0 & 0 & 0 & 0 & 0 & 0 & 0 & 1 & 0 & 0 & 0 & 0 & 0 & 0 & 0
   & 0 & 0 & 0 & 0 & 0 & 0 & 0 & 0 & 0 & 0 & 0 & 0 & 0 & 1 & 0 & -1
   & 0 & 0 & 0 & 0 & 0 & 0 & 0 & 0 & 0 & 0 & 0 & 0 & 0 & 0 & 0 & 0
   & 0 & 0 & 0 & 0 & 0 & 0 & 0 \\
 0 & 0 & 0 & 0 & 0 & 0 & 0 & 0 & -1 & 0 & 1 & 0 & 0 & 0 & 0 & 0 & 0
   & 0 & 0 & 0 & 0 & 0 & 0 & 0 & 0 & 0 & 0 & 0 & 0 & 0 & 0 & 0 & 1
   & 0 & -1 & 0 & 0 & 0 & 0 & 0 & 0 & 0 & 0 & 0 & 0 & 0 & 0 & 0 & 0
   & 0 & 0 & 0 & 0 & 0 & 0 & 0 \\
 0 & 0 & 0 & 0 & 0 & 0 & 0 & 0 & 0 & -1 & 0 & 1 & 0 & 0 & 0 & 0 & 0
   & 0 & 0 & 0 & 0 & 0 & -1 & 0 & 0 & 0 & 0 & 0 & 0 & 0 & 0 & 0 & 0
   & 0 & 1 & 0 & 0 & 0 & 0 & 0 & 0 & 0 & 0 & 0 & 0 & 0 & 0 & 0 & 0
   & 0 & 0 & 0 & 0 & 0 & 0 & 0 \\
 0 & 0 & 0 & 0 & 0 & 0 & 0 & 0 & 0 & 0 & -1 & 0 & 1 & 0 & 0 & 0 & 0
   & 0 & 0 & 0 & 0 & 0 & 1 & 0 & -1 & 0 & 0 & 0 & 0 & 0 & 0 & 0 & 0
   & 0 & 0 & 0 & 0 & 0 & 0 & 0 & 0 & 0 & 0 & 0 & 0 & 0 & 0 & 0 & 0
   & 0 & 0 & 0 & 0 & 0 & 0 & 0 \\
 0 & 0 & 0 & 0 & 0 & 0 & 0 & 0 & 0 & 0 & 0 & -1 & 0 & 1 & 0 & 0 & 0
   & 0 & 0 & 0 & 0 & 0 & 0 & 0 & 1 & 0 & -1 & 0 & 0 & 0 & 0 & 0 & 0
   & 0 & 0 & 0 & 0 & 0 & 0 & 0 & 0 & 0 & 0 & 0 & 0 & 0 & 0 & 0 & 0
   & 0 & 0 & 0 & 0 & 0 & 0 & 0 \\
 1 & 0 & 0 & 0 & 0 & 0 & 0 & 0 & 0 & 0 & 0 & 0 & -1 & 0 & 0 & 0 & 0
   & 0 & 0 & 0 & 0 & 0 & 0 & 0 & 0 & 0 & 1 & 0 & -1 & 0 & 0 & 0 & 0
   & 0 & 0 & 0 & 0 & 0 & 0 & 0 & 0 & 0 & 0 & 0 & 0 & 0 & 0 & 0 & 0
   & 0 & 0 & 0 & 0 & 0 & 0 & 0 \\
 0 & 0 & 0 & 0 & 0 & 0 & 0 & 0 & 0 & 0 & 0 & 0 & 0 & 0 & 0 & 1 & 0
   & 0 & 0 & 0 & -1 & 0 & 0 & 0 & 0 & 0 & 0 & 0 & 0 & 0 & 0 & 0 & 0
   & 0 & 0 & 0 & 0 & 0 & 0 & 0 & 0 & 0 & 0 & 0 & 0 & 0 & 0 & 0 & 0
   & 0 & 0 & 0 & 1 & -1 & 0 & 0 \\
 0 & 0 & 0 & 0 & 0 & 0 & 0 & 0 & 0 & 0 & 0 & 0 & 0 & 0 & -1 & 0 & 1
   & 0 & 0 & 0 & 0 & 0 & 0 & 0 & 0 & 0 & 0 & 0 & 0 & 0 & 0 & 0 & 0
   & 0 & 0 & 0 & 0 & 0 & 0 & 0 & 0 & 0 & 0 & 0 & 0 & 0 & 0 & 0 & 0
   & 0 & 0 & 0 & 0 & 1 & -1 & 0 \\
 0 & 0 & 0 & 0 & 0 & 0 & 0 & 0 & 0 & 0 & 0 & 0 & 0 & 0 & 0 & -1 & 0
   & 1 & 0 & 0 & 0 & 0 & 0 & 0 & 0 & 0 & 0 & 0 & 0 & 0 & 0 & 0 & 0
   & 0 & 0 & 0 & 0 & 0 & 0 & 0 & 0 & 0 & 0 & 0 & 0 & 0 & 0 & 0 & 0
   & 0 & 0 & 0 & 0 & 0 & 1 & -1 \\
 0 & 0 & 0 & 0 & 0 & 0 & 0 & 0 & 0 & 0 & 0 & 0 & 0 & 0 & 0 & 0 & -1
   & 0 & 1 & 0 & 0 & 0 & 0 & 0 & 0 & 0 & 0 & 0 & 0 & 0 & 0 & 0 & 0
   & 0 & 0 & 0 & 0 & 0 & 0 & 0 & 0 & 0 & 0 & 0 & 0 & 0 & 0 & 0 & 0
   & -1 & 0 & 0 & 0 & 0 & 0 & 1 \\
 0 & 0 & 0 & 0 & 0 & 0 & 0 & 0 & 0 & 0 & 0 & 0 & 0 & 0 & 0 & 0 & 0
   & -1 & 0 & 1 & 0 & 0 & 0 & 0 & 0 & 0 & 0 & 0 & 0 & 0 & 0 & 0 & 0
   & 0 & 0 & 0 & 0 & 0 & 0 & 0 & 0 & 0 & 0 & 0 & 0 & 0 & 0 & 0 & 0
   & 1 & -1 & 0 & 0 & 0 & 0 & 0 \\
 0 & 0 & 0 & 0 & 0 & 0 & 0 & 0 & 0 & 0 & 0 & 0 & 0 & 0 & 0 & 0 & 0
   & 0 & -1 & 0 & 1 & 0 & 0 & 0 & 0 & 0 & 0 & 0 & 0 & 0 & 0 & 0 & 0
   & 0 & 0 & 0 & 0 & 0 & 0 & 0 & 0 & 0 & 0 & 0 & 0 & 0 & 0 & 0 & 0
   & 0 & 1 & -1 & 0 & 0 & 0 & 0 \\
 0 & 0 & 0 & 0 & 0 & 0 & 0 & 0 & 0 & 0 & 0 & 0 & 0 & 0 & 1 & 0 & 0
   & 0 & 0 & -1 & 0 & 0 & 0 & 0 & 0 & 0 & 0 & 0 & 0 & 0 & 0 & 0 & 0
   & 0 & 0 & 0 & 0 & 0 & 0 & 0 & 0 & 0 & 0 & 0 & 0 & 0 & 0 & 0 & 0
   & 0 & 0 & 1 & -1 & 0 & 0 & 0 \\
 0 & 0 & 0 & 0 & 1 & -1 & 0 & 0 & 0 & 0 & 0 & 0 & 0 & 0 & 0 & 0 & 0
   & 0 & 0 & 0 & 0 & 0 & 0 & 1 & 0 & 0 & 0 & 0 & 0 & 0 & 0 & 0 & 0
   & -1 & 0 & 0 & 0 & 0 & 0 & 0 & 0 & 0 & 0 & 0 & 0 & 1 & -1 & 0 &
   0 & 0 & 0 & 0 & 0 & 0 & 0 & 0 \\
 0 & 0 & 0 & 0 & 0 & 0 & 0 & 0 & 0 & 0 & 1 & -1 & 0 & 0 & 0 & 0 & 0
   & 0 & 0 & 0 & 0 & 0 & 0 & 0 & 1 & 0 & 0 & 0 & 0 & 0 & 0 & 0 & 0
   & 0 & -1 & 0 & 0 & 0 & 0 & 1 & -1 & 0 & 0 & 0 & 0 & 0 & 0 & 0 &
   0 & 0 & 0 & 0 & 0 & 0 & 0 & 0 \\
 0 & 0 & 0 & 0 & 0 & 1 & -1 & 0 & 0 & 0 & 0 & 0 & 0 & 0 & 0 & 0 & 0
   & 0 & 0 & 0 & 0 & -1 & 0 & 0 & 0 & 1 & 0 & 0 & 0 & 0 & 0 & 0 & 0
   & 0 & 0 & 0 & 0 & 0 & 0 & 0 & 0 & 0 & 0 & 0 & 0 & 0 & 1 & -1 & 0
   & 0 & 0 & 0 & 0 & 0 & 0 & 0 \\
 0 & 0 & 0 & 0 & 0 & 0 & 0 & 0 & 0 & 0 & 0 & 1 & -1 & 0 & 0 & 0 & 0
   & 0 & 0 & 0 & 0 & 0 & -1 & 0 & 0 & 0 & 1 & 0 & 0 & 0 & 0 & 0 & 0
   & 0 & 0 & 0 & 0 & 0 & 0 & 0 & 1 & -1 & 0 & 0 & 0 & 0 & 0 & 0 & 0
   & 0 & 0 & 0 & 0 & 0 & 0 & 0 \\
 0 & 0 & 0 & 0 & 0 & 0 & 1 & -1 & 0 & 0 & 0 & 0 & 0 & 0 & 0 & 0 & 0
   & 0 & 0 & 0 & 0 & 0 & 0 & -1 & 0 & 0 & 0 & 1 & 0 & 0 & 0 & 0 & 0
   & 0 & 0 & 0 & 0 & 0 & 0 & 0 & 0 & 0 & 0 & 0 & 0 & 0 & 0 & 1 & -1
   & 0 & 0 & 0 & 0 & 0 & 0 & 0 \\
 0 & 0 & 0 & 0 & 0 & 0 & 0 & 0 & 0 & 0 & 0 & 0 & 1 & -1 & 0 & 0 & 0
   & 0 & 0 & 0 & 0 & 0 & 0 & 0 & -1 & 0 & 0 & 0 & 1 & 0 & 0 & 0 & 0
   & 0 & 0 & 0 & 0 & 0 & 0 & 0 & 0 & 1 & -1 & 0 & 0 & 0 & 0 & 0 & 0
   & 0 & 0 & 0 & 0 & 0 & 0 & 0 \\
 0 & -1 & 0 & 0 & 0 & 0 & 0 & 1 & 0 & 0 & 0 & 0 & 0 & 0 & 0 & 0 & 0
   & 0 & 0 & 0 & 0 & 0 & 0 & 0 & 0 & -1 & 0 & 0 & 0 & 1 & 0 & 0 & 0
   & 0 & 0 & 0 & -1 & 0 & 0 & 0 & 0 & 0 & 0 & 0 & 0 & 0 & 0 & 0 & 1
   & 0 & 0 & 0 & 0 & 0 & 0 & 0 \\
 -1 & 0 & 0 & 0 & 0 & 0 & 0 & 0 & 0 & 0 & 0 & 0 & 0 & 1 & 0 & 0 & 0
   & 0 & 0 & 0 & 0 & 0 & 0 & 0 & 0 & 0 & -1 & 0 & 0 & 0 & 1 & 0 & 0
   & 0 & 0 & -1 & 0 & 0 & 0 & 0 & 0 & 0 & 1 & 0 & 0 & 0 & 0 & 0 & 0
   & 0 & 0 & 0 & 0 & 0 & 0 & 0 \\
 0 & 1 & -1 & 0 & 0 & 0 & 0 & 0 & 0 & 0 & 0 & 0 & 0 & 0 & 0 & 0 & 0
   & 0 & 0 & 0 & 0 & 0 & 0 & 0 & 0 & 0 & 0 & -1 & 0 & 0 & 0 & 1 & 0
   & 0 & 0 & 0 & 1 & 0 & 0 & 0 & 0 & 0 & 0 & -1 & 0 & 0 & 0 & 0 & 0
   & 0 & 0 & 0 & 0 & 0 & 0 & 0 \\
 1 & 0 & 0 & 0 & 0 & 0 & 0 & 0 & -1 & 0 & 0 & 0 & 0 & 0 & 0 & 0 & 0
   & 0 & 0 & 0 & 0 & 0 & 0 & 0 & 0 & 0 & 0 & 0 & -1 & 0 & 0 & 0 & 1
   & 0 & 0 & 1 & 0 & -1 & 0 & 0 & 0 & 0 & 0 & 0 & 0 & 0 & 0 & 0 & 0
   & 0 & 0 & 0 & 0 & 0 & 0 & 0 \\
 0 & 0 & 1 & -1 & 0 & 0 & 0 & 0 & 0 & 0 & 0 & 0 & 0 & 0 & 0 & 0 & 0
   & 0 & 0 & 0 & 0 & 0 & 0 & 0 & 0 & 0 & 0 & 0 & 0 & -1 & 0 & 0 & 0
   & 1 & 0 & 0 & 0 & 0 & 0 & 0 & 0 & 0 & 0 & 1 & -1 & 0 & 0 & 0 & 0
   & 0 & 0 & 0 & 0 & 0 & 0 & 0 \\
 0 & 0 & 0 & 0 & 0 & 0 & 0 & 0 & 1 & -1 & 0 & 0 & 0 & 0 & 0 & 0 & 0
   & 0 & 0 & 0 & 0 & 0 & 0 & 0 & 0 & 0 & 0 & 0 & 0 & 0 & -1 & 0 & 0
   & 0 & 1 & 0 & 0 & 1 & -1 & 0 & 0 & 0 & 0 & 0 & 0 & 0 & 0 & 0 & 0
   & 0 & 0 & 0 & 0 & 0 & 0 & 0 \\
 0 & 0 & 0 & 1 & -1 & 0 & 0 & 0 & 0 & 0 & 0 & 0 & 0 & 0 & 0 & 0 & 0
   & 0 & 0 & 0 & 0 & 1 & 0 & 0 & 0 & 0 & 0 & 0 & 0 & 0 & 0 & -1 & 0
   & 0 & 0 & 0 & 0 & 0 & 0 & 0 & 0 & 0 & 0 & 0 & 1 & -1 & 0 & 0 & 0
   & 0 & 0 & 0 & 0 & 0 & 0 & 0 \\
 0 & 0 & 0 & 0 & 0 & 0 & 0 & 0 & 0 & 1 & -1 & 0 & 0 & 0 & 0 & 0 & 0
   & 0 & 0 & 0 & 0 & 0 & 1 & 0 & 0 & 0 & 0 & 0 & 0 & 0 & 0 & 0 & -1
   & 0 & 0 & 0 & 0 & 0 & 1 & -1 & 0 & 0 & 0 & 0 & 0 & 0 & 0 & 0 & 0
   & 0 & 0 & 0 & 0 & 0 & 0 & 0 \\
 0 & 0 & 0 & 0 & 0 & 0 & 0 & 0 & 0 & 0 & 0 & 0 & 0 & 0 & 0 & 0 & 0
   & 0 & 0 & 0 & 0 & 0 & 0 & 0 & 0 & 0 & 0 & 0 & 1 & 0 & -1 & 0 & 0
   & 0 & 0 & 0 & 0 & 1 & 0 & 0 & 0 & 0 & -1 & 0 & 0 & 0 & 0 & 0 & 0
   & 0 & 0 & 0 & 1 & -1 & 0 & 0 \\
 0 & 0 & 0 & 0 & 0 & 0 & 0 & 0 & 0 & 0 & 0 & 0 & 0 & 0 & 0 & 0 & 0
   & 0 & 0 & 0 & 0 & 0 & 0 & 0 & 0 & 0 & 0 & 1 & 0 & -1 & 0 & 0 & 0
   & 0 & 0 & 0 & 0 & 0 & 0 & 0 & 0 & 0 & 0 & 1 & 0 & 0 & 0 & 0 & -1
   & 0 & 0 & 0 & 1 & -1 & 0 & 0 \\
 0 & 0 & 0 & 0 & 0 & 0 & 0 & 0 & 0 & 0 & 0 & 0 & 0 & 0 & 0 & 0 & 0
   & 0 & 0 & 0 & 0 & 0 & 0 & 0 & 0 & 0 & 0 & 0 & 0 & 0 & 1 & 0 & -1
   & 0 & 0 & -1 & 0 & 0 & 1 & 0 & 0 & 0 & 0 & 0 & 0 & 0 & 0 & 0 & 0
   & 0 & 0 & 0 & 0 & 1 & -1 & 0 \\
 0 & 0 & 0 & 0 & 0 & 0 & 0 & 0 & 0 & 0 & 0 & 0 & 0 & 0 & 0 & 0 & 0
   & 0 & 0 & 0 & 0 & 0 & 0 & 0 & 0 & 0 & 0 & 0 & 0 & 0 & 0 & 0 & 1
   & 0 & -1 & 0 & 0 & -1 & 0 & 1 & 0 & 0 & 0 & 0 & 0 & 0 & 0 & 0 &
   0 & 0 & 0 & 0 & 0 & 0 & 1 & -1 \\
 0 & 0 & 0 & 0 & 0 & 0 & 0 & 0 & 0 & 0 & 0 & 0 & 0 & 0 & 0 & 0 & 0
   & 0 & 0 & 0 & 0 & 0 & -1 & 0 & 0 & 0 & 0 & 0 & 0 & 0 & 0 & 0 & 0
   & 0 & 1 & 0 & 0 & 0 & -1 & 0 & 1 & 0 & 0 & 0 & 0 & 0 & 0 & 0 & 0
   & -1 & 0 & 0 & 0 & 0 & 0 & 1 \\
 0 & 0 & 0 & 0 & 0 & 0 & 0 & 0 & 0 & 0 & 0 & 0 & 0 & 0 & 0 & 0 & 0
   & 0 & 0 & 0 & 0 & 0 & 1 & 0 & -1 & 0 & 0 & 0 & 0 & 0 & 0 & 0 & 0
   & 0 & 0 & 0 & 0 & 0 & 0 & -1 & 0 & 1 & 0 & 0 & 0 & 0 & 0 & 0 & 0
   & 1 & -1 & 0 & 0 & 0 & 0 & 0 \\
 0 & 0 & 0 & 0 & 0 & 0 & 0 & 0 & 0 & 0 & 0 & 0 & 0 & 0 & 0 & 0 & 0
   & 0 & 0 & 0 & 0 & 0 & 0 & 0 & 1 & 0 & -1 & 0 & 0 & 0 & 0 & 0 & 0
   & 0 & 0 & 0 & 0 & 0 & 0 & 0 & -1 & 0 & 1 & 0 & 0 & 0 & 0 & 0 & 0
   & 0 & 1 & -1 & 0 & 0 & 0 & 0 \\
 0 & 0 & 0 & 0 & 0 & 0 & 0 & 0 & 0 & 0 & 0 & 0 & 0 & 0 & 0 & 0 & 0
   & 0 & 0 & 0 & 0 & 0 & 0 & 0 & 0 & 0 & 1 & 0 & -1 & 0 & 0 & 0 & 0
   & 0 & 0 & 1 & 0 & 0 & 0 & 0 & 0 & -1 & 0 & 0 & 0 & 0 & 0 & 0 & 0
   & 0 & 0 & 1 & -1 & 0 & 0 & 0 \\
 0 & 0 & 0 & 0 & 0 & 0 & 0 & 0 & 0 & 0 & 0 & 0 & 0 & 0 & 0 & 0 & 0
   & 0 & 0 & 0 & 0 & 0 & 0 & 0 & 0 & 0 & 0 & 0 & 0 & 1 & 0 & -1 & 0
   & 0 & 0 & 0 & -1 & 0 & 0 & 0 & 0 & 0 & 0 & 0 & 1 & 0 & 0 & 0 & 0
   & 0 & 0 & 0 & 0 & 1 & -1 & 0 \\
 0 & 0 & 0 & 0 & 0 & 0 & 0 & 0 & 0 & 0 & 0 & 0 & 0 & 0 & 0 & 0 & 0
   & 0 & 0 & 0 & 0 & 0 & 0 & 0 & 0 & 0 & 0 & 0 & 0 & 0 & 0 & 1 & 0
   & -1 & 0 & 0 & 0 & 0 & 0 & 0 & 0 & 0 & 0 & -1 & 0 & 1 & 0 & 0 &
   0 & 0 & 0 & 0 & 0 & 0 & 1 & -1 \\
 0 & 0 & 0 & 0 & 0 & 0 & 0 & 0 & 0 & 0 & 0 & 0 & 0 & 0 & 0 & 0 & 0
   & 0 & 0 & 0 & 0 & -1 & 0 & 0 & 0 & 0 & 0 & 0 & 0 & 0 & 0 & 0 & 0
   & 1 & 0 & 0 & 0 & 0 & 0 & 0 & 0 & 0 & 0 & 0 & -1 & 0 & 1 & 0 & 0
   & -1 & 0 & 0 & 0 & 0 & 0 & 1 \\
 0 & 0 & 0 & 0 & 0 & 0 & 0 & 0 & 0 & 0 & 0 & 0 & 0 & 0 & 0 & 0 & 0
   & 0 & 0 & 0 & 0 & 1 & 0 & -1 & 0 & 0 & 0 & 0 & 0 & 0 & 0 & 0 & 0
   & 0 & 0 & 0 & 0 & 0 & 0 & 0 & 0 & 0 & 0 & 0 & 0 & -1 & 0 & 1 & 0
   & 1 & -1 & 0 & 0 & 0 & 0 & 0 \\
 0 & 0 & 0 & 0 & 0 & 0 & 0 & 0 & 0 & 0 & 0 & 0 & 0 & 0 & 0 & 0 & 0
   & 0 & 0 & 0 & 0 & 0 & 0 & 1 & 0 & -1 & 0 & 0 & 0 & 0 & 0 & 0 & 0
   & 0 & 0 & 0 & 0 & 0 & 0 & 0 & 0 & 0 & 0 & 0 & 0 & 0 & -1 & 0 & 1
   & 0 & 1 & -1 & 0 & 0 & 0 & 0 \\
 0 & 0 & 0 & 0 & 0 & 0 & 0 & 0 & 0 & 0 & 0 & 0 & 0 & 0 & 0 & 0 & 0
   & 0 & 0 & 0 & 0 & 0 & 0 & 0 & 0 & 1 & 0 & -1 & 0 & 0 & 0 & 0 & 0
   & 0 & 0 & 0 & 1 & 0 & 0 & 0 & 0 & 0 & 0 & 0 & 0 & 0 & 0 & -1 & 0
   & 0 & 0 & 1 & -1 & 0 & 0 & 0 \\
 0 & 0 & 0 & 0 & 0 & 0 & 0 & 0 & 0 & 0 & 0 & 0 & 0 & 0 & 0 & 0 & 0
   & 1 & -1 & 0 & 0 & 0 & 0 & 0 & 0 & 0 & 0 & 0 & 0 & 0 & 0 & 0 & 0
   & 0 & 0 & 0 & 0 & 0 & 0 & 1 & -1 & 0 & 0 & 0 & 0 & 1 & -1 & 0 &
   0 & 0 & 1 & 0 & 0 & 0 & 0 & -1 \\
 0 & 0 & 0 & 0 & 0 & 0 & 0 & 0 & 0 & 0 & 0 & 0 & 0 & 0 & 0 & 0 & 0
   & 0 & 1 & -1 & 0 & 0 & 0 & 0 & 0 & 0 & 0 & 0 & 0 & 0 & 0 & 0 & 0
   & 0 & 0 & 0 & 0 & 0 & 0 & 0 & 1 & -1 & 0 & 0 & 0 & 0 & 1 & -1 &
   0 & -1 & 0 & 1 & 0 & 0 & 0 & 0 \\
 0 & 0 & 0 & 0 & 0 & 0 & 0 & 0 & 0 & 0 & 0 & 0 & 0 & 0 & 0 & 0 & 0
   & 0 & 0 & 1 & -1 & 0 & 0 & 0 & 0 & 0 & 0 & 0 & 0 & 0 & 0 & 0 & 0
   & 0 & 0 & 0 & 0 & 0 & 0 & 0 & 0 & 1 & -1 & 0 & 0 & 0 & 0 & 1 &
   -1 & 0 & -1 & 0 & 1 & 0 & 0 & 0 \\
 0 & 0 & 0 & 0 & 0 & 0 & 0 & 0 & 0 & 0 & 0 & 0 & 0 & 0 & -1 & 0 & 0
   & 0 & 0 & 0 & 1 & 0 & 0 & 0 & 0 & 0 & 0 & 0 & 0 & 0 & 0 & 0 & 0
   & 0 & 0 & -1 & -1 & 0 & 0 & 0 & 0 & 0 & 1 & 0 & 0 & 0 & 0 & 0 &
   1 & 0 & 0 & -1 & 0 & 1 & 0 & 0 \\
 0 & 0 & 0 & 0 & 0 & 0 & 0 & 0 & 0 & 0 & 0 & 0 & 0 & 0 & 1 & -1 & 0
   & 0 & 0 & 0 & 0 & 0 & 0 & 0 & 0 & 0 & 0 & 0 & 0 & 0 & 0 & 0 & 0
   & 0 & 0 & 1 & 1 & -1 & 0 & 0 & 0 & 0 & 0 & -1 & 0 & 0 & 0 & 0 &
   0 & 0 & 0 & 0 & -1 & 0 & 1 & 0 \\
 0 & 0 & 0 & 0 & 0 & 0 & 0 & 0 & 0 & 0 & 0 & 0 & 0 & 0 & 0 & 1 & -1
   & 0 & 0 & 0 & 0 & 0 & 0 & 0 & 0 & 0 & 0 & 0 & 0 & 0 & 0 & 0 & 0
   & 0 & 0 & 0 & 0 & 1 & -1 & 0 & 0 & 0 & 0 & 1 & -1 & 0 & 0 & 0 &
   0 & 0 & 0 & 0 & 0 & -1 & 0 & 1 \\
 0 & 0 & 0 & 0 & 0 & 0 & 0 & 0 & 0 & 0 & 0 & 0 & 0 & 0 & 0 & 0 & 1
   & -1 & 0 & 0 & 0 & 0 & 0 & 0 & 0 & 0 & 0 & 0 & 0 & 0 & 0 & 0 & 0
   & 0 & 0 & 0 & 0 & 0 & 1 & -1 & 0 & 0 & 0 & 0 & 1 & -1 & 0 & 0 &
   0 & 1 & 0 & 0 & 0 & 0 & -1 & 0 \\
\end{array}
\right)$}

\newpage

\bibliographystyle{utphys}
\bibliography{orbifoldv2}

\end{document}